\documentclass[12pt,a4paper,oneside]{book}

\usepackage[T1]{fontenc}
\usepackage{ae}
\usepackage[ansinew]{inputenc}
\usepackage{subcaption}
\usepackage{slashed}
\usepackage{amsmath}  
\usepackage{amsfonts}
\usepackage{amssymb}
\usepackage{graphicx} 
\usepackage{graphics}
\usepackage{xspace}
\usepackage{color}
\usepackage{url}
\usepackage{longtable}
\usepackage{braket}
\usepackage{bbold}
\usepackage{epigraph}
\usepackage[titletoc]{appendix}

\usepackage[normalem]{ulem}
\usepackage{color}

\newcommand{\dx}[1]{\hspace{-0.4em}\ensuremath{\mathrm{d}#1}\,}

\newcommand{\eqn}[1]{Eq.~(\ref{#1})}

\def\etab{\eta_{b}}

\def\K0bar{\overline{K^0}}
\def\bge{\begin{equation}}
\def\ene{\end{equation}}
\def\bg{\begin{eqnarray}}
\def\en{\end{eqnarray}}

\def\d0bar{{\bar{D}^0}}

\def\tbf{\textbf}

\def\ket#1{{{\left.\right|} #1\rangle}}

\newcommand{\be}{\begin{equation}}
\newcommand{\ee}{\end{equation}}
\newcommand{\bea}{\begin{eqnarray}}
\newcommand{\eea}{\end{eqnarray}}

\usepackage[english]{babel}

\begin{document}

\begin{titlepage}
  \begin{center}
    \vspace{\fill}
    \LARGE{Guilherme Novais Zeminiani}
    \vspace{3.0cm}
    \par
    \LARGE{\bf $\Upsilon$ and $\eta_{b}$ mass shifts in nuclear matter and the nucleus bound states}
    \par\vfill
    \Large{S\~ao Paulo, December 2021}
  \end{center}
\end{titlepage}


\begin{titlepage}
  \begin{center}
    \large{\textsc{Universidade Cidade de S\~ao Paulo } \\
           \textsc{Programa de P\'os-Gradua\c{c}\~ao} \\ 
           \textsc{em Astrof\'isica e F\'isica Computacional} \\
          }
    \par\vfill
    \LARGE{Guilherme Novais Zeminiani}
    \par\vfill
    \LARGE{\bf $\Upsilon$ and $\eta_{b}$ mass shifts in nuclear matter and the nucleus bound states}
    \par\vfill
    \Large{S\~ao Paulo, December 2021}
  \end{center}
\end{titlepage}


\pagenumbering{roman}
\setcounter{page}{2}


\begin{center}
\Large{\textsc{\bf Guilherme Novais Zeminiani}}

\vfill

{\bf $\Upsilon$ and $\eta_{b}$ mass shifts in nuclear matter and the nucleus bound states}

\vspace{3.0cm}

\begin{flushright}
\begin{minipage}{0.45\textwidth}

{Texto apresentado ao Programa de P\'os-gradua\c{c}\~ao em Astrof\'isica e F\'isica Computacional da Universidade Cidade de S\~ao Paulo, para Defesa de Mestrado, sob a orienta\c{c}\~ao do Dr. Kazuo Tsushima.}

\end{minipage}
\end{flushright}

\vspace{3.0cm}

\vfill

S\~ao Paulo, December 2021.

\end{center}

\newpage


\begin{center}

Guilherme Novais Zeminiani

\vspace{1.0cm}

$\Upsilon$ and $\eta_{b}$ mass shifts in nuclear matter and the nucleus bound states

\vspace{1.0cm}

\begin{flushright}
\begin{minipage}{0.45\textwidth}

{Texto apresentado ao Programa de P\'os-gradua\c{c}\~ao em Astrof\'isica e F\'isica Computacional da Universidade Cidade de S\~ao Paulo, para Defesa de Mestrado, sob a orienta\c{c}\~ao do Dr. Kazuo Tsushima.}

\end{minipage}
\end{flushright}

\vfill

\begin{flushleft}

Aprovada em Dezembro de 2021.

\end{flushleft}

\vfill

BANCA EXAMINADORA

\vfill

\hrulefill \\Prof. Dr. Kazuo Tsushima - Orientador\\UNIVERSIDADE CIDADE DE S\~AO PAULO\\

\vfill

\hrulefill \\Prof. Dr. Jo\~ao Pacheco Bicudo Cabral de Melo\\UNIVERSIDADE CIDADE DE S\~AO PAULO\\

\vfill

\hrulefill \\Prof. Dr. Jes\'us Javier Cobos-Mart\'inez\\UNIVERSIDADE DE SONORA\\

\vfill

S\~ao Paulo, December 2021.

\end{center}

\newpage


\begin{flushright}
\begin{minipage}{0.5\textwidth}

\vspace{15.0cm} 

{\it To God, source of all knowledge, to Holy Mary, throne of wisdom, and Saint Josemaria Escriv\`a,
who taught me to see the study and erudition as a way of redemption and servitude.
To my family and friends.}

\end{minipage}
\end{flushright}


\chapter*{Acknowledgements}
\addcontentsline{toc}{chapter}{Acknowledgements}

\thispagestyle{myheadings}

\noindent

My sincere thanks to Kazuo Tsushima, for his supervision
and guidance throughout these years. 
I also would like to thank Jes\'us Javier Cobos-Mart\'inez, for the useful and
helpful discussions, as well as for providing some of the
results presented in this thesis.
I would like to express my sincere gratitude to Jo\~ao Pacheco Bicudo Cabral de Melo,
Bruno Omar El-Bennich and Gilberto Ramalho for the many
inspiring lectures, advices, and for the help to improve this thesis.
I would also like to extend my thanks to Tomoi Koide for his insightful comments and suggestions.
I give many thanks to the Laborat\'{o}rio de F\'{i}sica Te\'{o}rica e Computacional (LFTC), Universidade Cidade de S\~ao Paulo (UNICID) and Universidade
Cruzeiro do Sul (UNICSUL) for offering me this opportunity.
Many thanks to my friends and family for their support.
And my thanks to CAPES for financing this project.

\newpage
\thispagestyle{empty}

\vspace*{\fill}
\epigraph{Eatur quo deorum ostenta et inimicorum iniquitas vocat. Iacta alea est.}{C. IVLII CAESARIS, \textit{Vita Divi Iuli}}


\listoffigures
\addcontentsline{toc}{chapter}{List of Figures}

\thispagestyle{myheadings}


\listoftables
\addcontentsline{toc}{chapter}{List of Tables}

\thispagestyle{myheadings}


\chapter*{List of Publications}
\addcontentsline{toc}{chapter}{List of Publications}

\begin{itemize}

\item $\varUpsilon $ and $\eta _b$ mass shifts in nuclear matter,
Eur. Phys. J. A \textbf{57}, 259 (2021).

\item[] [1]Title: $\Upsilon$ and $\eta_b$ mass shifts in nuclear matter and the $^{12}$C nucleus bound states, [2]Title: $\Upsilon$ and $\eta_b$ mass shifts in nuclear matter and the nucleus bound states,
[arXiv:2109.08636 [hep-ph]]. 

\end{itemize}

\newpage

\tableofcontents

\thispagestyle{myheadings}


\chapter*{Resumo}
\addcontentsline{toc}{chapter}{Resumo}

\thispagestyle{myheadings}

\noindent

Os desvios de massa (potenciais escalares) dos m\'esons $\Upsilon$ e $\eta_b$, assim como o 
do m\'eson $B^{*}$, s\~ao calculados pela primeira vez em mat\'eria nuclear sim\'etrica.
O ponto principal \'e saber se a for\c{c}a das intera\c{c}\~oes bottomonium-mat\'eria nuclear e 
charmonium-mat\'eria nuclear s\~ao similares ou muito diferentes, dentro de um intervalo de algumas 
dezenas de MeV \`a densidade de satura\c{c}\~ao da mat\'eria nuclear.
Isso porqu\^e, cada grupo ($\Upsilon,J/\Psi$) e ($\eta_c,\eta_b$) \'e geralmente assumido como tendo
propriedades muito similares, baseado nas massas pesadas dos quarks charm e bottom.
A estimativa para o m\'eson $\Upsilon$ \'e feita usando uma Lagrangiana efetiva em SU(5),
estudando as contribui\c{c}\~oes dos loops $BB$, $BB^{*}$ e $B^{*}B^{*}$
\`a suas auto-energias no v\'acuo e em meio nuclear.
Como resultado, apenas a contribui\c{c}\~ao do loop $BB$ \'e incluida como nossa previs\~ao m\'inima
para o m\'eson $\Upsilon$.
As massas em meio nuclear dos m\'esons $B$ e $B^{*}$ presentes nos loops da auto-energia s\~ao
calculadas pelo modelo de acoplamento m\'eson-quark.
Fatores de forma s\~ao usados para regularizar as integrais das auto-energias, com uma ampla
gama de valores para as massas de corte.
Uma an\'alise detalhada das contribui\c{c}\~oes dos loops $BB$, $BB^{*}$ e $B^{*}B^{*}$ ao
desvio de massa do m\'eson $\Upsilon$ \'e feita comparando-as com suas respectivas contribui\c{c}\~oes
correspondentes dos loops $DD$, $DD^{*}$ e $D^{*}D^{*}$ ao desvio de massa do m\'eson $J/\Psi$.
Baseado na an\'alise feita para o m\'eson $\Upsilon$, a previs\~ao para o desvio de massa do m\'eson
$\eta_b$ \'e feita nos mesmos moldes do m\'eson $\Upsilon$, isto \'e, incluindo apenas o loop m\'inimo $BB^{*}$.
O desvio de massa do m\'eson $\Upsilon$ \'e previsto de estar entre -16 a -22 MeV \`a densidade de satura\c{c}\~ao
da mat\'eria nuclear, com os valores da massa de corte no intervalo de 2000 - 6000 MeV usando a constante
de acoplamento de $\Upsilon BB$, determinada pelo modelo de domin\^ancia de m\'eson vetorial com dados experimentais,
enquanto o desvio de massa do m\'eson $\eta_b$ \'e previsto entre -75 a -82 MeV, com a constante de acoplamento
universal para SU(5) determinada pela constante de acoplamento de $\Upsilon BB$ para o mesmo intervalo de valores
das massas de corte.
Os resultados sugerem que ambos, $\Upsilon$ e $\eta_b$, devem formar estados ligados com diversos n\'ucleos
considerados neste estudo, para os quais foram calculadas as energias de liga\c{c}\~ao   
$\Upsilon$-n\'ucleo e $\eta_b$-n\'ucleo.
Os resultados tamb\'em mostram uma diferen\c{c}a consider\'avel entre a for\c{c}a das intera\c{c}\~oes
bottomonium-mat\'eria nuclear e charmonium-mat\'eria nuclear.
Tamb\'em s\~ao estudados os desvios de massa de $\Upsilon$ e $\eta_b$ em uma simetria de quarks pesados
(m\'esons pesados), ou seja, calculando seus desvios de massa usando a mesma constante de acoplamento
usada para estimar os desvios de massa dos m\'esons $J/\Psi$ e $\eta_c$.
Para o desvio de massa de $\eta_b$, o caso de quebra da simetria SU(5) tamb\'em \'e estudado nesse limite.
As previs\~oes para esses casos \`a densidade de mat\'eria nuclear s\~ao -6 a -9 MeV para $\Upsilon$,
-31 a -38 MeV para $\eta_b$ e -8 a -11 MeV para $\eta_b$ com a simetria SU(5) quebrada, onde os respectivos
correspondentes no setor de charm s\~ao -5 a -21 MeV para $J/\Psi$, -49 a -87 MeV para $\eta_c$ e -17 a -51 MeV para
$\eta_c$ com a simetria SU(4) quebrada.
Al\'em disso, um estudo inicial foi feito para investigar a influ\^encia da escolha do fator de forma
nas nossas previs\~oes. N\'os testamos um fator de forma diferente, mais sens\'ivel \`a massa de corte,
mas que retorna um desvio de massa similar \`aqueles tomados como nossas previs\~oes.

Palavras-chave: F\'isica Hadr\^onica; Estrutura Hadr\^onica; Quark-Gl\'uon Plasma e Mat\'eria Hadr\^onica; Mat\'eria nuclear e de Quarks.   


\chapter*{Abstract}
\addcontentsline{toc}{chapter}{Abstract}

\thispagestyle{myheadings}

\noindent

The $\Upsilon$ and $\eta_b$ as well as $B^*$ meson  
mass shifts (scalar potentials) are estimated for the first time
in symmetric nuclear matter. 
The main interest is, whether or not the strengths of the 
bottomonium-nuclear matter and charmonium-nuclear matter  
interactions are similar or very different, in the range of a few tens of MeV 
at the nuclear matter saturation density. This is because, each ($\Upsilon,J/\Psi$) and 
($\eta_c,\eta_b$) meson group is usually assumed to have very similar properties 
based on the heavy charm and bottom quark masses.
The estimate for the $\Upsilon$ is made using an SU(5) effective Lagrangian density, 
by studying the $BB$, $BB^*$, and $B^*B^*$ meson loop contributions  
for the self-energy in free space and in nuclear medium. 
As a result, only the $BB$ meson loop contribution 
is included as our minimal prediction.
As for the $\eta_b$, is included only the $BB^*$ meson loop contribution  
in the self-energy, to be consistent with the minimal prediction 
for the $\Upsilon$.
The in-medium masses of the $B$ and $B^{*}$ mesons appearing in the  
self-energy loops are calculated by the quark-meson coupling model. 
Form factors are used to regularize the loop integrals with 
a wide range of the cutoff mass values. 
A detailed analysis on the $BB$, $BB^{*}$, and $B^{*}B^{*}$ meson 
loop contributions for the $\Upsilon$ mass shift is made by comparing with
the respectively corresponding $DD, DD^*$, and $D^*D^*$ meson loop contributions 
for the $J/\Psi$ mass shift. Based on the analysis for the $\Upsilon$, 
the prediction for the $\eta_b$ mass shift is made on the same footing 
as that for the $\Upsilon$, namely including only the minimal $BB^*$ meson loop. 
The $\Upsilon$ mass shift is predicted to be -16 to -22 MeV at the nuclear 
matter saturation density with the cutoff mass values in the range of 2000 - 6000 MeV  
using the $\Upsilon BB$ coupling constant determined by the vector 
meson dominance model with the experimental data, 
while the $\eta_b$ mass shift is predicted to be -75 to -82 MeV with 
the SU(5) universal coupling constant determined by the $\Upsilon BB$ 
coupling constant for the same range of the cutoff mass values.
The results suggest that both $\Upsilon$ and $\eta_b$ should form bound states with
a variety of nuclei considered in this study, for which the $\Upsilon$-nucleus and 
$\eta_b$-nucleus bound state energies are calculated.
The results also show an appreciable difference between the 
bottomonium-nuclear matter and charmonium-nuclear matter interaction strengths. 
Are also studied the $\Upsilon$ and $\eta_b$ mass shifts in a 
heavy quark (heavy meson) symmetry limit, namely, by calculating 
their mass shifts using the same coupling constant value that was used 
to estimate the $J/\Psi$ and $\eta_c$ mass shifts. 
For the $\eta_b$ mass shift an SU(5) symmetry breaking case is also studied in this limit. 
The predictions for these cases at nuclear matter saturation density are, 
-6 to -9 MeV for $\Upsilon$, -31 to -38 MeV for $\eta_b$, 
and -8 to -11 MeV for $\eta_b$ with a broken SU(5) symmetry, 
where the corresponding charm sector ones are, 
-5 to -21 MeV for $J/\Psi$, -49 to -87 MeV for $\eta_c$, and 
-17 to -51 MeV for $\eta_c$ with a broken SU(4) symmetry.
In addition, an initial study was done to investigate the influence of the choice of the form factor
on our predictions. We have tested a different form factor, which is more sensitive to
the cutoff mass value, but gives mass shifts similar to those regarded as our predictions.

Keywords: Hadron Physics; Hadron Structure; Quark-Gluon Plasma and Hadronic Matter; Nuclear and Quark Matter.


\pagebreak
\pagenumbering{arabic}


\chapter{Introduction}

\thispagestyle{empty}

\noindent

The 12 GeV upgrade of the Continuous Electron Beam Accelerator Facility (CEBAF) at the Jefferson Lab made it possible to produce low-momentum heavy-quarkonia in an atomic nucleus. 
In a recent experiment~\cite{Ali:2019lzf}, a photon beam was 
used to produce a $J/\Psi$ meson near-threshold, which was identified by the decay into an 
electron-positron pair. Also with the construction of the Facility for Antiproton and Ion Research 
(FAIR) in Germany, heavy and heavy-light mesons will be produced copiously by the annihilation of antiprotons on nuclei~\cite{Durante:2019hzd}.  

The production of heavy quarkonium in nuclei is one of the most useful methods for studying the
interaction of the heavy quarkonium with nucleon, in particular,
for probing its gluonic properties. 
We can, thus advance in understanding the hadron properties and their interactions based on 
quantum chromodynamics (QCD).  
Since the heavy quarkonium interacts with nucleon primarily via gluons, 
its production in a nuclear medium can be of great relevance 
to explore the roles of gluons. 

In the past few decades, many attempts 
were made~\cite{Hosaka:2016ypm,Krein:2016fqh,Metag:2017yuh,Krein:2017usp}  
to find alternatives to the meson-exchange mechanism for 
the (heavy-quarkonium)-nucleon interaction.
Some works employed charmed meson 
loops~\cite{Ko:2000jx,Krein:2010vp,Tsushima:2011kh,Tsushima:2011fg,Krein:2013rha}, 
others were based on QCD sum 
rules~\cite{Klingl:1998sr,Hayashigaki:1998ey,Kim:2000kj,Kumar:2010hs},  
phenomenological potentials~\cite{Belyaev:2006vn,Yokota:2013sfa}, 
the charmonium color polarizability~\cite{Peskin:1979va,Kharzeev:1995ij}, 
and van der Waals type 
forces~\cite{Ko:2000jx,Kaidalov:1992hd,Luke:1992tm,deTeramond:1997ny,Brodsky:1997gh,
Sibirtsev:2005ex,Voloshin:2007dx,TarrusCastella:2018php}.

Furthermore, lattice QCD simulations for charmonium-nucleon 
interaction in free space were performed in the last 
decade~\cite{Yokokawa:2006td,Liu:2008rza,Kawanai:2010ev,Kawanai:2010ru,Skerbis:2018lew}.
More recently, studies for the binding of charmonia with 
nuclear matter and finite nuclei, 
as well as light mesons and baryons, were performed in lattice QCD 
simulations~\cite{Beane:2014sda,Alberti:2016dru}. 
These simulations, however, used unphysically heavy pion masses.

In addition, medium modifications of charmed and bottom hadrons were 
studied in 
Refs.~\cite{Sibirtsev:1999js,Sibirtsev:1999jr,Tsushima:1998ru,Tsushima:2002cc,Tsushima:2002sm,
Tsushima:2003dd} based on the quark-meson coupling (QMC) model~\cite{Guichon:1987jp}, 
on which we will partly rely in this study.
The QMC model is a phenomenological, but very successful quark-based relativistic mean 
field model for nuclear matter, nuclear structure, and hadron properties 
in a nuclear medium. 
The model relates the relativistically moving  
confined light $u$ and $d$ quarks in the nucleon bags with the scalar-isoscalar ($\sigma$), 
vector-isoscalar ($\omega$), and vector-isovector ($\rho$) mean fields self-consistently 
generated by the light quarks in the nucleons~\cite{Guichon:1987jp}.
The QMC model assumes that the mean fields do not couple to the $s$ quark,
because the density dependence of the $s$-quark condensate is much smaller than that of the $u$ 
and $d$ quarks, in the sense that the $s$-quark condensate changes very little as baryon density increases, by virtue of the heavier current mass of the $s$ quark compared to the quarks $u$ and 
$d$~\cite{Gubler:2018ctz}. 

Based on the $D$ and $D^{*}$ meson mass modifications 
in symmetric nuclear matter calculated by the QMC model, 
the mass shift of $J/\Psi$ meson was predicted to be -16 to -24 MeV~\cite{Krein:2010vp} 
at the symmetric nuclear matter saturation density ($\rho_0 = 0.15$ fm$^{-3}$).
However, because of the unexpected contribution from the heavier $D^*D^*$ meson loop  
for the $J/\Psi$ self-energy, the authors G.~Krein, A.~W.~Thomas and K.~Tsushima, updated the prediction for the $J/\Psi$ mass shift by including only the $DD$ meson loop~\cite{Krein:2017usp}. 
This gives the prediction of -3.0 to -6.5 MeV downward shift of the $J/\Psi$ mass at $\rho_0$.
In the QMC model the internal structure of hadrons changes in medium    
by the strong nuclear mean fields directly interacting with 
the light quarks $u$ and $d$ (in this thesis, light quarks mean only $u$ and $d$, but not the strange 
(s) quark), the present case in $D$ and $D^*$ mesons, 
and the dropping of these meson masses enhances  
the self-energy of $J/\Psi$ more than that in free space, 
resulting in an attractive $J/\Psi$-nucleus potential  
(negative mass shift)~\cite{Tsushima:2011kh}.

As for the $\eta_c$ meson, experimental studies of the production 
in heavy ion collisions at the LHC were  
performed~\cite{Aaij:2019gsn,Tichouk:2020dut,Tichouk:2020zhh,Goncalves:2018yxc,Klein:2018ypk}.
However, nearly no experiments were aimed to produce the $\eta_c$ 
at lower energies, probably due to the difficulties to perform experiment.
Furthermore, only recently the in-medium properties of $\eta_c$ meson were renewed 
theoretically~\cite{Cobos-Martinez:2020ynh}.

When it comes to the bottomonium sector on which we focus here, 
studies were made for $\Upsilon$ photoproduction at EIC 
(Electron-Ion Collider)~\cite{Xu:2020uaa,Gryniuk:2020mlh},  
$\Upsilon$ production in $p$Pb collisions~\cite{Aaij:2018scz}, 
and $\Upsilon(nl)$ (excited state) decay into $ B^{(*)} \bar B^{(*)}$~\cite{Liang:2019geg}. 
By such studies, we can further improve our understanding of the heavy 
quarkonium properties.
QCD predicts that chiral symmetry would be partially restored in a nuclear medium, and 
the effect of the restoration is expected to change the properties of hadrons in medium, 
particularly those hadrons that contain nonzero light quarks $u$ and $d$, 
because the reductions of the light quark $u$ and $d$ condensates are expected to be 
faster than those of the heavier quarks as nuclear density increases.  
Thus, usually the light quark condensates are regarded as the order parameters of the 
(dynamical) chiral symmetry. 
Some studies support the faster reduction of the light-quark condensates in medium as nuclear 
density increases than those of the heavier quarks: (i) based on the NJL 
model~\cite{Tsushima:1991fe,Maruyama:1992ab} 
for the light and strange quark condensates in nuclear matter (more comment on the in-medium strange quark condensate will be given on page 4), 
and (ii) the result that the heavy quark condensates are proportional 
to the gluon condensate obtained by 
the operator product expansion~\cite{Shifman:1978bx} and also by a world-line effective 
action-based study~\cite{Antonov:2012ud}, together with the result of the model independent 
estimate that the gluon condensate at nuclear matter saturation density decreases only 
about 5\% by the QCD trace anomaly and Hellman-Feynman theorem~\cite{Cohen:1991nk}. 

The frequently considered interactions between the heavy quarkonium and 
the nuclear medium are QCD van der Waals type 
interactions~\cite{Ko:2000jx,Kaidalov:1992hd,Luke:1992tm, 
deTeramond:1997ny,Brodsky:1997gh,Sibirtsev:2005ex,Voloshin:2007dx, 
TarrusCastella:2018php}. Naively, this must occur by the exchange of gluons 
in the lowest order, since heavy quarkonium has no light quarks,   
whereas the nuclear medium is composed of light quarks, and thus the light-quark or light-flavored 
hadron exchanges do not occur in this order. 
Another possible mechanism for the heavy quarkonium interaction with the nuclear medium 
is through the excitation of the intermediate state hadrons which contain light quarks. 

One of the simple, but interesting questions may be, 
whether or not the strengths of the charmonium-(nuclear matter)  
and bottomonium-(nuclear matter) interactions are indeed similar, 
since one often expects the similar properties of charmonium and bottomonium  
based on the heavy charm and bottom quark masses. 

\begin{figure}[htb]
\centering 
  \includegraphics[scale=0.9]{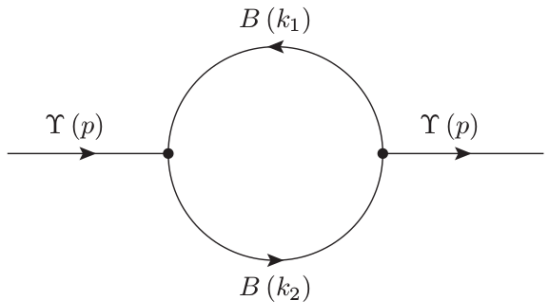}
 \caption{BB meson loop contribution for the $\Upsilon$ self-energy.}
 \label{fig1}
\end{figure}

In this thesis, after calculating the in-medium $B$ and $B^*$ meson masses, 
we estimate first the mass shift of $\Upsilon$ meson in terms of 
the excitations of intermediate state hadrons with light quarks  
in the self-energy. As an example we show in 
Fig.~\ref{fig1} the $BB$ meson loop contribution for 
the $\Upsilon$ self-energy --- we will also study the $BB^*$ and $B^*B^*$ meson loop contributions.
Next, we also estimate the mass shift of the pseudoscalar 
quarkonium, $\eta_b$ meson, which is the lightest $b\bar{b}$ bound state. 
The estimates will be made using an SU(5) effective 
Lagrangian density (hereafter we will denote simply by ''Lagrangian'') 
which contains both the $\Upsilon$ and $\eta_b$ mesons with one universal coupling constant, 
and the anomalous coupling one respecting an SU(5) 
symmetry in the coupling constant.
Then, the present study can also provide information 
on the SU(5) symmetry breaking. 

Upon expanding the SU(5) effective Lagrangian with minimal substitutions, 
we get the interaction Lagrangians for calculating the $\Upsilon$  
self-energy, i.e., the $BB$, and $B^*B^*$ meson loops   
as well as for the the $\eta_b$ self-energy, the $BB^*$ meson loop 
($\Upsilon BB^*$ and $\eta_bB^*B^*$ interaction Lagrangians are  
anomalous coupling ones,  
not derived from the SU(5) effective Lagrangian). 
Thus, we need to have better knowledge on the in-medium properties (Lorentz-scalar  
and Lorentz-vector potentials) of the $B$ and  $B^{*}$ mesons.
For this purpose we use the QMC model invented by Guichon~\cite{Guichon:1987jp},  
which has been successfully applied for various 
studies~\cite{Krein:2017usp,Krein:2010vp,Tsushima:2002cc,Guichon:1995ue,Tsushima:1997df, 
Saito:1996sf,Tsushima:2019wmq,Saito:2005rv}.
Note that, the in-medium $B^*$ meson mass is estimated and presented for the first 
time in this study, calculated by the QMC model. 

We analyze the $BB, BB^*$ and $B^*B^*$ meson loop contributions    
for the $\Upsilon$ self-energy. After a detailed analysis, 
our predictions for the $\Upsilon$ and $\eta_b$ mass shifts
are made by including only the lowest order $BB$ meson loop contribution for 
the $\Upsilon$, 
and only the $BB^*$ meson loop contribution for the $\eta_b$,  
where the in-medium masses of the $B$ and $B^*$ mesons are calculated by the QMC model.
In addition, a detailed comparison is made between the $\Upsilon$ and $J/\Psi$ meson self-energies, 
in order to get a better insight into the cutoff mass values used in the form factors, 
as well as the form factors themselves.

This thesis is organized as follows.
In Chap.~\ref{Chapter1} some important and basic knowledge that is going to be needed
throughout the next chapters.
We then discuss in Chap.~\ref{Chapter2} possible mechanisms for the production of heavy-quarkonia, 
as well as the possibilities of experimental productions.
The production of a heavy-quarkonium vector meson is described by both photo- and electro-production,
with the former process being presented by two different models.
While the pseudo-scalars $\eta_c$ and $\eta_b$ mesons being covered in another section, with their production described as a production of a quark-antiquark pair.
Chap.~\ref{Chapter3} covers the quark-meson coupling (QMC) model. It contains the model's description
of nuclear matter and its properties, as well as the properties of hadrons in nuclear matter and
the description of finite nuclei. In addition, we estimate in this chapter the $B$ and $B^*$ mesons 
in symmetric nuclear matter.
Next, in Chap.~\ref{Chapter4} we estimate the $\Upsilon$ meson mass shift in nuclear matter and compare
the results with the ones for the $J/\Psi$ mass shift. The $\Upsilon$-nucleus potentials for various nuclei are then obtained and used to calculate the $\Upsilon$-nucleus bound state energies.
The Chap.~\ref{Chapter5} is organized in a similar way, presenting the $\eta_b$ mass shift in nuclear matter, its nuclear potentials and bound state energies for the same set of nuclei as that for $\Upsilon$.
In Chap.~\ref{Chapter6} we consider a heavy quark (heavy meson) symmetry limit 
for the $\Upsilon$ and $\eta_b$, and a broken SU(5) symmetry for the $\eta_b$ 
in this limit, and also give predictions for these cases.
In addition, We perform an initial study for the effects of the form factor  
on the $\Upsilon$ and $\eta_b$ mass shifts using a different form factor. 
Lastly, summary and conclusion are given in Chap.~\ref{Chapter7}.


\chapter{Building blocks} 
\label{Chapter1} 
\thispagestyle{empty} 

\noindent

\section{Standard Model}
\label{Ch1Sc1}

\thispagestyle{myheadings}

\noindent

The basic and elementary constituents of the visible matter in the Universe are described by
the Standard Model (SM), a theory of interacting fields that divides the elementary particles in
two main groups: Fermions with half-integer spins, which are the fundamental building blocks of ordinary matter, and bosons with integer spins, particles which mediate the interactions among fields.

These fields are of four types, being the Electromagnetic, Weak, Strong and Gravitational fields.
The electromagnetic interaction acts between electrically charged elementary particles, with the mediated field quanta being the photons. 
As for the weak interaction, the quanta of the exchanged fields are the $W^{+}$ and  $W^{-}$
charged bosons, and $Z$, which is electrically neutral. Unlike the Electromagnetic field, which has
a massless field quanta, the weak interaction is short ranged due to the mass of the bosons exchanged.
Regarding the quanta of the strong interaction field, the gluons (g), just like the photons they are assumed to be massless, but unlike the electromagnetic force it has a very short range.
Gluons and quarks, which interact through gluons, are confined in a finite region under a normal condition.
In addition to the four vector fields mentioned above, the SM also describes a building block particle, called the Higgs boson. The masses of elementary particles in the SM are induced through the interaction with the Higgs boson.
Since the gravitational forces are of insignificant influence on the scales of particle physics, the Standard Model excludes it from the framework.
Furthermore, it has not yet been successfully quantized.

The fermions, on the other hand, comes in two different types: quarks and leptons. 
While quarks can interact via all the three forces mentioned (ignoring gravitational force), the leptons can only interact through the electromagnetic and weak interactions.

There are six types of quarks $(q)$ called flavors: up $(u)$, down $(d)$, charm $(c)$, top $(t)$ and bottom $(b)$ and their antiparticles are called antiquarks $(\overline{q})$. 
They combine themselves together to form more complex structures known as hadrons. 
Combinations of three quarks $3q$ forms barions, while combinations of two $q\overline{q}$ forms mesons. 
Quarks are not observed in isolated states in nature, although tetraquarks $2q2\overline{q}$ and pentaquarks $3q(q\overline{q})$ are admitted and confirmed experimentally. Some examples of observed
tetra and pentaquarks are the $Z$ and $P_{c}$ states.

\begin{figure}[htb]
\centering 
  \hspace{-2cm}  
  \includegraphics[scale=0.5]{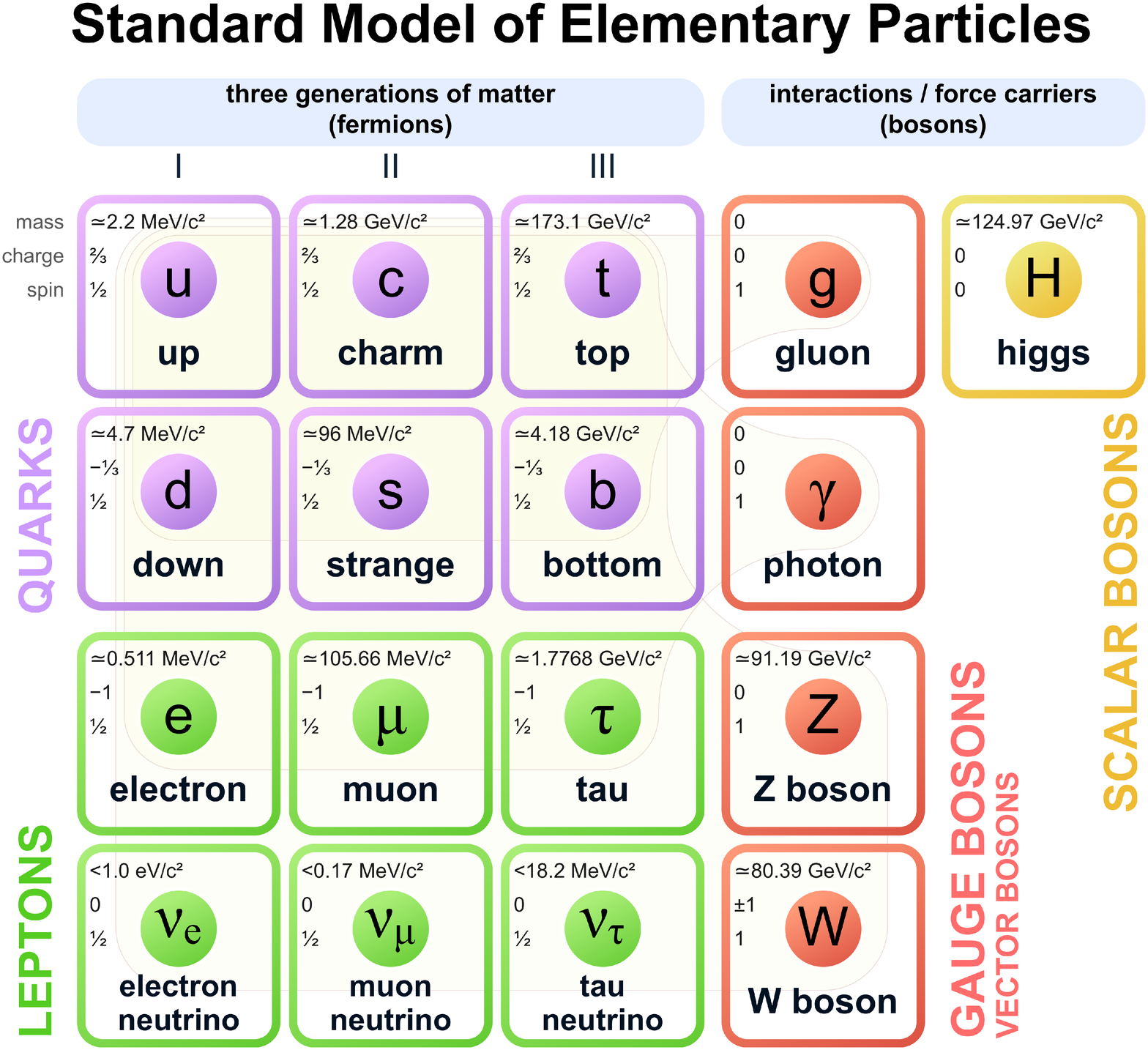}
 \caption{Particles of the Standard Model with their corresponding masses, charges and spin.}
 \label{smtab}
\end{figure}

The particles of the SM are presented, with their masses, charges and spin in Fig.~\ref{smtab}.
\newpage

\newpage
\section{Feynman diagram loops}
\label{Ch1Sc2}

\noindent

\begin{figure}[!htb]
 \includegraphics[scale=1.3]{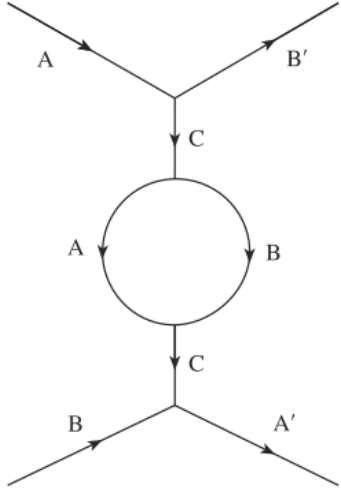}
 \caption{Example of a Feynman diagram with first order loops in the process
$A + B \rightarrow A+B$ in a propagator $C$ in free space.
The particles $A$ and $B$ involved in the scattering process have
masses $m_A$ and $m_B$, respectively, with incoming four-momenta
$P_A$ and $P_B$, and outgoing four-momenta $P^{'}_{A}$ and $P^{'}_{B}$.
The exchanged particle $C$ have mass $m_C$, and four-momentum
$q = P_A - P^{'}_{B} = P^{'}_{A} - P_B$. Since $q$ flows in and out of the loop,
it os not specified how it is to be shared between $A$ and $B$. So if
we say that the four-momentum of $A$ is $k$, then the one of $B$ will
be $q - k$. For the definitions of the four-momentum see appendix~\ref{apx-conv}.}
 \label{figl}
\end{figure}

Loops arise due to the interactions, and modify the propagator, as shown in a Feynman diagram 
in~\ref{figl}.
Considering the process $A + B \rightarrow A' + B'$, representing the scattering of spinless
arbitrary particles with a propagator $C$. The lowest order loop in this propagator occur 
as shown in Fig.~\ref{figl}.
 
\begin{figure}[!htb]
 \includegraphics[width=\linewidth]{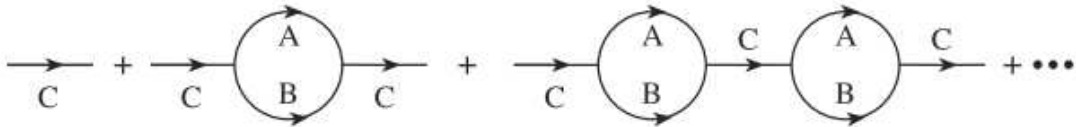}
 \caption{Series of loops in the propagator $C$.}
 \label{figls}
\end{figure}

The corresponding amplitude for this diagram is~\cite{Aitchison:2003tq}
\begin{equation}
\left( -ig \right) ^{2} \left( 2 \pi  \right) ^{4} \delta ^{4} \left( P_{A}^{'}+P_{B}^{'}-P_{A}-P_{B} \right)
\frac{i}{q^{2}-m_{C}^{2}+i \epsilon } \left( -i \Sigma  
 \left( q^{2} \right)  \right) \frac{i}{q^{2}-m_{C}^{2}+i \epsilon },
\label{ampli}
\end{equation}
where $g$ is the coupling constant, the two terms $\frac{i}{q^{2}-m_{C}^{2}+i \epsilon }$ represent the two propagators $C$ and $-i \Sigma \left( q^{2} \right)$ is the term 
associated with the loop, which is given by 
\begin{equation}
-i \Sigma \left( q^{2} \right) = \left( -ig \right) ^{2} \int _{}^{}\frac{d^{4}k}{ \left( 2 \pi  \right) ^{4}}\frac{i}{k^{2}-m_{A}^{2}  +i \epsilon }\frac{i}{ \left( q-k \right) ^{2}-m_{B}^{2}+i \epsilon },
\end{equation}
with $k$ being the momentum of $A$ and $q-k$ the one of $B$.
A new term
\begin{equation} 
\frac{i}{q^{2}-m_{C}^{2}+i \epsilon } \left( -ig^{2} \Sigma \left( q^{2} \right)\right) \frac{i}{q^{2}-m_{C}^{2}+i \epsilon },
\end{equation}
is added at the amplitude for each new loop in the propagator.
Considering a series of loops in the propagator $C$, as shown in Fi.~\ref{figls}, the form of the
propagator will be
\begin{equation}
\frac{i}{q^{2}-m_{C}^{2}+i \epsilon } \left( 1+r+r^{2}+ \ldots  \right),
\end{equation}
with
\begin{equation}
r=\frac{ g^{2} \Sigma \left( q^{2} \right) }{q^{2}-m_{C}^{2}+i \epsilon },
\end{equation}
where $\Sigma \left( q^{2} \right)$ is called the irreducible self-energy.

\section{Quantum chromodynamics}
\label{Ch1Sc3}

\noindent

In this section we shall introduce basic common knowledge of Quantum Chromodynamics,
in a way to be minimally self-contained.

Aside from the flavor, explained in Sec.~\ref{Ch1Sc1}, the quarks also posses another property called color charge.
Quarks come in three primary colors: red ($R$), green ($G$) and blue ($B$). In the same way,
antiquarks come in anticolors called antired ($\overline{R}$), antigreen ($\overline{G}$), 
and antiblue ($\overline{B}$). Each quark (or antiquark) can take only one of three values or charges.

Gluons, on the other hand, have their color charge constitued by a mixture of two colors 
(or anticolors).
They can either be in a color singlet state

\begin{equation}
\label{singlu}
(R \overline{R} + G \overline{G} + B \overline{B})/\sqrt{3},
\end{equation}
or in a color octet 

\begin{eqnarray}
&&(R \overline{G} + G \overline{R})/\sqrt{2}
\qquad
-i(R \overline{G} - G \overline{R})/\sqrt{2}
\nonumber \\
&&(R \overline{B} + B \overline{R})/\sqrt{2}
\qquad
-i(R \overline{B} - B \overline{R})/\sqrt{2}
\nonumber \\
&&(G \overline{B} + B \overline{G})/\sqrt{2}
\qquad
-i(G \overline{B} - B \overline{G})/\sqrt{2}
\nonumber \\
&&(R \overline{R} - G \overline{G})/\sqrt{2}
\qquad
(R \overline{R} + G \overline{G} - 2 B \overline{B})/\sqrt{6},
\end{eqnarray}
where

\begin{equation}
R =  
\begin{pmatrix}
1 \\
0 \\
0 \\
\end{pmatrix},
\qquad
G = 
\begin{pmatrix}
0 \\
1 \\
0 \\
\end{pmatrix},
\qquad
B = 
\begin{pmatrix}
0 \\
0 \\
1 \\
\end{pmatrix}.
\end{equation}

But the strong force is of very short range, only eight kinds of gluons exist, excluding the singlet gluon of Eq.~(\ref{singlu}), because if a singlet gluon exists, it would be exchanged in a long-range interaction~\cite{Griffiths:2008zz}.
In the representation of the SU(3) group, explained in Appendix~\ref{apx-gluon},
there are $3^{2} - 1 = 8$ generators, so 
these eight color states are equivalent to the Gell-Mann matrices.
\subsection{QCD Lagrangian}
\label{Ch1Sc2-1}

\noindent

We begin writing down the QCD Lagrangian as~\cite{Fayyazuddin:2012qfa}
\begin{equation}
{\cal L}_{QCD} = {\cal L}_{q} + {\cal L}_{Q} + {\cal L}_{g},
\end{equation}
where ${\cal L}_{q}$ contains the light-flavor quark Dirac fields
$q = \left(u, d, s\right)^{T}$, ${\cal L}_{Q}$ the heavy-flavor quark Dirac fields
$Q = \left(c, b, t\right)^{T}$ and ${\cal L}_{g}$ the pure-gluon part, with the lagrangians
being
\begin{eqnarray}
{\cal L}_{q} =&& \overline{q}(x)\left(i \slashed{D} - m_{q}\right)q(x), \nonumber \\
{\cal L}_{Q} =&& \overline{Q}(x)\left(i \slashed{D} - m_{Q}\right)Q(x), \nonumber \\
{\cal L}_{g} =&& -\frac{1}{4} G^{a}_{\mu \nu}(x) G^{a \mu \nu}(x).
\end{eqnarray}
with $\slashed{D} = \gamma^{\mu} D_{\mu}$, where $D_{\mu}$ is the gauge-covariant derivative
\begin{equation}
D_{\mu} = \partial_{\mu} + ig A_{\mu}(x),
\end{equation}
and $G^{a}_{\mu \nu}$ is the gluon field strength tensor
\begin{equation}
G^{a}_{\mu \nu}(x) = \partial_{\mu}A^{a}_{\mu}(x) - g f^{abc}A^{b}_{\mu}(x) A^{c}_{\nu}(x),
\end{equation}
where $A_{\mu}(x)$ is the gluon field in the adjoint representation of SU(3) (explained in appendix~\ref{apx-gluon}).
The gluon field represents the gluon propagation in the strong interaction between quarks,
being given by~\cite{Aitchison:2004cs}

\begin{equation}
 \ A_{ \mu } \left( x \right) =T^{a}A_{ \mu }^{a} \left( x \right),
\end{equation}

with each $A_{ \mu }^{a}$ being a vector field and the index $a$ representing each of the
eight possible combination of colors for the gluons
$a=1, 2, \ldots , 8$.

The light and heavy quark mass matrices are 
\begin{equation}
m_q = \begin{pmatrix}
m_u & 0 & 0\\
0 & m_d & 0\\
0 & 0 & m_s
\end{pmatrix}, \qquad
m_Q=\begin{pmatrix}
m_c & 0 & 0\\
0 & m_b & 0\\
0 & 0 & m_t
\end{pmatrix}.       
\end{equation}


\subsection{Chiral symmetry breaking}
\label{Ch1Sc2-2}

\noindent

The mass scale $\Lambda_{QCD}$ is the value at which the theory becomes strongly coupled and 
nonperturbative and its value depends on the renormalization scheme used and on the number of active flavors (when the energy-momentum involved in the process allows to make active only a certain number of flavors, which means that heavier flavors will be suppressed even including the virtual state excitations).
A common choice for the renormalization scheme is the Modified Minimal Subtraction scheme 
($\overline{\text{MS}}$). 
For the $\overline{\text{MS}}$ scheme, the values (in MeV) for different flavor numbers $N_{f}$ are

\begin{equation}
  \Lambda_{QCD} =
    \begin{cases}
      332~\text{for}~N_{f} &= 3\\
      292  &= 4\\
      210  &= 5\\
       89  &= 6.
    \end{cases}       
\end{equation}

Since the masses of the light quarks $u$, $d$ and $s$ are small compared with $\Lambda_{QCD}$ when
just tree flavors are active (which is the case we consider), we may consider an approximation to QCD where the masses of these quarks are set to zero and do perturbation theory in $m_{q}$ about this limit. The limit $m_{q} \rightarrow 0$ is known as the chiral limit, where the light quark Lagrangian can be rewritten in terms of the operators
$q_{L}(x) = 1/2 \left( 1 - \gamma_{5} \right) q(x)$ and
$q_{R}(x) = 1/2 \left( 1 + \gamma_{5} \right) q(x)$, 
(see appendix~\ref{apx-conv} for definitions on the gamma matrices) being given by~\cite{Manohar:2000dt}

\begin{equation}
{\cal L}_{q} = \overline{q}(x) i \slashed{D} q(x) = \overline{q}_{L}(x) i \slashed{D} q_{L}(x) 
+ \overline{q}_{R}(x) i \slashed{D} q_{R}(x),
\end{equation}
and has a $\text{SU}(3)_{L} \times \text{SU}(3)_{R}$ chiral symmetry under which the right and left-handed 
quark fields transform differently. Also, this Lagrangian is invariant under unitary transformations of the type

\begin{equation}
\label{eqtra}
q_{L}(x) \rightarrow q'_{L} = L q_{L}(x), \qquad q_{R}(x) \rightarrow q'_{R} = R q_{R}(x),
\end{equation}
where $L$ and $R$ are independent SU(2) unitary transformation matrices satisfying
$L^{\dagger} L = R^{\dagger} R = 1$.

The dynamical breaking of the $\text{SU}(3)_{L} \times \text{SU}(3)_{R}$ symmetry can be written in terms of the nonzero value of the vacuum expectation value (v.e.v.) 
$\braket{\overline{q}^{f}(x) q^{f'}(x)}$ known as the quark condensate

\begin{equation}
\left\langle \overline{q}^{f}_{R}(x)q^{f'}_{L}(x) + \overline{q}^{f}_{L}(x)q^{f'}_{R}(x) \right\rangle = B \delta^{ff'},
\end{equation}
where $\overline{q}^{f}(x) q^{f'}(x) = \overline{q}^{f}_{R} q^{f'}_{L}(x) + \overline{q}^{f}_{L} q^{f'}_{R}(x)$,
$B \sim \Lambda^{3}_{QCD}$ and $f$ and $f'$ are flavor indices. A nonzero value of the condensate
implies a dynamically generated quark mass, since the dynamical mass is a function of the quark condensate~\cite{Politzer:1976tv}.


\chapter{Heavy-Quarkonia production} 

\label{Chapter2} 

\thispagestyle{empty} 

\noindent

This chapter aims to describe the production of heavy quarkonia
by the models of photo-production featured in~\cite{Xu:2020uaa}.
The process $\gamma p \rightarrow (Q\overline{Q}) p$ can be explained by
various models, of which we focus here on the photon-gluon fusion (PGF) model
and the pomeron exchange (PMEX) model, to be explained below. 
A separate section is dedicated to treat the specific case of $\eta_{c,b}$ production. 
This chapter also covers the estimate of heavy quarkonia
production at an Electron-Ion Collider (EIC) and a brief mention of the possible
production of heavy and heavy-light mesons at the Facility for Antiproton and Ion Research (FAIR).


\section{Photo-production}
\label{Ch2Sc1}

\thispagestyle{myheadings}

\noindent

The photo-production of heavy quarkonia is a process where a bound $Q\overline{Q}$ pair is produced by 
the interaction of a proton with a high energy photon $\gamma p \rightarrow (Q\overline{Q}) p$. 
Since the proton is a complex system of valence quarks (the ones that contribute to a hadron's quantum numbers), sea quarks (which are virtual quark-antiquark pairs ($q \overline{q}$) formed when a gluon of the hadron's color field splits) and gluons, it is difficult to explicitly describe this process.
Both models presented in this section are attempts to provide an explanation for this process. 

\subsection{Photo-gluon fusion model}
\label{Ch2Sc1-1}

\noindent

\begin{figure}[htb]
\vspace{4ex}
\centering
 \includegraphics[scale=0.6]{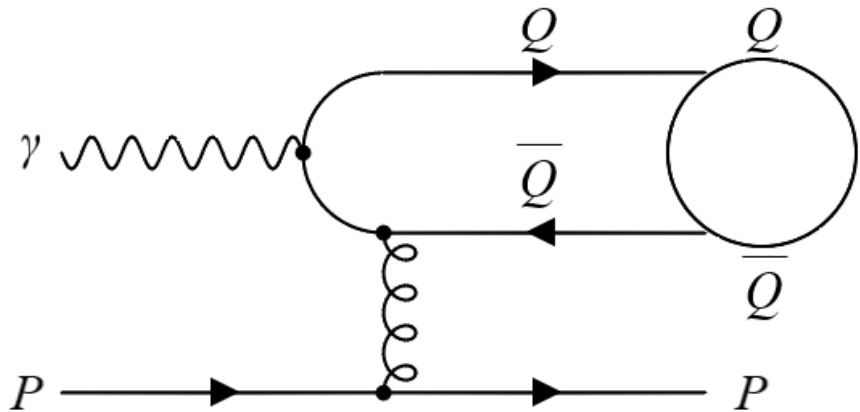}
 \caption{Schematic diagram of the photon-gluon fusion model.}
 \label{phogl}
\end{figure}

In the photon-gluon fusion (PGF) model the photon fuses with a gluon emitted from the proton
and split into a $Q\overline{Q}$ pair, as shown in Fig.~\ref{phogl}. The model depends on the
photon-gluon cross section, which is similar to (QCD analogue of) the pair production 
in photon-photon collisions, being given by~\cite{Sibirtsev:2004ca}

\begin{equation}
\sigma_{\gamma g \rightarrow Q\overline{Q}} = \frac{2 \pi \alpha e^{2}_{Q} \alpha_{s}}{\overline{s}^{3}}
\left( \left[ \overline{s}^{2} + 4m_{Q}^{2} \left( \overline{s}^{2} - 2m_{Q}^{2} \right) \right]
ln \left( \frac{1 + \beta}{1 - \beta} \right)
- \beta \left[ \overline{s}^{2} + 4 \overline{s} m_{Q}^{2} \right] \right),
\end{equation}
where $\alpha = 1/137$ is the electromagnetic fine structure constant, $e_{Q}$ and $m_{Q}$ are are the electric charge and the mass of 
the heavy quark respectively, $\alpha_{s} = 0.2$ is the strong coupling 
at $\mu^{2} = m^{2}_{b\overline{b}}$ ($\alpha_{s} = 0.5$ at $\mu^{2} = m^{2}_{c\overline{c}}$), 
$s = m^{2}_{p} + 2m_{p}E_{\gamma}$, 
where $m_{p}$ is the proton mass and $E_{\gamma}$ is the photon energy, 
$\overline{s} = (k_{\gamma} + k_{\text{gluon}})^{2}$ 
and $\beta$ is defined as 
$\beta \equiv \sqrt{1 - 4m^{2}_{Q}/\overline{s}}$.

The cross section of heavy quarkonia photo-production is then given by

\begin{equation}
\sigma_{\gamma p \rightarrow Q\overline{Q} p} = f \int^{4m^{2}_{D,B}}_{4m^{2}_{Q}}
\frac{d\overline{s}}{\overline{s}} \sigma_{\gamma g \rightarrow Q\overline{Q}}
g \left( x = \frac{\overline{s}}{s}, m^{2}_{Q\overline{Q}} \right),
\end{equation}
with $g \left(x, m^{2}_{Q\overline{Q}} \right)$ being the gluon distribution of the proton at the mass scale $\mu^{2} = m^{2}_{Q\overline{Q}}$, and $f$ is an adjustable parameter that accounts for the fraction of the specific quarkonium bound states available in the mass region between $2m_{Q}$ and $2m_{D,B}$, and $D$ and $B$ are the intermediate state hadrons for $\sigma_{\gamma p \rightarrow c\overline{c} p}$
and $\sigma_{\gamma p \rightarrow b\overline{b} p}$, respectively.

\subsection{Pomeron-exchange model}
\label{Ch2Sc1-2}

\noindent

\begin{figure}[htb]
\vspace{4ex}
\centering
 \includegraphics[scale=0.6]{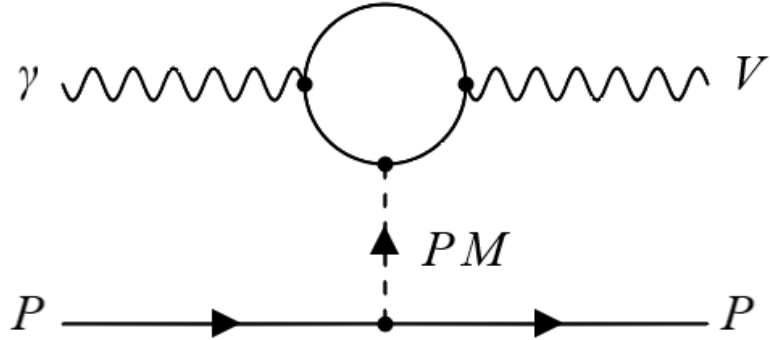}
 \caption{Schematic diagram of the pomeron exchange model.}
 \label{pomre}
\end{figure}

The phenomenological pomeron exchange (PMEX) model is used to study the photo-production of vector mesons (V), which is a process of the type $\gamma p \rightarrow V p$, represented by the Feynman diagram shown in Fig.~\ref{pomre}. A pomeron (PM) is a particle of a pre-QCD approach to describe hadron interactions, known as Regge theory (see Appendix~\ref{apx-regge}). In this theory, the pomerons (see Appendix~\ref{apx-pomeron}) are the objects exchanged by the interacting hadrons.

Bringing this description to a QCD view, one can interpret a two-gluon exchange (or even several gluon ladders) as if the gluons are ``reggeised'', meaning that it can be taken as a pomeron exchange \cite{Lipatov:1976zz}.

The $\gamma p \rightarrow V p$ is parameterized as a function of the photon-proton ($\gamma p$)
center-of-mass energy $W_{\gamma p}$~\cite{Klein:2016yzr}, and for the $\Upsilon$ and 
$J/\Psi$ cases we have

\begin{equation}
\sigma \left(\gamma p \rightarrow V p \right) = \sigma_{p}
\left[1 - \frac{ \left(m_{p} + m_{V} \right)^{2}}{W^{2}_{\gamma p}} \right]^{2} W_{\gamma p},
\end{equation}
where $W_{\gamma p}$ contains only the parameter for the photon-pomeron center of mass 
(no meson exchange is considered), and the free parameter $\sigma_{p}$ is experimentally determined.

\section{Electro-production}
\label{Ch2Sc2}

\noindent

\begin{figure}[htb]
\vspace{4ex}
\centering
 \includegraphics[scale=0.35]{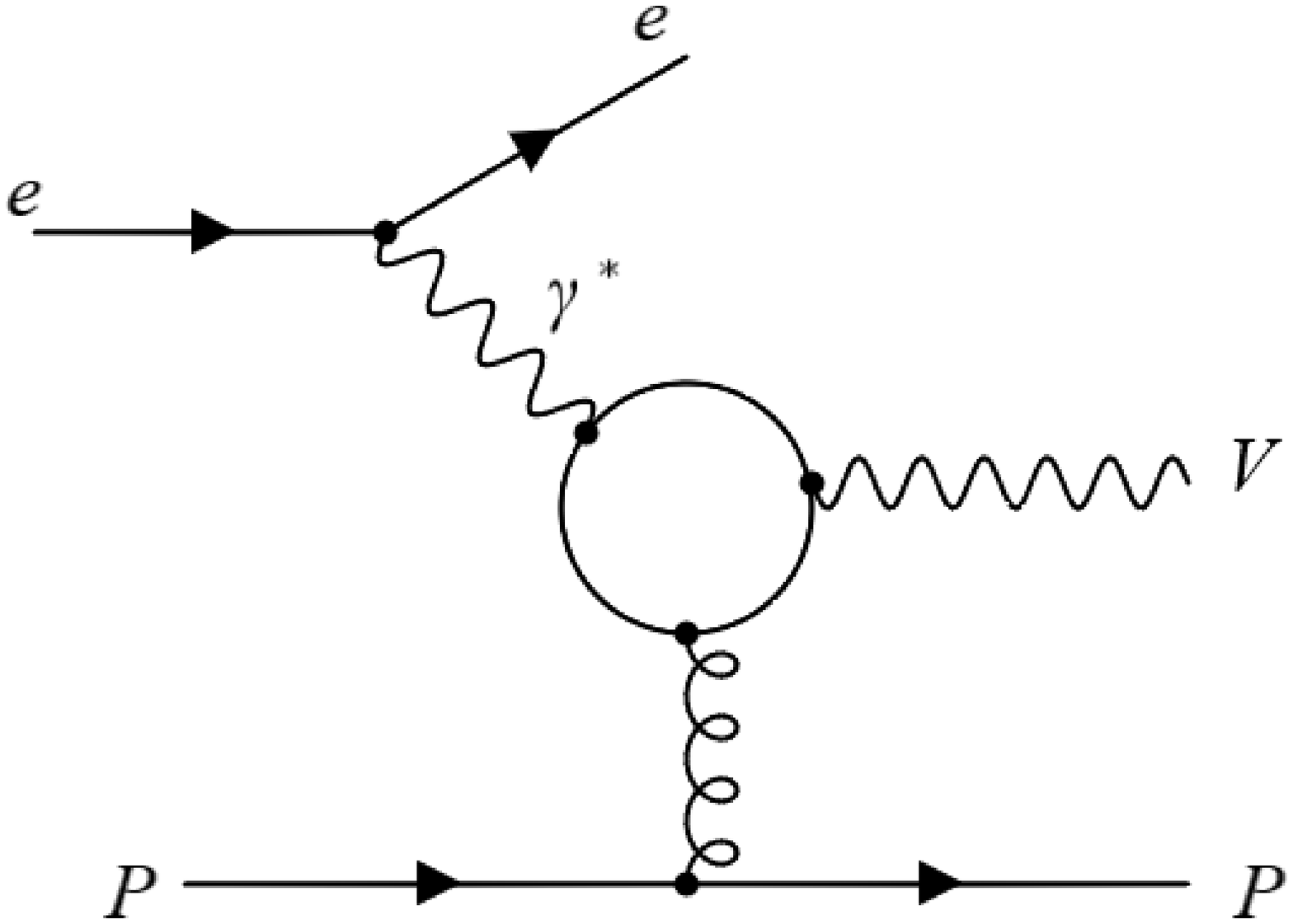}
\hspace{2ex}
 \includegraphics[scale=0.35]{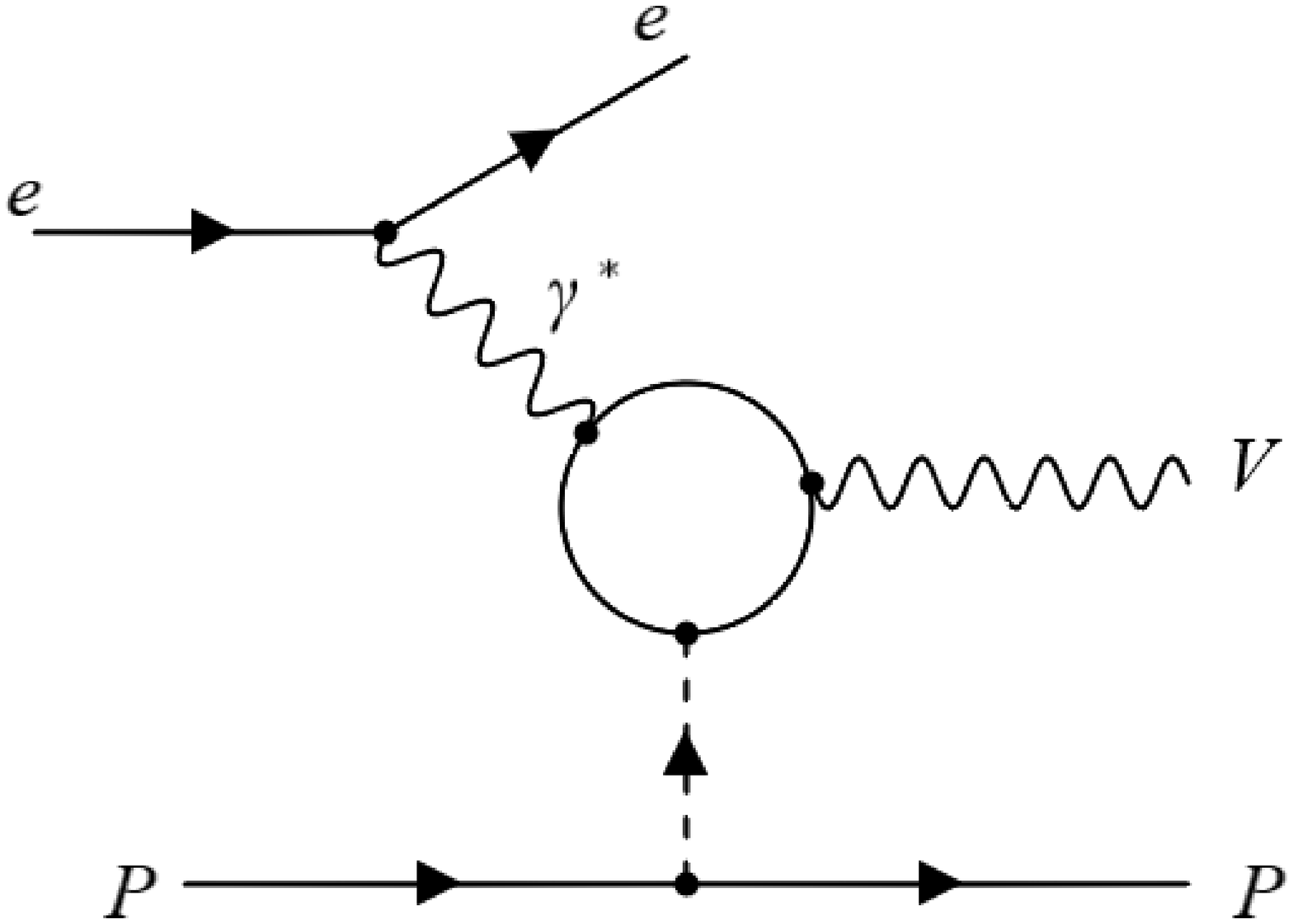}
 \caption{Schematic diagram of the electro-production of a vector meson, represented
by the photo-gluon fusion model (left) and the pomeron-exchange model (right).}
 \label{elecpro}
\end{figure}

Now we discuss the electro-production of vector mesons, with the process 
represented in Fig.~\ref{elecpro} for both the PGF and PMEX models. 
We express the cross section of $ep \rightarrow epV$ as

\begin{equation}
\sigma \left(ep  \rightarrow epV \right) = \int^{}_{}dk \int^{}_{}dQ^{2}
\frac{d^{2} N_{\gamma}}{dkdQ^{2}}
\sigma_{\gamma^{*}p \rightarrow Vp} \left(W, Q^{2} \right),
\end{equation}
where $\sigma_{\gamma^{*}p \rightarrow Vp}$ is the cross section for
the vector meson production by a virtual photon, $W$ is the final state invariant 
mass~\cite{H1:1999pji} and

\begin{equation}
\frac{d^{2} N_{\gamma}}{dkdQ^{2}} = \frac{\alpha}{\pi kQ^{2}}
\left[1 - \frac{k}{E_{e}} + \frac{k^{2}}{2E^{2}_{e}}
- \left(1 - \frac{k}{E_{e}} \right) \frac{Q^{2}_{min}}{Q^{2}} \right],
\end{equation}
which is the probability of electron emitting virtual photons with
energy $k$ and $Q^{2} = -q^{2}$, where $q$ is the four-momentum of the intermediate photon, and minimum 
$Q^{2}_{min} = m^{2}_{e} k^{2}/ \left[ E_{e} \left( E_{e} - k \right) \right]$~\cite{Budnev:1974de}.

Since the vector meson production by the virtual photon is $Q^{2}$-dependent, the models described above may be used to calculate the electro-production cross-sections of heavy quarkonia.

\section{$\eta_{c, b}$ Production}
\label{Ch2Sc3}

\noindent

\begin{figure}[htb]
\vspace{4ex}
\centering
 \includegraphics[scale=0.5]{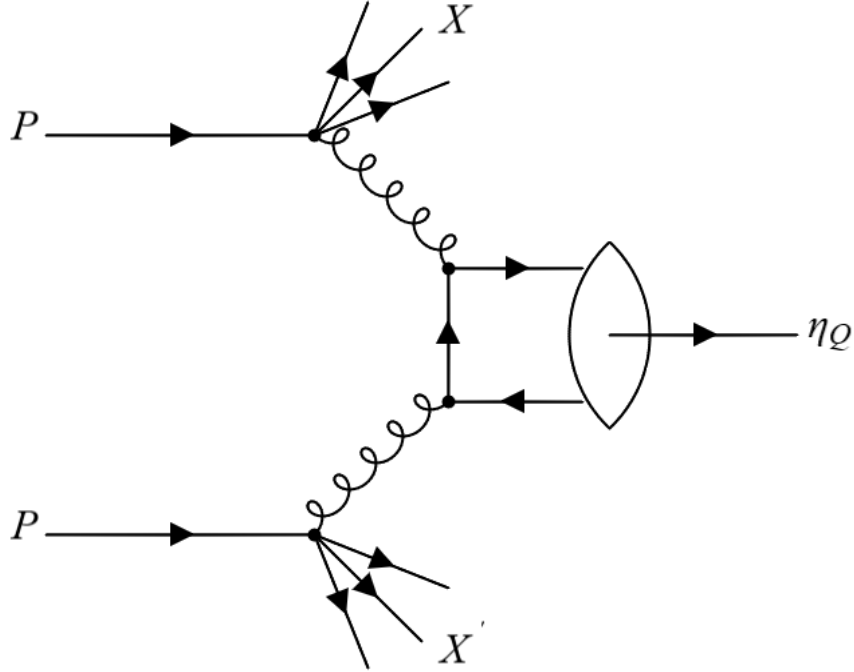}
 \caption{Schematic diagram of the inclusive production of $\eta_{Q}$.}
 \label{etaQprod}
\end{figure}

This section covers the methods presented in~\cite{Likhoded:2014fta} to obtain
the cross section of the inclusive $\eta_{Q}$ ($Q = c,b$) production
$pp \rightarrow \eta_{Q} + X$, represented in Fig.~\ref{etaQprod}.
Considering now the production of a pseudo-scalar $\eta_{c, b}$ meson, 
it can be described as a production of heavy quark-antiquark pair $Q\overline{Q}$.

For a general Fock state (see Appendix~\ref{apx-fock}), where we consider the state of 
the $Q\overline{Q}$ pair in the various Fock-state components (e.g. $\ket{Q\overline{Q}g}$, 
which includes a dynamical gluon),
the $Q\overline{Q}$ pair  can be in either a 
color-singlet state or a color-octet state, and its angular-momentum state denoted by
$^{2S + 1}L_{J}$, where $S = 0,1$ is the total spin of the quark and antiquark, $L = 0,1,2,...$
(or $L = S,P,D,...$) is the orbital angular momentum, and $J$ is the total angular momentum.
In the Fock state $\ket{Q\overline{Q}}$, the $Q\overline{Q}$ pair must be 
in a color-singlet state and in an angular momentum state $^{2S + 1}L_{J}$, because the color-octet 
$Q\overline{Q}$ state evolves non-perturbatively into a physical color-singlet state by emission of one or more gluons~\cite{Biswal:2010xk}.

Using a nonrelativistic QCD (NRQCD) approach, one needs to take into account the contributions of 
the color singlet (CS) and color octet (CO) components, with unknown nonperturbative long-distance matrix
elements (LDME), that are treated as free parameters.
LDMEs are organized into a hierarchy according to their scaling with $v$, the typical velocity of the
heavy quark.
The calculation of the cross section involves colored heavy quark-antiquark pairs in $L = S =0$
and $L = S = 1$ configurations, represented by the $Q\overline{Q} \left[^{2S + 1} L^{(1,8)}_{J}\right]$ sates, where the indexes (1,8) represent whether the components are color siglet or color octet.  

The cross section for production of a quarkonium state $H$ can be factorised as

\begin{equation}
\sigma (H) = \sum_{n={S,L,J}} \frac{F_{n}}{M^{d_{n}-4}_{Q}} 
\braket{\mathcal{O}^{H}_{n} \left( ^{2S + 1}L_{J} \right)},
\end{equation}
where the $F_{n}$s are short-distance coefficients, $\mathcal{O}^{H}_{n}$ are operators of dimension
$d_{n}$ describing the long-distance effects and $M_{Q}$ is the heavy quark mass.

To study the $\eta_{Q}$ production cross section, we write the Fock space expansion (See Appendix 
\ref{apx-fock}) for the physical $\eta_{Q}$~\cite{Cho:1995ce}, with the main
contributions to the cross section being the terms~\cite{Biswal:2010xk}

\begin{equation}
\ket{\eta_{Q}} = \mathcal{O}(1) \ket{Q\overline{Q} \left[^{1} S^{(1)}_{0}\right]}
+ \mathcal{O}(v^{2}) \ket{Q\overline{Q} \left[^{1} P^{(8)}_{1}\right]g}
+ \mathcal{O}(v^{4}) \ket{Q\overline{Q} \left[^{3} S^{(8)}_{1}\right]g}
+ \cdots .
\end{equation}

The color-octet state $^{1} P^{(8)}_{1}$ ($^{3} S^{(8)}_{1}$) becomes a physical
$\eta_{Q}$ by emitting a gluon ($g$), then the case considered is 

\begin{equation}
gg \rightarrow Q \overline{Q} \left[^{2S + 1} L_{J}\right]g,
\end{equation}
so the differential cross section for the $\eta_{Q}$ production with specific 
angular momentum and colour states is given by~\cite{Likhoded:2014fta}

\begin{eqnarray}
\frac{d \sigma}{d p_{T}} \left(pp \rightarrow Q\overline{Q} 
\left[^{2S + 1} L^{(1,8)}_{J}\right]X \right)  = &&
\frac{d \sigma (gg \rightarrow \eta_{Q}g)}{d p_{T}} =
|R(0)|^{2} \frac{d \sigma \left(gg \rightarrow Q\overline{Q} 
\left[^{1} S^{(1)}_{0}\right]g \right)}{d p_{T}} \nonumber \\
&& + \braket{O_{S}}
\frac{d \sigma \left(gg \rightarrow Q\overline{Q} 
\left[^{3} S^{(8)}_{1}\right]g \right)}{d p_{T}} \nonumber \\
&& + \braket{O_{P}} \frac{d \sigma \left(gg \rightarrow Q\overline{Q} 
\left[^{1} P^{(8)}_{1}\right]g \right)}{d p_{T}},
\end{eqnarray}
where $p_{T}$ is the transverse momentum of the final quarkonium, $R(0)$ is the value of the heavy quarkonum wave function in the CS state at the origin, and the following notations 
for CO LDME are used~\cite{Meijer:2007eb}

\begin{eqnarray}
\braket{O_{S}} =&& \braket{R_{\eta_{Q}}\left[^{3} S^{(8)}_{1}\right]}
= \frac{\pi}{6} \braket{0|\mathcal{O}^{\eta_{Q}}_{8} \left[^{3} S_{1}\right]|0},\\
\braket{O_{P}} =&& \braket{R_{\eta_{Q}}\left[^{1} P^{(8)}_{1}\right]}
= \frac{\pi}{18} \braket{0|\mathcal{O}^{\eta_{Q}}_{8} \left[^{1} P_{1}\right]|0}.
\end{eqnarray}

\section{Production at EIC}
\label{Ch2Sc4}

\noindent

In nuclear physics, many questions arise from the current fundamental understanding of QCD.
How quarks and gluons inside the proton combine their spins to generate the proton's overall spin 
(1/2), and how these objects are distributed in space and momentum inside the nucleon? Why quarks or gluons must remain confined?
There is a boundary between the saturated gluon density and the more dilute quark-gluon matter regions? And if so, how does the quark-gluon distribution changes from each other?

\begin{figure}[htb]
\vspace{4ex}
\centering
 \includegraphics[scale=0.3]{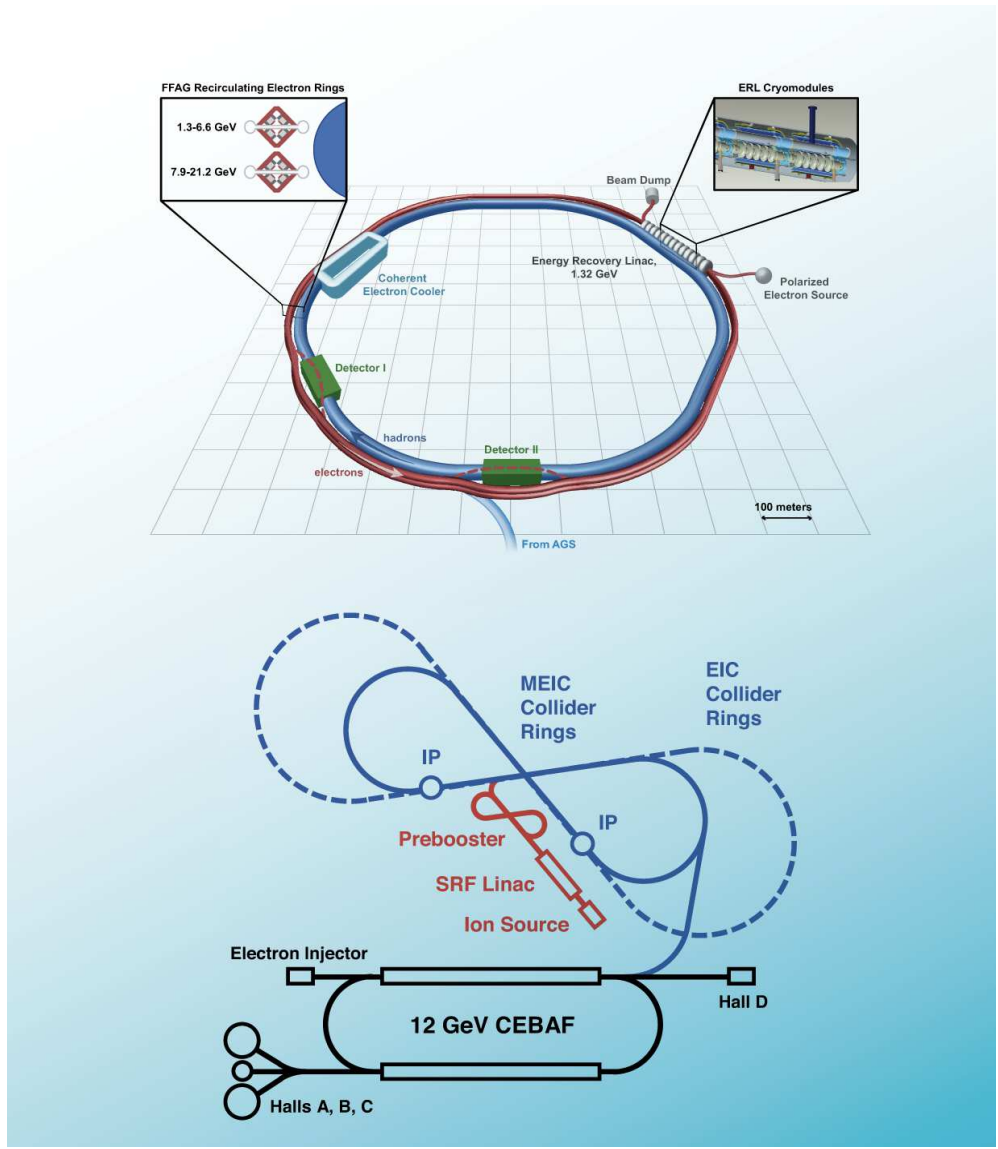}
 \caption{Top: The schematic of eRHIC at BNL, which would require construction of an
electron beam facility (red) to collide with the RHIC blue beam at up to three interaction points.
Bottom: The schematic of the Medium energy Electron Ion Collider (MEIC) at JLab, which would require construction of an ion linac (red), and an electron-ion collider ring (blue) with at least two interaction points, around the
12 GeV CEBAF~\cite{Accardi:2012qut}.}
 \label{cebaf}
\end{figure}

To provide answers to these questions, electron-ion colliders (EIC) is being constructed.
In addition, the electro-production of heavy quarkonia are among the key reactions that are going to be measured on electron-ion colliders. In the USA, two independent designs for a EIC are being
considered using already existing infrastructure and facilities.
At the  Jefferson Laboratory (JLab) a new electron and ion collider ring complex
will be added together with the already existing 12GeV upgraded Continuous Electron Beam Accelerator Facility (CEBAF), which in a recent experiment ~\cite{Ali:2019lzf}, a photon beam was used to produce a $J/\Psi$ meson near-threshold, which was identified by its decay into an electron-positron pair.
While at Brookhaven National Laboratory (BNL), the eRHIC design utilizes a new electron beam facility 
based on an Energy Recovery Linear Particle Accelerator (ERL) to be built inside the (Relativistic Heavy Ion Collider) RHIC tunnel to collide with RHICs existing high-energy polarized proton and nuclear beams.
The schematics of both colliders are shown in Fig.~\ref{cebaf}. More information on this can be found 
in~\cite{Accardi:2012qut}.

\section{Production at FAIR}
\label{Ch2Sc5}

\noindent

\begin{figure}[htb]
\vspace{4ex}
\centering
 \includegraphics[scale=0.6]{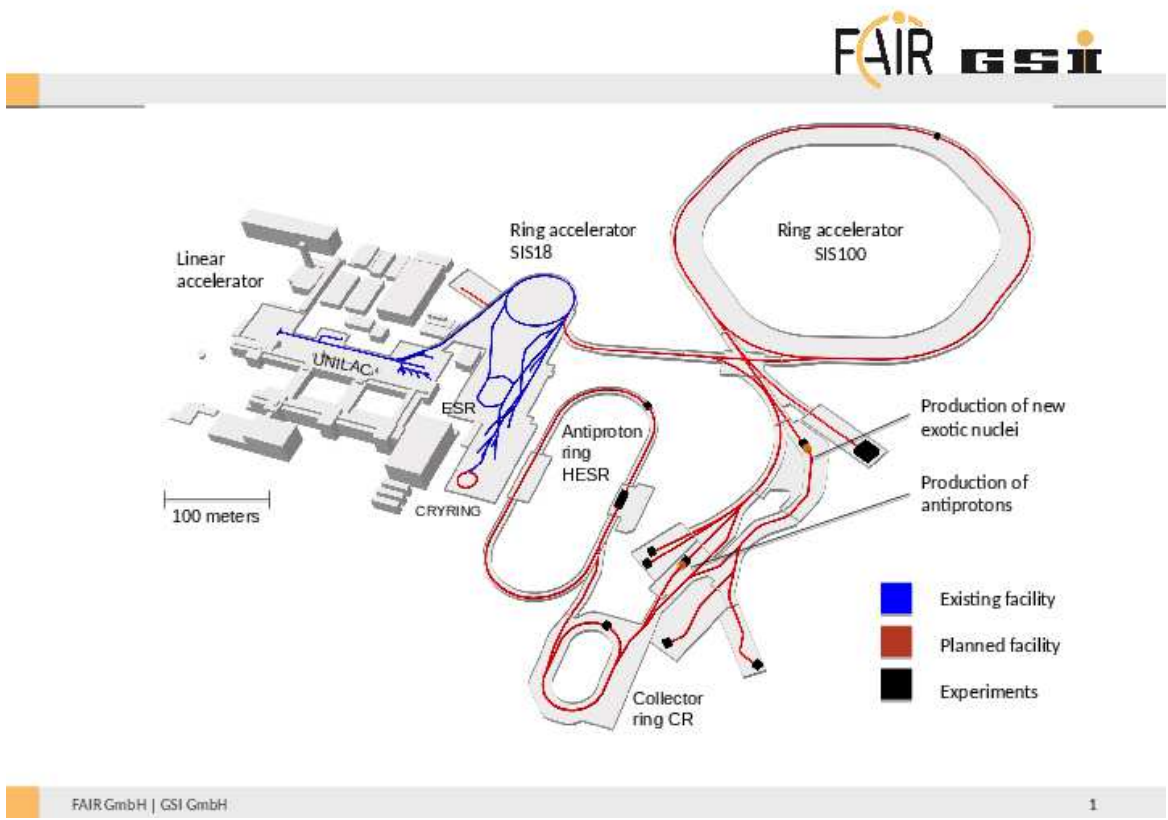}
 \caption{The FAIR accelerator complex~\cite{Durante:2019hzd}.}
 \label{fair}
\end{figure}

The Facility for Antiproton and Ion Research (FAIR) is currently under construction as an 
international facility at the campus of the GSI Helmholtzzentrum for Heavy-Ion Research in Darmstadt, Germany.
It will be an accelerator-based research facility in many basic sciences and their applications. FAIR will cover researches on the structure and evolution of matter on both the microscopic and the
cosmic scale. It will be able to explore hadron structure and dynamics exploiting proton-antiproton annihilation.
With the construction of the FAIR facility, heavy and heavy-light mesons will be produced copiously
by the annihilation of antiprotons on nuclei~\cite{Durante:2019hzd}. The facility layout is shown in Fig.~\ref{fair}.

\chapter{Quark-meson coupling model} 

\label{Chapter3} 

\thispagestyle{empty}

\noindent

The quark-meson coupling (QMC) model is a quark-based model for nuclear matter and
finite nuclei by describing the internal structure of the nucleon  
using the MIT bag (original version), and the binding of nucleons by the self-consistent 
couplings of the confined light quarks $(u, d)$ to the scalar-$\sigma$, 
isoscalar-vector-$\omega$ and isovector-vector-$\rho$ 
meson fields generated by the confined light quarks
in the nucleons~\cite{Guichon:1987jp,Guichon:1995ue,Saito:1996sf}.
In a nuclear medium, the hadrons with light quarks 
are expected to change their properties, and thus affect the interaction with nucleons, 
what makes the QMC model a useful model to describe the changes of the internal structure of 
hadrons in a nuclear medium.

Although the QMC model was invented by Guichon~\cite{Guichon:1987jp}, a similar model for nuclear matter
was studied around the same time by Frederico et al.~\cite{Frederico:1987xy}, where the nucleon on nuclear matter properties is investigated within a $\sigma$-$\omega$-$q$ model,  
and the quark confinement explained by a quark potential which takes the form of a harmonic oscillator.
Also, in the early 90's, Banerjee~\cite{Banerjee:1991nm} studied the changes in the structure of a nucleon when it is placed in nuclear matter, Naar and Birse~\cite{Naar:1993cy} studied nucleon
properties in nuclear matter by treating it within a color-dielectric model, and 
Mishra et al.~\cite{Mishra:1992zz} investigated nuclear matter in the relativistic Hartree approximation
using a $\sigma$-$\omega$ model. 

In this chapter, the basis of this model will be set by describing the MIT bag model and
its mathematical structure.
Based on that, the description of the QMC model will be followed by details on the calculation
of the in-medium masses of the $B$ and $B^{*}$ mesons, with the latter being obtained for the first time in symmetric nuclear matter.

\section{MIT bag model}

\thispagestyle{myheadings}

\noindent
Initially, a kind of bag model was proposed by P.N. Bogoliubov in 1967~\cite{Bogolioubov:1968} to model the atomic nucleus as a system of independent quarks in a confining potential.
The Bogoliubov model considered three massless quarks in a vacuum cavity of radius
chosen so as to relate the energy of the 3 quarks to the mass of the hadron. This cavity
contained a finite, spherical, square well potential, which finally was set to infinity.

This model was then improved seven years later by a group of five scientists from the Massachussets Institute of Technology (MIT)~\cite{Chodos:1974je, Johnson:1975zp}, that stated that strongly interacting quarks are in a finite region of space to which fields are confined. This region is called a bag (hence the name MIT bag model).
It also provided a natural explanation for the confinement phenomenon, which is accomplished
within the model by surrounding the finite region with a constant (potential) energy per unit volume, $B$.

Now we describe the mathematics of the MIT bag, emphasizing some of its aspects that will be of use further ahead on the study of the QMC model, that is, to include the influence of external fields to the bag.

\begin{figure}[htb]
\vspace{4ex}
\centering
 \includegraphics[scale=0.35]{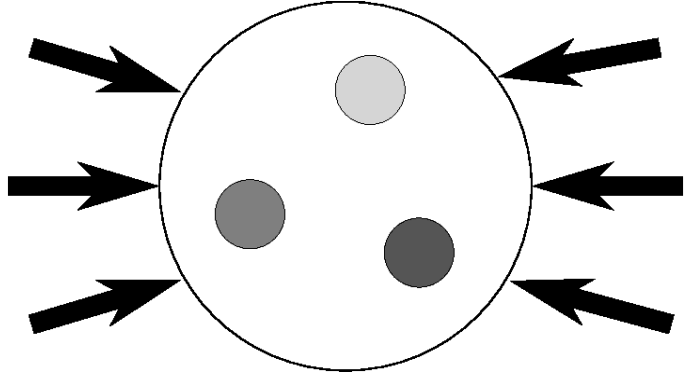}
 \caption{Representation of a baryon in the MIT-Bag Model. The quarks are confined in a cavity (bag) of finite volume due to a constant external pressure $B_p$ represented by the arrows around the bag.}
 \label{bagmodl}
\end{figure}

We start by considering a system of three quarks, each with mass $m_q$, in a scalar potential $U$, where
these quarks are confined in a cavity (in reality by the strong interactions among quarks mediated by the (exchanged) gluon fields) and under external fields.
The enclosed region in which the quarks can move has an radius $R$, and is subjected to a pressure $B_p$.
The bag radius is determined by the stability equation
$d E_{b}/d R =0$, where $E_{b}$ is the bag energy.
This system is represented here in Fig.~\ref{bagmodl}.

The quarks inside the cavity may interact with external fields of
scalar-$\sigma$ and vector-$\omega$, $\rho$ types.
This means that $\sigma$, $\omega$ and $\rho$ meson exchanges may occur. 

The Dirac equation which describes this system has the form

\begin{equation}
\left[ \gamma_{\mu} \left(P^{\mu} + V^{\mu}_{\nu}\right)
- \left(m_{q} + V_{s} + U\right)\right]\psi = 0,
\end{equation}
with $P^{\mu} \equiv \left(i \frac{\partial}{\partial t}, -i\bf{\nabla}\right)$.
So the Dirac equation containing real fields is

\begin{equation}
\left[ \gamma_{\mu} \left(P^{\mu} + V^{\mu}_{\omega} + \frac{1}{2} \tau^{z} V^{\mu}_{\rho}\right)
- \left(m_{q} - V_{\sigma} + U\right)\right]\psi = 0,
\end{equation}
where $1/2~\tau^{z}$ is the isospin third component and $V_{\sigma} = g_{\sigma}\sigma$, $V^{\mu}_{\omega} = g_{\omega}\omega^{\mu}$ and $V^{\mu}_{\rho} = g_{\rho}b^{\mu}$ describe the interactions between quarks and the fields $\sigma$, $\omega$ and $b (\equiv \rho)$ through the quark-meson coupling constants 
$g_{\sigma}$, $g_{\omega}$ and $g_{\rho}$.

\section{Quark-meson coupling model}

\noindent
The quark-meson coupling model is a quark-based model that describes nuclear matter by nucleons and mesons, where the quark structure of the nucleon is modified by the surrounding nucleons and meson fields~\cite{Guichon:1987jp}.
The infinite nuclear matter at normal density (or at a not very higher density, up to about 3 times the nuclear matter saturation  density, since at higher densities the model description with nonoverlapping MIT bags for the nuclear medium may not be proper, and also quark-hadron mixed phase is expected to appear.) 
is a uniform distribution of nucleons interacting through the exchange of mesons which are coupled
directly to the quarks, and only the $\sigma$-exchange, the (time component of the) $\omega$-exchange  ($\omega^{\mu} = (\omega,0)$) and the $\rho$-exchange ($\rho^{\mu} = (b, 0)$, in the isospin asymmetric nuclear matter, where $b$ is determined by the difference in proton and neutron densities, 
$b \sim \rho_p - \rho_n$) contributes to the isospin asymmetric nuclear matter. 

\subsection{QMC description of nuclear matter}
\label{Ch3Sc2-1}

\noindent

We start the description of the internal structure of the nucleon by treating it classically 
(at first). The descriptions of this subsection, subsections \ref{Ch3Sc2-2} and \ref{Ch3Sc2-5} are 
from~\cite{Guichon:1995ue}.
It is then useful to take the coordinates in the rest frame of the nucleus (NRF), denoted by $(t, \mathbf{r})$.
In this frame the nucleon follows a classical trajectory, $\mathbf{R}(t)$, and the instantaneous velocity of the nucleon is given by $\mathbf{v} = d\mathbf{R}/dt$. 
We then define an instantaneous rest frame for a nucleon at each time $t$ (IRF), which is denoted with primes $(t',\mathbf{r}')$, and are related to the NRF coordinates by the Lorentz transformation 
(with $c$ = 1):

\begin{eqnarray}
r_{L} &=& r'_{L}~ \text{cosh}~\xi + t'~ \text{sinh}~\xi, \nonumber \\
\textbf{r}_{\perp} &=& \textbf{r}'_{\perp}, \nonumber \\
t &=& t'~ \text{cosh}~\xi + r'_{L}~ \text{sinh}~\xi,
\end{eqnarray}
where $r_{L}$ and $\mathbf{r}_{\perp}$ are, respectively, the parallel and transverse components to the velocity and $\xi$ is the rapidity defined by $\text{tanh}~\xi = |\mathbf{v}(t)|$.

We can thus describe the internal structure of the nucleon in the IRF, adopting the the static spherical cavity approximation to the MIT bag. The appropriate Lagrangian density in the IRF is then
~\cite{Guichon:1995ue,Saito:2005rv}

\begin{equation}
{\cal L}_{0} = \bar{\psi}_{q}' \left(i \gamma^{\mu} \partial_{\mu} -m_{q} \right) 
\psi_{q}' - B_{p}V_{B}, \qquad \text{for} \quad |\mathbf{u}'| \leq R_{N}
\end{equation}
with $B_{p}$ being the bag constant, $V_{B}$ the bag volume, $R_{N}$ the bag radius for the nucleon and
$\mathbf{u}'$ the position of the quark from the center of the bag in the IRF, where the
4-vector $u'$ is denoted as $u' = (t', \mathbf{u}') = (t', \mathbf{r}' - \mathbf{R}')$.

Now to include the effects of the interactions with the surrounding nucleons, we incorporate the
$\sigma$ and $\omega$ fields generated by them. In the NRF they are functions of position, so we 
express them in the IRF by Lorentz transformation

\begin{eqnarray}
\sigma_{I}(t', \mathbf{u}') &=& \sigma(\mathbf{r}), \nonumber \\
\omega_{I}(t', \mathbf{u}') &=& \omega(\mathbf{r})~ \text{cosh}~\xi, \nonumber \\
\mathbf{\omega}_{I}'(t', \mathbf{u}') &=& - \omega(\mathbf{r}) \hat{v}~ \text{sinh}~\xi,
\end{eqnarray}
where the subscript $I$ stands for the IRF. Now we can write the interaction as
$(|\mathbf{u}'| \leq R_{N})$

\begin{equation}
{\cal L}_{I} = g^{q}_{\sigma} \bar{\psi}_{q}' \psi_{q}'(u') \sigma_{I}(u')
- g^{q}_{\omega} \bar{\psi}_{q}' \gamma_{\mu} \psi_{q}'(u') \omega^{\mu}_{I}(u'),
\end{equation}
with the coupling constants $g^{q}_{\sigma}$ and $g^{q}_{\omega}$. We do not include, however, the 
effect of the $\rho$ meson field (at least for now).

To construct the hamiltonian in the IRF, we shall evaluate the interaction term at equal time $t'$
for all $\mathbf{u}'$ in the bag. Supposing that at time $t'$ the bag center is located at $\mathbf{R}'$ in the IRF, in the NRF it will be located at $\mathbf{R}$ at time $T$, defined by

\begin{eqnarray}
r_{L} &=& R'_{L}~ \text{cosh}~\xi + t'~ \text{sinh}~\xi + u'_{L}~ \text{cosh}~\xi, \nonumber \\
&=& R_{L} + u'_{L}~ \text{cosh}~\xi, \nonumber \\
\mathbf{r}_{\perp} &=& \mathbf{R}'_{\perp} + \mathbf{u}'_{\perp}.
\end{eqnarray}

We then express the $\sigma$ and $\omega$ fields as

\begin{eqnarray}
\sigma_{I}(t',\mathbf{u}') &=& \sigma \left(R_{L}(T) + u'_{L}~ \text{cosh}~\xi, 
\mathbf{R}'_{\perp}(T) + \mathbf{u}'_{\perp} \right), \nonumber \\
\omega^{\mu}_{I}(t',\mathbf{u}') &=& \eta^{\mu} \omega \left(R_{L}(T) + u'_{L}~ \text{cosh}~\xi, 
\mathbf{R}'_{\perp}(T) + \mathbf{u}'_{\perp} \right),
\end{eqnarray}
where $\eta^{\mu} = (\text{cosh}~\xi, - \hat{v}~ \text{sinh}~\xi)$.

Thus, in the IRF the interaction Lagrangian is

\begin{equation}
{\cal L}_{I} = g^{q}_{\sigma} \bar{\psi}_{q}' \psi_{q}' (u') \sigma(\mathbf{R} + \mathbf{u}') - g^{q}_{\omega} \bar{\psi}_{q}' (t', \mathbf{u}') \left[\gamma_{0}~ \text{cosh}~\xi
+ \mathbf{\gamma} \cdot \hat{v}~ \text{sinh}~\xi \right] \psi_{q}' (t', \mathbf{u}')
\omega (\mathbf{R} + \mathbf{u}'),
\end{equation}
and the corresponding Hamiltonian expressed in two pieces $H = H_{0} + H_{1}$:

\begin{eqnarray}
&H_{0} = \int_{V_{B}} d\mathbf{u}' \bar{\psi}_{q}' \left[-i \mathbf{\gamma} \cdot \mathbf{\nabla} + m_{q} - g^{q}_{\sigma} \sigma(\mathbf{R}) \right. \nonumber \\
&+ \left. g^{q}_{\omega} (\gamma_{0}~ \text{cosh}~\xi + \mathbf{\gamma} \cdot \hat{v}~ \text{sinh}~\xi)
\omega(\mathbf{R}) \right] \psi_{q}'(t', \mathbf{u}') + B_{p} V_{B}, \nonumber \\
&H_{1} = \int_{V_{B}} d\mathbf{u}' \bar{\psi}_{q}' \left[- g^{q}_{\sigma} \left( \sigma (\mathbf{R} + \mathbf{u}') 
- \sigma(\mathbf{R}) \right) \right. \nonumber \\
& \left. + g^{q}_{\omega} (\gamma_{0}~ \text{cosh}~\xi + \mathbf{\gamma} \cdot \hat{v}~ \text{sinh}~\xi)
\left( \omega (\mathbf{R} + \mathbf{u}') -\omega (\mathbf{R}) \right) \right]
\psi_{q}' (t', \mathbf{u}'),
\end{eqnarray}
where $H_{1}$ is treated as a perturbation.

We then write a complete orthogonal set of eigenfunctions for the quark field, denoted by
$\phi^{\alpha}$, where $\alpha$ is a collective symbol to label the quantum numbers

\begin{eqnarray}
h \phi^{\alpha} (\mathbf{u}') &\equiv& (-i \gamma^{0} \mathbf{\gamma} \cdot \mathbf{\nabla} + m^{*}_{q} \gamma^{0})
\phi^{\alpha} (\mathbf{u}') = \frac{\Omega_{\alpha}}{R_{B}} \phi^{\alpha} (\mathbf{u}'), \nonumber \\
(1 + i \mathbf{\gamma} \cdot u') \phi^{\alpha} (\mathbf{u}') &=& 0, \quad \text{at} 
\quad |\mathbf{u}'| = R_{B}, \nonumber \\
\int_{V_{B}} d\mathbf{u}' \phi^{\alpha \dagger} \phi^{\beta} &=& \delta^{\alpha \beta}, 
\end{eqnarray}
with $m^{*}_{q}$ being a parameter. The lowest positive eigenfunction is given by

\begin{equation}
\phi^{0m} (t', \mathbf{u}') = \frac{{\cal N}}{\sqrt{4 \pi}} 
\begin{pmatrix}
j_{0} (x u'/ R_{B}) \\
i \beta_{q} \mathbf{\sigma} \cdot u' j_{1} (xu'/R_{B}) 
\end{pmatrix}
\chi_{m},
\end{equation}
with $u' = |\mathbf{u}'|$, $\chi_{m}$ being the (Pauli 2-component spinor) spin function and

\begin{eqnarray}
\Omega_{0} &=& \sqrt{x^{2} + (m^{*}_{q} R_{B})^{2}}, 
\qquad \beta_{q} = \sqrt{\frac{\Omega_{0} - m^{*}_{q} R_{B}}{\Omega_{0} + m^{*}_{q} R_{B}}},
\nonumber \\
{\cal N}^{-2} &=& 2R^{3}_{B} j^{2}_{0}(x) \left[ \Omega_{0}(\Omega_{0} - 1)
+ m^{*}_{q} R_{B}/2 \right] / x^{2},
\end{eqnarray}
where $x$ is the eigenvalue for the lowest mode, which satisfies the boundary condition at the bag surface, $j_{0}(x) = \beta_{q}j_{1}(x)$, where $j_{0}$ and $j_{1}$ are the speherical Bessel functions.

The quark field is then expanded as 

\begin{equation}
\psi_{q}'(t', \mathbf{u}') = \sum_{\alpha} e^{-i \mathbf{k} \cdot \mathbf{u}'} 
\phi^{\alpha} (\mathbf{u}') b_{\alpha}(t'),
\end{equation}
with $b_{\alpha}$ being the annihilation operator for the quark and 
$\mathbf{k} = g^{q}_{\omega} \omega(\mathbf{R}) \hat{v}~ \text{sinh}~\xi$.

Considering now just the free Hamiltonian, we choose the effective quark mass as
$m^{*}_{q} = m_{q} - g^{q}_{\sigma} \sigma (\mathbf{R})$, and so we find the leading
part of the energy and momentum operators in the IRF to be~\cite{Guichon:1995ue,Saito:2005rv}

\begin{eqnarray}
H^{I}_{0} &=& \sum_{\alpha} \frac{\Omega_{\alpha}(\mathbf{R})}{R_{B}} b^{\dagger}_{\alpha} b_{\alpha}
+ \hat{N}_{q} g^{q}_{\omega} \omega (\mathbf{R})~ \text{cosh}~\xi + B_{p}V_{B}, \nonumber \\
\mathbf{P}^{I} &=& \sum_{\alpha \beta} \braket{\alpha | -i \mathbf{\nabla} | \beta} 
b^{\dagger}_{\alpha} b_{\alpha}
- \hat{N}_{q} g^{q}_{\omega} \omega(\mathbf{R}) \hat{v}~ \text{sinh}~\xi,
\end{eqnarray}
where the frequency $\Omega$ depends on $\mathbf{R}$ because the effective quark mass varies, depending 
on position through the $\sigma$ field. The nucleon is supposed to be described in terms of the 
three quarks in the lowest mode ($\alpha = 0$), and since it should remain in that configuration
as $\mathbf{R}$ changes, the gradient term in the momentum operator becomes
zero because of parity conservation. Then, the energy and momentum in the IRF can be written as

\begin{eqnarray}
E^{I}_{0} &=& M^{*}_{N} (\mathbf{R}) + 3 g^{q}_{\omega} \omega(\mathbf{R})~ \text{cosh}~\xi, \nonumber \\
\mathbf{P}^{I} &=& -3 g^{q}_{\omega} \omega(\mathbf{R}) \hat{v}~ \text{sinh}~\xi,
\end{eqnarray}
with the effective nucleon mass

\begin{equation}
M^{*}_{N} (\mathbf{R}) = \frac{3 \Omega_{0}(\mathbf{R})}{R_{B}} + B_{p}V_{B},
\end{equation}
with the bag radius $R_{B}$ given by the stability equation

\begin{equation}
\frac{dM^{*}_{N}(\mathbf{R})}{dR_{B}} = 0.
\end{equation}

Now we use the Lorentz transformation to express these terms in the NRF as

\begin{eqnarray}
E_{0} &=& M^{*}_{N} (\mathbf{R})~ \text{cosh}~\xi + 3 g^{q}_{\omega} \omega(\mathbf{R}), \nonumber \\
\mathbf{P} &=& M^{*}_{N} (\mathbf{R}) \hat{v}~ \text{sinh}~\xi.
\end{eqnarray}

So the leading term in the energy can be written

\begin{equation}
E_{0} = \sqrt{M^{*2}_{N} (\mathbf{R}) + \mathbf{P}^{2}} + 3 g^{q}_{\omega} \omega(\mathbf{R}).
\end{equation}
with the effective nucleon mass in matter taking the form

\begin{equation}
\label{qmceq5}
M^{*}_{N} (\mathbf{R}) = \frac{3 \Omega_{0} (\mathbf{R}) -z_{0}}{R_{B}} + B_{p}V_{B},
\end{equation}
where we parameterize the sum of the center of mass (c.m.) correction and gluon fluctuation corrections
to the bag energy by $-z_{0}/R_{B}$, with these fluctuations and change in the center of mass arising due to the movement of the valence quarks and gluons inside the nucleon bag, where $z_{0}$ is assumed to be independent of the nuclear density,
with the parameters $B_p$ and $z_{0}$ fixed by the free nucleon mass 
($M_{N}$ = 939 MeV)~\cite{Saito:2005rv}.

This is justified by the Born-Oppenheimer approximation, according to which the internal structure of
the nucleon has enough time to adjust the varying external field so as to stay in its ground state.

Now we estimate the perturbation term $H_{1}$ by expanding $\sigma (\mathbf{R} + \mathbf{u}')$ and
$\omega (\mathbf{R} + \mathbf{u}')$ in powers of $\mathbf{u}'$

\begin{eqnarray}
\sigma (\mathbf{R} + \mathbf{u}') &=& \sigma (\mathbf{R}) + \mathbf{u}' \cdot \mathbf{\nabla}_{R} \sigma (\mathbf{R}) + \cdots ,
\nonumber \\
\omega (\mathbf{R} + \mathbf{u}') &=& \omega (\mathbf{R}) + \mathbf{u}' \cdot \mathbf{\nabla}_{R} \omega (\mathbf{R}) + \cdots ,
\end{eqnarray}
and computing the effect to first order. To this order several terms give
zero because of parity and one is left with

\begin{equation}
\braket{(0)^{3} | H_{1} | (0)^{3}} = g^{q}_{\omega}
\sum_{\alpha \beta} \braket{(0)^{3} | b^{\dagger}_{\alpha} b_{\beta} | (0)^{3}}
\braket{ \alpha | \gamma_{0} \mathbf{\gamma} \cdot \hat{v} \mathbf{u}'~ \text{sinh}~\xi | \beta }
\cdot \mathbf{\nabla}_{R} \omega{\mathbf{R}},
\end{equation}
where ${\alpha} = {0,m_{\alpha}}$ is setted, with $m_{\alpha}$ the spin projection of the quark
in the lowest mode ${0}$. Then

\begin{eqnarray}
\braket{ \alpha | \gamma_{0} \mathbf{\gamma} \cdot \hat{v} \mathbf{u}' | \beta }
&=& \int d \mathbf{u}' \phi^{0m_{\alpha}*} \gamma_{0} (\mathbf{\gamma} \cdot \hat{v}) \mathbf{u}' \phi^{0m_{\beta}},
\nonumber \\
&=& -I(\mathbf{R}) \braket{m_{\alpha}| \frac{\mathbf{\sigma}}{2}| m_{\beta}} \times \hat{v},
\end{eqnarray}
with

\begin{equation}
\label{qmceq1}
I(\mathbf{R}) = \frac{4}{3} \int du' u'^{3} f(u') g(u'),
\end{equation}
where $f$ and $g$ are the upper and lower components of the quark wave function,
respectively. And using the wave function of the MIT bag 

\begin{equation}
I(\mathbf{R}) = \frac{R_{B}}{3} \left[\frac{4\Omega_{0} + 2m^{*}_{q} R_{B} - 3}
{2\Omega_{0} (\Omega_{0} -1) + m^{*}_{q} R_{B}} \right].
\end{equation}

The integral $I(\mathbf{R})$ depends on $\mathbf{R}$ through the implicit dependence of $R_{B}$ and $x$ on
the local scalar field. Its value in the free case, $I_{0}$, may then be expressed in terms of the nucleon 
isoscalar magnetic moment: $I_{0} = 3 \mu_{s}/M_{N}$ with $\mu_{s} = \mu_{p} + \mu_{n}$ and 
$\mu_{p}$ = 2.79 and $\mu_{n}$ = -1.91 the experimental values (all in units of $\mu_{N}$).
Using this, $H_{1}$ may be written by

\begin{equation}
\braket{(0)^{3} | H_{1} | (0)^{3}} = \mu_{s}
\left[ \frac{I(\mathbf{R}) M^{*}_{N}(\mathbf{R})}{I_{0} M_{N}} \right]
\frac{1}{M^{*2}_{N}(\mathbf{R}) R}
\left[ \frac{d}{dR} 3 g^{q}_{\omega} \omega(\mathbf{R}) \right]
\mathbf{S} \cdot \mathbf{L},
\end{equation}
with $\mathbf{S}$ and $\mathbf{L}$ being respectively the nucleon spin and angular momentum operators.

The interaction of the magnetic moment of the nucleon with the
magnetic field of the $\omega$ meson seen from the rest frame of the nucleon induces a rotation
of the spin as a function of time. But even when $\mu_{s}$ is zero, the spin even so would rotate
because of the Thomas precession, which is independent of the structure and other effects~\cite{Jackson:1998nia}

\begin{equation}
H_{prec} = - \frac{1}{2} \mathbf{v} \times \frac{d \mathbf{v}}{dt} \cdot \mathbf{S},
\end{equation}
where

\begin{equation}
\frac{d \mathbf{v}}{dt} = - \frac{1}{M^{*2}_{N}(\mathbf{R})} \mathbf{\nabla}
\left[ M^{*}_{N}(\mathbf{R}) + 3 g^{q}_{\omega} \omega(\mathbf{R}) \right].
\end{equation}

The total spin-orbit interaction comes from combining this precession with the effect of $H_{1}$

\begin{equation}
H_{1} + H_{prec} = V_{s.o.} (\mathbf{R}) \mathbf{S} \cdot \mathbf{L},
\end{equation}
where

\begin{equation}
V_{s.o.}(\mathbf{R}) = - \frac{1}{2M^{*2}_{N}(\mathbf{R}) R}
\left[ \left( \frac{d}{dR} M^{*}_{N} (\mathbf{R}) \right)
+ \left( 1 - 2\mu_{s} \eta_{s} (\mathbf{R}) \right)
\left( \frac{d}{dR} 3 g^{q}_{\omega} \omega(\mathbf{R}) \right) \right],
\end{equation}
and 

\begin{equation}
\eta_{s} (\mathbf{R}) = \frac{I(\mathbf{R}) M^{*}_{N} (\mathbf{R})}{I_{0} M_{N}}.
\end{equation}

For completeness, we shall introduce the effect of the $\rho$ meson field.
We do this by adding the following interaction term to ${\cal L}_{I}$

\begin{equation}
{\cal L}^{\rho}_{I} = - g^{q}_{\rho} \bar{\psi}_{q}' \gamma_{\mu}
\frac{\tau_{\alpha}}{2} \psi_{q}'(u') \rho^{\mu, \alpha}_{I}(u'),
\end{equation}
where $\rho^{\mu, \alpha}_{I}$ is the $\rho$-meson field with isospin component $\alpha$ in the IRF,
and $\tau_{\alpha}$ are the Pauli matrices acting on the quark fields. In the mean field approximation
only $\alpha = 3$ contributes (the $\rho$ mean field is of a isovector type, so it must be in the third direction of isospin (proportional to $\rho_p - \rho_n$)). If the mean field value of the time component of the $\rho$ field is denoted by $b(\mathbf{R})$ in the NRF, the results can be transposed for the $\omega$ field. The difference coming
from the trivial isospin form factors

\begin{equation}
3 g^{q}_{\omega} \rightarrow g^{q}_{\rho} \frac{\tau^{N}_{3}}{2}, \quad
\mu_{s} \rightarrow \mu_{v} = \mu_{p} - \mu_{n},
\end{equation}
where $\tau^{N}_{3}/2$ is the nucleon isospin operator.

In the NRF, the energy and momentum of the nucleon moving in the meson fields are

\begin{eqnarray}
E &=& M^{*}_{N} (\mathbf{R})~ \text{cosh}~\xi + V (\mathbf{R}) \nonumber \\
\mathbf{P} &=& M^{*}_{N} (\mathbf{R}) \hat{v}~ \text{sinh}~\xi,
\end{eqnarray}
with

\begin{eqnarray}
\label{qmceq6}
V (\mathbf{R}) &=& V_{c} (\mathbf{R}) + V_{s.o.} (\mathbf{R}) \mathbf{S} \cdot \mathbf{L}, \nonumber \\
V_{c} (\mathbf{R}) &=& 3 g^{q}_{\omega} \omega(\mathbf{R}) 
+ g^{q}_{\rho} \frac{\tau^{N}_{3}}{2} b (\mathbf{R}) \nonumber \\
V_{s.o.} (\mathbf{R}) &=& - \frac{1}{2M^{*2}_{N}(\mathbf{R})R}
\left[\Delta_{\sigma} + \left( 1-2 \mu_{s} \eta_{s} (\mathbf{R}) \right) \Delta_{\omega} \right.
\nonumber \\
&+& \left.\left( 1-2 \mu_{v} \eta_{v} (\mathbf{R}) \frac{\tau^{N}_{3}}{2} \Delta_{\rho} \right) \right],
\end{eqnarray}
and

\begin{equation}
\Delta_{\sigma} = \frac{d}{dR} M^{*}_{N} (\mathbf{R}), \quad
\Delta_{\omega} = \frac{d}{dR} 3 g^{q}_{\omega} \omega(\mathbf{R}), \quad
\Delta_{\rho} = \frac{d}{dR} g^{q}_{\rho} b(\mathbf{R}).
\end{equation}

To finally quantize the model, the following Lagrangian is found by realizing that
the above energy-momentum expressions can be derived from it

\begin{equation}
L(\mathbf{R}, \mathbf{v}) = - M^{*}_{N} (\mathbf{R}) \sqrt{1-v^{2}} - V_{c} (\mathbf{R}).
\end{equation}

Then, the non-relativistic expression of the Lagrangian may be

\begin{equation}
L_{nr}(\mathbf{R}, \mathbf{v}) = \frac{1}{2}  M^{*}_{N} (\mathbf{R}) v^{2}
-  M^{*}_{N} (\mathbf{R}) - V_{c} (\mathbf{R}),
\end{equation}
which leads to the Hamiltonian

\begin{equation}
H_{nr} (\mathbf{R}, \mathbf{P}) = \mathbf{P} \cdot \frac{1}{2M^{*}_{N} (\mathbf{R})} \mathbf{P}
+ M^{*}_{N} (\mathbf{R}) + V (\mathbf{R}),
\end{equation}
where the spin-orbit interaction is reintroduced in the potential $V (\mathbf{R})$.
Thus, the nuclear, quantum Hamiltonian for the nucleus with atomic number $A$ is given by

\begin{equation}
\label{qmceq7}
H_{nr} = \sum^{A}_{i=1} H_{nr} (\mathbf{R}_{i}, \mathbf{P}_{i}), \quad \mathbf{P}_{i} = -i \mathbf{\nabla}_{i}.
\end{equation}

Now we present the equations for the meson fields. For the meson-field operators
$(\hat{\sigma}, \hat{\omega}^{\nu}, \hat{\rho}^{\nu} )$, the equations of motion
are given by

\begin{eqnarray}
\label{combeq1}
\left( \partial_{\mu} \partial^{\mu} + m^{2}_{\sigma} \right) \hat{\sigma}
&=& g^{q}_{\sigma} \bar{\psi}_{q} \psi_{q}, \nonumber \\
\left( \partial_{\mu} \partial^{\mu} + m^{2}_{\omega} \right) \hat{\omega}^{\nu}
&=& g^{q}_{\omega} \bar{\psi}_{q} \gamma^{\nu} \psi_{q}, \nonumber \\
\left( \partial_{\mu} \partial^{\mu} + m^{2}_{\rho} \right) \hat{\rho}^{\nu}
&=& g^{q}_{\rho} \bar{\psi}_{q} \gamma^{\nu} \frac{\tau_{3}}{2} \psi_{q}, 
\end{eqnarray}
where the masses of $\sigma$, $\omega$ and $\rho$ mesons are $m_{\sigma}$, $m_{\omega}$ 
and $m_{\rho}$, respectively.

To apply the mean field approximation to these meson fields, the mean fields are calculated as
the expectation values with respect to the nuclear ground state $\ket{A}$:

\begin{eqnarray}
\label{combeq2}
\braket{A|\hat{\sigma}(t, \mathbf{r}) |A} &=& \sigma(\mathbf{r}), \nonumber \\
\braket{A|\hat{\omega}^{\nu}(t, \mathbf{r}) |A} &=& \delta_{\nu , 0} \omega(\mathbf{r}), \nonumber \\
\braket{A|\hat{\rho}^{\nu}(t, \mathbf{r}) |A} &=& \delta_{\nu , 0} b(\mathbf{r}),
\end{eqnarray}
then we need the expectation values of the sources

\begin{equation}
\braket{A| \bar{\psi}_{q} \psi_{q} (t, \mathbf{r}) |A}, \quad
\braket{A| \bar{\psi}_{q} \gamma^{\nu} \psi_{q} (t, \mathbf{r}) |A}, \quad
\braket{A| \bar{\psi}_{q} \gamma^{\nu} \frac{\tau_{3}}{2} \psi_{q} (t, \mathbf{r}) |A}.
\end{equation}

In the mean field approximation the sources are the sums of the sources created
by each nucleon. We shall not treat now the source for the $\rho$ meson for simplicity.
Thus we write

\begin{equation}
\bar{\psi}_{q} \psi_{q} (t, \mathbf{r}) = \sum_{i} \braket{\bar{\psi}_{q} \psi_{q} (t, \mathbf{r})}_{i},
\quad
\bar{\psi}_{q} \gamma^{\nu} \psi_{q} (t, \mathbf{r}) = \sum_{i} \braket{\bar{\psi}_{q} \gamma^{\nu} \psi_{q} (t, \mathbf{r})}_{i},
\end{equation}
where $\braket{\cdots}_{i}$ is the matrix element in the nucleon $i$ located at $\mathbf{R}$ at time $t$.
In the Born-Oppenheimer approximation, which is used in this model, the nucleon structure is described
by 3 quarks in the lowest mode. In the IRF

\begin{eqnarray}
\braket{\bar{\psi}_{q}' \psi_{q}' (t', \mathbf{r}')}_{i} 
&=& 3 \sum_{m} \bar{\phi}^{0, m}_{i} (\mathbf{u}') \phi^{0,m}_{i} (\mathbf{u}') 
\equiv 3s_{i} (\mathbf{u}'), \nonumber \\
\braket{\bar{\psi}_{q}' \gamma^{\nu} \psi_{q}' (t', \mathbf{r}')}_{i}
&=& 3 \delta_{\nu ,0} \sum_{m} \phi^{\dagger 0,m}_{i} (\mathbf{u}') \phi^{0,m}_{i} (\mathbf{u}')
\equiv 3 \delta_{\nu ,0} w_{i} (\mathbf{u}').
\end{eqnarray}

In the NRF, the sources are given by

\begin{eqnarray}
\label{nrfsauce}
\braket{\bar{\psi}_{q} \psi_{q} (t, \mathbf{r})}_{i} 
&=& 3s_{i} \left( \left( r_{L} - R_{i,L} \right)~ \text{cosh}~\xi_{i}, 
\mathbf{r}_{\perp} - \mathbf{R}_{i, \perp} \right), \nonumber \\
\braket{\bar{\psi}_{q} \gamma^{0} \psi_{q} (t, \mathbf{r})}_{i} 
&=& 3w_{i} \left( \left( r_{L} - R_{i,L} \right)~ \text{cosh}~\xi_{i}, 
\mathbf{r}_{\perp} - \mathbf{R}_{i, \perp} \right)~ \text{cosh}~\xi_{i}, \nonumber \\
\braket{\bar{\psi}_{q} \mathbf{\gamma} \psi_{q} (t, \mathbf{r})}_{i} 
&=& 3w_{i} \left( \left( r_{L} - R_{i,L} \right)~ \text{cosh}~\xi_{i}, 
\mathbf{r}_{\perp} - \mathbf{R}_{i, \perp} \right) \hat{v}~ \text{sinh}~\xi_{i}.
\end{eqnarray}

These sources can be rewritten in the form

\begin{eqnarray}
\braket{\bar{\psi}_{q} \psi_{q} (t, \mathbf{r})}_{i} 
&=& \frac{3}{(2 \pi)^{3}~ \text{cosh}~\xi_{i}}
\int d \mathbf{k} e^{i \mathbf{k} \cdot (\mathbf{r} - \mathbf{R}_{i})} S( \mathbf{k}, \mathbf{R}_{i} ),
\nonumber \\
\braket{\bar{\psi}_{q} \gamma^{0} \psi_{q} (t, \mathbf{r})}_{i}
&=& \frac{3}{(2 \pi)^{3}} \int d \mathbf{k} e^{i \mathbf{k} \cdot (\mathbf{r} 
- \mathbf{R}_{i})} W( \mathbf{k}, \mathbf{R}_{i} ), \nonumber \\
\braket{\bar{\psi}_{q} \mathbf{\gamma} \psi_{q} (t, \mathbf{r})}_{i}
&=& \frac{3}{(2 \pi)^{3}} \hat{v} \int d \mathbf{k} e^{i \mathbf{k} \cdot (\mathbf{r} 
- \mathbf{R}_{i})} W( \mathbf{k}, \mathbf{R}_{i} ),
\end{eqnarray}
with the sources in momentum space being

\begin{eqnarray}
S( \mathbf{k}, \mathbf{R}_{i} ) &=& \int d\mathbf{u} 
e^{-i \left( \mathbf{k}_{\perp} \cdot \mathbf{u}_{\perp}
+ k_{L} u_{L} / \text{cosh}~\xi_{i} \right) } s_{i} (\mathbf{u}), \nonumber \\
W( \mathbf{k}, \mathbf{R}_{i} ) &=& \int d\mathbf{u} 
e^{-i \left( \mathbf{k}_{\perp} \cdot \mathbf{u}_{\perp}
+ k_{L} u_{L} / \text{cosh}~\xi_{i} \right) } w_{i} (\mathbf{u}).
\end{eqnarray}

The mean field expressions for the meson sources are then given by

\begin{eqnarray}
\braket{A| \bar{\psi}_{q} \psi_{q} (t, \mathbf{r}) |A} &=& \frac{3}{(2 \pi)^{3}}
\int d \mathbf{k} e^{i \mathbf{k} \cdot \mathbf{r}}
\braket{A| \sum_{i} (\text{cosh}~\xi_{i})^{-1} 
e^{-i \mathbf{k} \cdot \mathbf{R}_{i}} S( \mathbf{k}, \mathbf{R}_{i} ) |A} \nonumber \\
\braket{A| \bar{\psi}_{q} \gamma^{0} \psi_{q} (t, \mathbf{r}) |A} &=& \frac{3}{(2 \pi)^{3}}
\int d \mathbf{k} e^{i \mathbf{k} \cdot \mathbf{r}}
\braket{A| \sum_{i} e^{-i \mathbf{k} \cdot \mathbf{R}_{i}} W( \mathbf{k}, \mathbf{R}_{i} ) |A} \nonumber \\
\braket{A| \bar{\psi}_{q} \boldsymbol{\gamma} \psi_{q} (t, \mathbf{r}) |A} &=& 0
\end{eqnarray}
where the last equation follows from the fact that the velocity vector averages to zero
(see Eq.~(\ref{nrfsauce})).
This can be simplified even further to~\cite{Guichon:1995ue}

\begin{eqnarray}
\label{combeq3}
\braket{A| \bar{\psi}_{q} \psi_{q} (t, \mathbf{r}) |A} 
&=& 3S(\mathbf{r}) \rho_{s}(\mathbf{r}),
\nonumber \\
\braket{A| \bar{\psi}_{q} \gamma^{\nu} \psi_{q} (t, \mathbf{r}) |A}
&=& 3 \delta_{\nu ,0} \rho_{B} (\mathbf{r}),
\nonumber \\
\braket{A| \bar{\psi}_{q} \gamma^{\nu} \frac{\tau_{3}}{2} \psi_{q} (t, \mathbf{r}) |A}
&=& \delta_{\nu ,0} \rho_{3} (\mathbf{r}),
\end{eqnarray}
with the scalar ($\rho_{s}$), baryon ($\rho_{B}$) and isospin ($\rho_{3}$) 
densities of the nucleon in the nucleus

\begin{eqnarray}
\label{qmceq3}
\rho_{s}(\mathbf{r}) &=& \braket{A| \sum_{i} \frac{M^{*}_{N}(\mathbf{R}_{i})}
{E_{i} - V(\mathbf{R}_{i})} \delta(\mathbf{r} - \mathbf{R}_{i}) |A}, \nonumber \\
\rho_{B} (\mathbf{r}) &=& \braket{A| \sum_{i} \delta(\mathbf{r} - \mathbf{R}_{i}) |A}, \nonumber \\
\rho_{3} (\mathbf{r}) &=& \braket{A| \sum_{i} \frac{\tau_{3}}{2} \delta(\mathbf{r} - \mathbf{R}_{i}) |A},
\end{eqnarray}
where the $S(\mathbf{r})$ ($=S(\mathbf{r}, \sigma(\mathbf{r}))$) in the first equation, for the 
$i$-th nucleon at $\mathbf{r}$ is given by

\begin{equation}
\label{qmceq2}
S(\mathbf{r}) = S(\mathbf{0}, \mathbf{r}) = \int d \mathbf{u} s_{i} (\mathbf{u})
= \frac{\Omega_{0}/2 + m^{*}_{q} R_{B} (\Omega_{0} - 1)}
{\Omega_{0}(\Omega_{0} - 1) + m^{*}_{q} R_{B}/2},
\end{equation} 
in the MIT bag model. Then, combining the Eqs.~(\ref{combeq1}), (\ref{combeq2})
and (\ref{combeq3}), we finally get the equations
for $\sigma (\mathbf{r})$, $\omega (\mathbf{r})$ and $b(\mathbf{r})$

\begin{eqnarray}
\label{qmceq4}
\left(- \nabla^{2}_{r} + m^{2}_{\sigma} \right) \sigma (\mathbf{r})
&=& g_{\sigma} C(\mathbf{r}) \rho_{s} (\mathbf{r}), \nonumber \\
\left(- \nabla^{2}_{r} + m^{2}_{\omega} \right) \omega (\mathbf{r})
&=& g_{\omega} \rho_{B} (\mathbf{r}), \nonumber \\
\left(- \nabla^{2}_{r} + m^{2}_{\rho} \right) \rho (\mathbf{r})
&=& g_{\rho} \rho_{3} (\mathbf{r}),
\end{eqnarray}
where $C(\mathbf{r})$ ($=C(\mathbf{r}, \sigma(\mathbf{r}))$) and the meson-nucleon coupling constants are defined by

\begin{equation}
\label{eqsig}
C(\mathbf{r}, \sigma(\mathbf{r})) = S(\mathbf{r}, \sigma(\mathbf{r}))/S(\sigma = 0), \quad
g_{\sigma} = 3 g^{q}_{\sigma} S(\sigma = 0), \quad
g_{\omega} = 3g^{q}_{\omega}, \quad
g_{\rho} = g^{q}_{q}.
\end{equation}

The energy carried by the mean fields is

\begin{equation}
E_{meson} = \frac{1}{2} \int d (\mathbf{r}) \left[ \left( \mathbf{\nabla} \sigma \right)^{2}
+ m^{2}_{\sigma} \sigma^{2} - \left( \mathbf{\nabla} \omega \right)^{2} - m^{2}_{\omega} \omega^{2}
- \left( \mathbf{\nabla} b \right)^{2} - m^{2}_{\rho} b^{2} \right].
\end{equation}

In order to make the calculation for finite nuclei self-consistent, the following steps must be taken:

\begin{enumerate}
\item When using the MIT bag as the quark model for the nucleon, choose a quark mass $m_{q}$ and fix the bag parameters $B_p$ and $z_{0}$ to fit the nucleon mass and bag radius.

\item Assume that the coupling constants and the masses of the mesons are known.

\item Calculate $I(\sigma)$ (Eq.~(\ref{qmceq1})) and $S(\sigma)$
(Eq.~(\ref{qmceq2})) at a given value of $\sigma$.

\item Guess initial forms of the densities $\rho_{s}(\mathbf{r})$, 
$\rho_{B}(\mathbf{r})$ and $\rho_{3}(\mathbf{r})$ in Eqs.~(\ref{qmceq3}).

\item Fixed $\rho_{s}(\mathbf{r})$, 
$\rho_{B}(\mathbf{r})$ and $\rho_{3}(\mathbf{r})$, solve Eqs.~(\ref{qmceq4}) for the meson fields.

\item Evaluate the effective nucleon mass in the nucleus $M^{*}_{N}$ (Eq.~(\ref{qmceq5})) and the potential $V(\mathbf{r})$ given in Eq.~(\ref{qmceq6}).

\item Solve the eigenvalue problem given by the nuclear Hamiltonian (Eq.~(\ref{qmceq7})) and compute $\rho_{s}(\mathbf{r})$, $\rho_{B}(\mathbf{r})$ and $\rho_{3}(\mathbf{r})$.

\item Go to step 5 and iterate until self-consistency is achieved.
\end{enumerate}
\subsection{Relativistic model}
\label{Ch3Sc2-2}

\noindent

Now we express the model in the framework of relativisitic field theory at the mean field level.
The Lagrangian density for the symmetric nuclear system may be written as~\cite{Saito:2005rv}

\begin{equation}
{\cal L} = \bar{\psi} \left[ i \gamma \cdot \partial - M^{*}_{N} (\sigma)
- g_{\omega} \hat{\omega}^{\mu} \gamma_{\mu} \right] \psi + {\cal L}_{meson},
\end{equation}
where $\psi$, $\hat{\sigma}$ and $\hat{\omega}$ are the nucleon, $\sigma$ and $\omega$ operators, respectively,
and the free meson Lagrangian in the last term is given by

\begin{equation}
{\cal L}_{meson} = \frac{1}{2} \left( \partial_{\mu} \hat{\sigma} \partial^{\mu} \hat{\sigma}
- m^{2}_{\sigma} \hat{\sigma}^{2} \right) - \frac{1}{2} \partial_{\mu} \hat{\omega}_{\nu}
\left( \partial^{\mu} \hat{\omega}^{\nu} - \partial^{\nu} \hat{\omega}^{\mu} \right)
+ \frac{1}{2} m^{2}_{\omega} \hat{\omega}^{\mu} \hat{\omega}_{\mu}.
\end{equation}

The nucleon mass can be separated as

\begin{equation}
M^{*}_{N} (\hat{\sigma}) = M_{N} - g_{\sigma} (\hat{\sigma}) \hat{\sigma},
\end{equation}
and so the Lagrangian density can be rewritten as

\begin{equation}
\label{intqmcL}
{\cal L} = \bar{\psi} \left[ i \gamma \cdot \partial - M_{N} + g_{\sigma} (\hat{\sigma}) \hat{\sigma}
- g_{\omega} \hat{\omega}^{\mu} \gamma_{\mu} \right] \psi + {\cal L}_{meson}.
\end{equation}

In the mean field approximation, we can replace the meson-field operators by their expectation 
values in the Lagrangian as $\hat{\sigma} \rightarrow \sigma(\mathbf{r})$ and 
$\hat{\omega}^{\mu}(\mathbf{r}) \rightarrow \delta_{\mu ,0}\omega(\mathbf{r})$.
Then, the nucleon and meson fields satisfy the equations:

\begin{eqnarray}
\label{eqmotqmc}
\left( i \gamma \cdot \partial - M^{*}_{N} (\sigma) 
- g_{\omega} \gamma_{0} \omega \right) \psi &=& 0, \nonumber \\
\left( - \nabla^{2}_{r} + m^{2}_{\sigma} \right) \sigma (\mathbf{r})
&=& - \left( \frac{\partial}{\partial \sigma} M^{*}_{N} (\sigma ) \right)
\braket{A| \bar{\psi} \psi (\mathbf{r}) |A}, \nonumber \\
\left( - \nabla^{2}_{r} + m^{2}_{\omega} \right) \omega (\mathbf{r})
&=& g_{\omega} \braket{A| \psi^{\dagger} \psi (\mathbf{r}) |A},
\end{eqnarray}
where

\begin{equation}
\frac{d}{d \sigma} M^{*}_{N} (\sigma ) = - C (\sigma ) g_{\sigma} (\sigma = 0)
= - \frac{d}{d \sigma} \left( g_{\sigma} (\sigma) \sigma \right).
\end{equation}

To include now the $\rho$-meson contribution, and also to the Coulomb interaction, 
the following interaction Lagrangian must be added to Eq.~(\ref{intqmcL})~\cite{Saito:2005rv}

\begin{equation}
{\cal L}_{\rho + \gamma} = - \bar{\psi}_{q} \left[ g^{q}_{\rho} \frac{\tau_{3}}{2}
\gamma_{\mu} \rho^{\mu} - e \left( \frac{1}{6} + \frac{\tau_{3}}{2} \right)
\gamma_{\mu} A^{\mu} \right] \psi_{q}.
\end{equation}

The expressions for the $\rho$ and Coulomb fields can be achieved by including trivial isospin factors
$3g^{q}_{\omega} \omega (\mathbf{r}) \rightarrow g_{\rho} \left( \tau^{N}_{3}/2 \right) b (\mathbf{r})$
or $ \rightarrow (e/2)(1 + \tau^{N}_{3}) A(\mathbf{r})$. The effective Lagrangian density for the QMC model in the mean field approximation is given by

\begin{eqnarray}
\label{Lqmc}
{\cal L}_{QMC} &=& \bar{\psi} \left[ i \gamma \cdot \partial - M^{*}_{N} \left( \sigma (\mathbf{r}) \right)
- g_{\omega} \omega (\mathbf{r}) \gamma_{0} - g_{\rho} \frac{\tau^{N}_{3}}{2} b (\mathbf{r})
\gamma_{0} - \frac{e}{2} \left( 1 + \tau^{N}_{3} \right) A (\mathbf{r}) \gamma_{0} \right] \psi
\nonumber \\
&-& \frac{1}{2} \left[ \left(\mathbf{\nabla} \sigma (\mathbf{r})\right)^{2} + m^{2}_{\sigma} \sigma (\mathbf{r})^{2} \right]
+ \frac{1}{2} \left[ \left(\mathbf{\nabla} \omega (\mathbf{r})\right)^{2} + m^{2}_{\omega} \omega (\mathbf{r})^{2} \right]
\nonumber \\
&+& \frac{1}{2} \left[ \left(\mathbf{\nabla} b (\mathbf{r})\right)^{2} + m^{2}_{\rho} b (\mathbf{r})^{2} \right]
+ \frac{1}{2} \left( \mathbf{\nabla} A (\mathbf{r}) \right)^{2}.
\end{eqnarray}

\subsection{Nuclear matter properties}
\label{Ch3Sc2-3}

\noindent

\begin{figure}[htb]
\vspace{4ex}
\centering
 \includegraphics[scale=0.35]{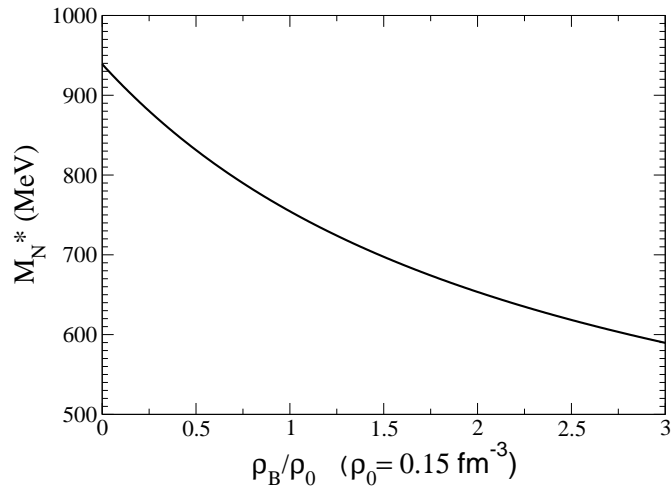}
 \caption{Effective nucleon mass as a function of baryon density~\cite{deMelo:2014gea}.}
 \label{nucmass}
\end{figure}

Considering the rest frame of infinitely large, symmetric nuclear matter,
we set to zero the Coulomb field $A (\mathbf{r})$ in Eq.~(\ref{Lqmc}) Lagrangian,
and drop all the terms with any derivatives of the fields, and so it is done in
the equations of motion in Eq.~(\ref{eqmotqmc}).
So in the nuclear matter limit, within the Hartree mean-field approximation,
the baryon and scalar densities with the nucleon Fermi momentum $k_F$ are respectively 
given by~\cite{Saito:2005rv, Tsushima:2018goq, Walecka:1974qa}

\begin{eqnarray}
\rho_{B} &=& \frac{4}{(2\pi)^{3}} \int d \mathbf{k} \theta (k_{F} -k)
= \frac{2k^{3}_{F}}{3 \pi^{2}}, \nonumber \\
\rho_{s} &=& \frac{4}{(2\pi)^{3}} \int d \mathbf{k} \theta (k_{F} -k)
\frac{M^{*}_{N}}{\sqrt{M^{*2}_{N} + \mathbf{k}^{2}}},
\end{eqnarray}
where $M^{*}_{N}$ is the (constant) value of the effective nucleon mass at a given density,
shown in Fig.~\ref{nucmass}~\cite{deMelo:2014gea}.
The meson mean fields $\sigma$ and $\omega$ are given by Eq.~(\ref{eqmotqmc}) with the
derivatives of the fields set to zero

\begin{eqnarray}
\omega &=& \frac{g_{\omega} \rho_{B}}{m^{2}_{\omega}}, \nonumber \\
\sigma &=& \frac{g_{\sigma}}{m^{2}_{\sigma}} C_{N} (\sigma) 
\frac{4}{(2\pi)^{3}} \int d \mathbf{k} \theta (k_{F} -k)
\frac{M^{*}_{N}}{\sqrt{M^{*2}_{N} + \mathbf{k}^{2}}},
\end{eqnarray}
where $C_{N} (\sigma)$ is the constant value of the scalar density ratio (see Eq.~\ref{eqsig}).

The total energy per nucleon can be evaluated after one solves the self-consistency for
the $\sigma$ field

\begin{equation}
E^{tot}/A = \frac{4}{(2\pi)^{3} \rho_{B}} \int d \mathbf{k} \theta (k_{F} -k)
\sqrt{M^{*2}_{N} + \mathbf{k}^{2}} + \frac{m^{2}_{\sigma} \sigma^{2}}{2 \rho_{B}}
+ \frac{g^{2}_{\omega} \rho_{B}}{2m^{2}_{\omega}}.
\end{equation}

\begin{figure}[htb]
\vspace{4ex}
\centering
 \includegraphics[scale=0.35]{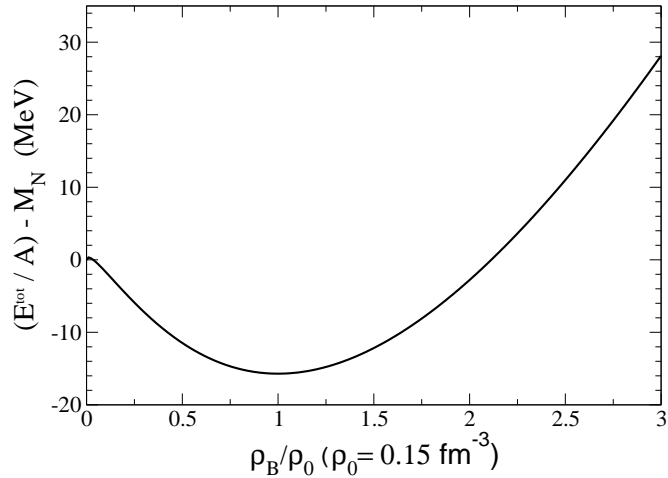}
 \caption{Energy per nucleon for symmetric nuclear matter $(E^{tot}/A) - M_{N}$ 
for ($m_{q}$ = 5 MeV and $R_{N}$ = 0.8 fm)~\cite{Tsushima:2019wmq}.}
 \label{figennuc}
\end{figure}

The coupling constants $g_{\sigma}$ and $g_{\omega}$ are determined so as to fit the binding energy 
(-15.7 MeV) per nucleon at the saturation density, $\rho_{0}$ = 0.15 fm$^{-3}$, for symmetric nuclear matter.
The coupling constants at the nucleon level are $(g^{N}_{\sigma})/4\pi$ = 5.39 and $g^{2}_{\omega}/4\pi$ = 5.30.
We show in Fig.~\ref{figennuc} the calculated values of the total energy per nucleon, obtained
using the determined quark-meson coupling constants~\cite{Tsushima:2019wmq}.

\begin{figure}[htb]
\vspace{4ex}
\centering
 \includegraphics[scale=0.35]{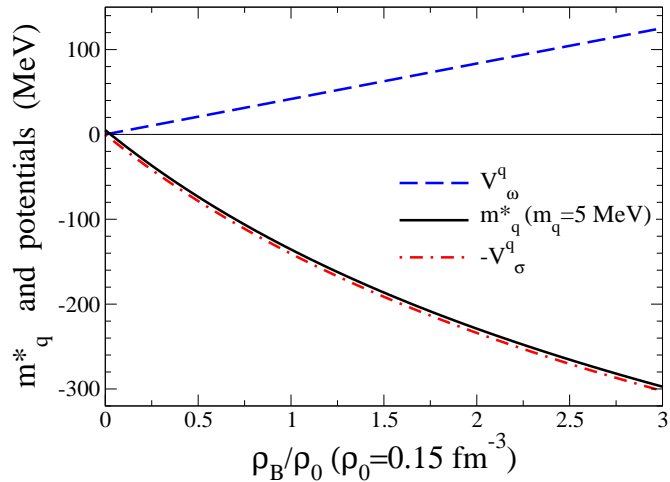}
 \caption{The effective light-quark mass $m^{*}_{q}$, and the scalar $(-V^{q}_{\sigma})$  and vector 
$V^{q}_{\omega}$ potentials felt by the light quarks~\cite{Tsushima:2019wmq}, introduced first
by~\cite{Tsushima:1997df}.}
 \label{lqmass}
\end{figure}

The effective light-quark mass in nuclear medium is given by 
$m^{*}_{q} = m_{q} - g^{q}_{\sigma} \sigma = m_{q} - V^{q}_{\sigma}$, with $V^{q}_{\sigma}$ being the 
interaction of the light-quark with the scalar field. The effective light-quark mass, together with
the corresponding mean field potentials felt by the light-quarks, $V^{q}_{\sigma}$ and 
$V^{q}_{\omega}$ , are shown in Fig.~\ref{lqmass}~\cite{Tsushima:2019wmq}. One should have in mind that the negative value of $m^{*}_{q}$ only reflects the strength of the attractive scalar potential, and thus the interpretation that the mass for a physical particle is positive should not be applied here.

\subsection{Hadron properties in nuclear matter}
\label{Ch3Sc2-4}

\noindent

The original version of the QMC model is based on the MIT bag, and so, the Dirac equations
for the quarks and antiquarks in nuclear matter, in bags of hadrons $h$, 
($q = u$ or $d$, and $Q = s, c,$ or $b$) neglecting the Coulomb force
are given by ($|x| \leq$ bag radius)

\begin{eqnarray}
&&\left[i\gamma \cdot \partial_{x} - \left(m_{q} - V^{q}_{\sigma}\right)
\mp \gamma^{0} \left(V^{q}_{\omega} + \frac{1}{2}V^{q}_{\rho}\right)\right]
\begin{pmatrix}
        \psi_{u}\left(x\right)\\
        \psi_{\overline{u}}\left(x\right)
       \end{pmatrix} = 0,\\
&&\left[i\gamma \cdot \partial_{x} - \left(m_{q} - V^{q}_{\sigma}\right)
\mp \gamma^{0} \left(V^{q}_{\omega} - \frac{1}{2}V^{q}_{\rho}\right)\right]
\begin{pmatrix}
        \psi_{d}\left(x\right)\\
        \psi_{\overline{d}}\left(x\right)
       \end{pmatrix} = 0,\\
&&\left[i\gamma \cdot \partial_{x} - m_{Q}\right]\psi_{Q, \overline{Q}}\left(x\right) = 0.
\end{eqnarray}

The (constant) mean-field potentials for the light $u$ and $d$ quarks in nuclear
matter are defined by $V^{q}_{\sigma} \equiv g^{q}_{\sigma}\sigma$, 
$V^{q}_{\omega} \equiv g^{q}_{\omega}\omega = g^q_\omega\, \delta^{\mu,0} \omega^\mu$, 
$V^{q}_{\rho} \equiv g^{q}_{\rho}b = g^q_\rho\, \delta^{i,3} \delta^{\mu,0} \rho^{i,\mu}$, 
with the $g^{q}_{\sigma}$, $g^{q}_{\omega}$ and
$g^{q}_{\rho}$ being the corresponding quark-meson coupling constants, as before.

The static solution for the ground state quarks (antiquarks) with flavor
$f$ in the hadron $h$ is written as 
$\psi_{f}\left(x\right) = N_{f}e^{-i\epsilon_{f}t/R^{*}_{h}}\psi_{f}\left(\textbf{r}\right)$, 
with normalization 
factor $N_{f}$ and $\psi_{f}\left(\textbf{r}\right)$ as the corresponding spin and spatial part of 
the wave function.
The eigenenergies for the quarks and antiquarks in $h$ in units of 
$1/R^{*}_{h}$ are~\cite{Tsushima:1997df}

\begin{eqnarray}
&&\begin{pmatrix}
        \epsilon_{u}\left(x\right)\\
        \epsilon_{\overline{u}}\left(x\right)
       \end{pmatrix} = \Omega^{*}_{q} \pm R^{*}_{h} 
\left(V^{q}_{\omega} + \frac{1}{2}V^{q}_{\rho}\right),\\
&&\begin{pmatrix}
        \epsilon_{d}\left(x\right)\\
        \epsilon_{\overline{d}}\left(x\right)
       \end{pmatrix} = \Omega^{*}_{q} \pm R^{*}_{h} 
\left(V^{q}_{\omega} - \frac{1}{2}V^{q}_{\rho}\right),\\
&&\epsilon_{Q} = \epsilon_{\overline{Q}} = \Omega_{Q}.
\end{eqnarray}

The light quark field is modified by the term $V_{s} = g^{q}_{\sigma} \sigma$ which acts as 
$m_{q} - V^{q}_{\sigma}$. 
Then, the mass modification is expected to occur to 
the hadrons containing those light quarks inside their bags.
The mass of a hadron $h$ in a nuclear medium $m^{*}_{h}$ is calculated by

\begin{eqnarray}
\label{mstar}
&& m^{*}_{h} = \sum_{j=q,\overline{q},Q,\overline{Q}}
\frac{n_{j}\Omega^{*}_{j}- Z_{h}}{R^{*}_{h}} + \frac{4}{3} \pi R^{*3}_{h}B_{p},
\\
&&\left. \frac{d m^{*}_{h}}{d R_{h}}\right|_{R_{h} = R^{*}_{h}} = 0.
\end{eqnarray}

with $\Omega^{*}_{q} = \Omega^{*}_{\overline{q}} = \left[x^{2}_{q} + \left(R^{*}_{h} 
m^{*}_{q}\right)^{2}\right]^{\frac{1}{2}}$, where $m^{*}_{q} = m_{q} - g^{q}_{\sigma}\sigma$
and $\Omega^{*}_{Q} = \Omega^{*}_{\overline{Q}} = \left[x^{2}_{Q} + \left(R^{*}_{h} 
m_{Q}\right)^{2}\right]^{\frac{1}{2}}$, with $x_{q,Q}$ being the lowest mode bag eigenfrequencies. 
$B_{p}$ is the bag constant, and $n_{q,Q}$ ($n_{\overline{q},\overline{Q}}$) are the lowest 
mode quark (antiquark) numbers for the quark flavors $q$ and $Q$ in the hadron $h$, 
and the $Z_{h}$ parameterize the sum of the center-of-mass 
and gluon fluctuation effects and are assumed to be independent of density~\cite{Guichon:1995ue}.
Since $x_{q}$ and $R_{h}$ are modified in medium when $h$ contains light quarks, $m_{h^{*}}$ is
also modified in medium.

The bag radius of the nucleon in free space input value chosen is $R_{N}$ = 0.8 fm. The parameter $B_p$ is determined by the nucleon mass $M_N =$ 939 MeV with $R_N = $0.8 fm, and $Z_h$ is fixed to reproduce the hadron mass, while the quark-meson coupling constants, 
$g^{q}_{\sigma}$, $g^{q}_{\omega}$ and $g^{q}_{\rho}$ were determined by the fit 
to the saturation energy (-15.7 MeV) at the saturation density ($\rho_0 = 0.15$ fm$^{-3}$) of 
symmetric nuclear matter for $g^{q}_{\sigma}$ and $g^{q}_{\omega}$ , and the bulk symmetry energy 
(35 MeV) for $g^{q}_{\rho}$~\cite{Guichon:1987jp,Saito:2005rv}
(see Eq.~\ref{eqsig}).

\subsection{QMC description of finite nuclei}
\label{Ch3Sc2-5}

\noindent

The description of a finite nucleus with different numbers of protons and neutrons ($Z \ne N$)
needs to include the contributions of the $\rho$ meson, as well the Coulomb force.
The following equations for static, spherically symmetric nuclei, comes from the variation of 
${\cal L}_{QMC}$~\cite{Saito:2005rv}

\begin{eqnarray}
\frac{d^{2}}{dr^{2}} \sigma (r) + \frac{2}{r} \frac{d}{dr} \sigma (r) - m^{2}_{\sigma} \sigma (r)
&=& - g_{\sigma} C_{N} \left( \sigma (r) \right) \rho_{s} (r) \nonumber \\
&\equiv & - g_{\sigma} C_{N} \left( \sigma (r) \right) \sum^{occ}_{\alpha} d_{\alpha} (r)
\left( |G_{\alpha} (r)|^{2} - |F_{\alpha} (r)|^{2} \right), \\
\frac{d^{2}}{dr^{2}} \omega (r) + \frac{2}{r} \frac{d}{dr} \omega (r) - m^{2}_{\omega} \omega (r)
&=& - g_{\omega} \rho_{B} (r) \nonumber \\
&\equiv & - g_{\omega} \sum^{occ}_{\alpha} d_{\alpha} (r)
\left( |G_{\alpha} (r)|^{2} + |F_{\alpha} (r)|^{2} \right), \\
\frac{d^{2}}{dr^{2}} b(r) + \frac{2}{r} \frac{d}{dr} b(r) - m^{2}_{\rho} b(r)
&=& -\frac{g_{\rho}}{2} \rho_{3} (r) \nonumber \\
&\equiv & -\frac{g_{\rho}}{2} \sum^{occ}_{\alpha} d_{\alpha} (r) (-)^{t_{\alpha} - 1/2}
\left( |G_{\alpha} (r)|^{2} + |F_{\alpha} (r)|^{2} \right), \\
\frac{d^{2}}{dr^{2}} A(r) + \frac{2}{r} \frac{d}{dr} A(r) &=& -e \rho_{p} (r) \nonumber \\
&\equiv & -e \sum^{occ}_{\alpha} d_{\alpha} (r) (t_{\alpha} + \frac{1}{2})
\left( |G_{\alpha} (r)|^{2} + |F_{\alpha} (r)|^{2} \right),
\end{eqnarray}
where $d_{\alpha} (r) = (2j_{\alpha} + 1)/4\pi^{2}$ and

\begin{eqnarray}
\frac{d}{dr} G_{\alpha} (r) + \frac{k}{r} G_{\alpha} (r) 
- \left[ \epsilon_{\alpha} - g_{\omega} \omega (r) - t_{\alpha} g_{\rho} b(r) \right.
&-& (t_{\alpha} + \frac{1}{2} e A(r) + M_{N} \nonumber \\
&-& \left. g_{\sigma} \left( \sigma (r) \right) \sigma (r) \right] F_{\alpha} (r) =0, \\
\frac{d}{dr} F_{\alpha} (r) - \frac{k}{r} F_{\alpha} (r) 
+ \left[ \epsilon_{\alpha} - g_{\omega} \omega (r) - t_{\alpha} g_{\rho} b(r) \right.
&-& (t_{\alpha} + \frac{1}{2} e A(r) - M_{N} \nonumber \\
&+& \left. g_{\sigma} \left( \sigma (r) \right) \sigma (r) \right] G_{\alpha} (r) =0,
\end{eqnarray}
with $\alpha$ labeling the quantum numbers, $\epsilon$ being the energy, and $iG_{\alpha} (r)/r$ 
and $-F_{\alpha}(r)/r$ being, respectively, the radial part of the upper and 
the lower components of the solution to the Dirac equation for the 
nucleon~\cite{Saito:1996yb,Horowitz:1981xw}

\begin{equation}
\psi (\mathbf{r}) =  
\begin{pmatrix}
i \left[ G_{\alpha}(r)/r \right] \Phi_{\kappa} \\
- \left[ F_{\alpha}(r)/r \right] \Phi_{-\kappa} 
\end{pmatrix}
\chi_{t_{\alpha}},
\end{equation}
under the normalization condition

\begin{equation}
\int dr \left( |G_{\alpha} (r)|^{2} + |F_{\alpha}|^{2} \right) = 1,
\end{equation}
where $\chi_{t_{\alpha}}$ is a two-component spinor and $\Phi_{\kappa, -\kappa}$ is the angular part wave functions.

The angular quantum numbers are specified by $\kappa$ and the eigenvalue of the 
isospin operator by $t_{\alpha}$.
Also, it is well parametrized the $\sigma$-dependent coupling $g_\sigma (\sigma(\mathbf{r}))$ as,

\begin{equation}
g_{\sigma} \left( \sigma (\mathbf{r}) \right) = g_{\sigma} 
\left[ 1 - \frac{a_{N}}{2} g_{\sigma} \sigma (\mathbf{r}) \right],
\end{equation}
where $a_{N}$ is a slope parameter for the nucleon, ranging $a_N$ = 8.97 $\sim$ 9.01 $\times$ 10$^{-4}$ (MeV$^{-1}$), 
depending on the parameterization for the $\sigma$ mass in matter.
The total energy of the system is then given by

\begin{eqnarray}
E_{tot} &=& \sum^{occ}_{\alpha} (2j_{\alpha} + 1) \epsilon_{\alpha}
- \frac{1}{2} \int d \mathbf{r} \left[ -g_{\sigma} C_{N} \right. 
\left( \sigma (r) \right) \sigma (r) \rho_{s}(r) \nonumber \\
&+& g_{\omega} \omega (r) \rho_{B}(r) + \frac{1}{2} g_{\rho} b(r) \rho_{3}(r)
\left. + eA(r) \rho_{p}(r) \right].
\end{eqnarray}


\section{$B$ and $B^{*}$ meson in-medium masses}

\noindent

In this section we focus on the properties of $B$ and $B^*$ mesons 
in nuclear matter, calculating their effective masses using the 
QMC model. This is enough, since the vector potentials appearing  
in the $BB, BB^*$ and $B^*B^*$ loops in the self-energy calculation,  
or in the energy-contour integral for each loop, cancel out, 
and this is consistent with the baryon number conservation 
at the quark level.

\begin{figure}[htb]
\vspace{4ex}
\centering
 \includegraphics[scale=0.35]{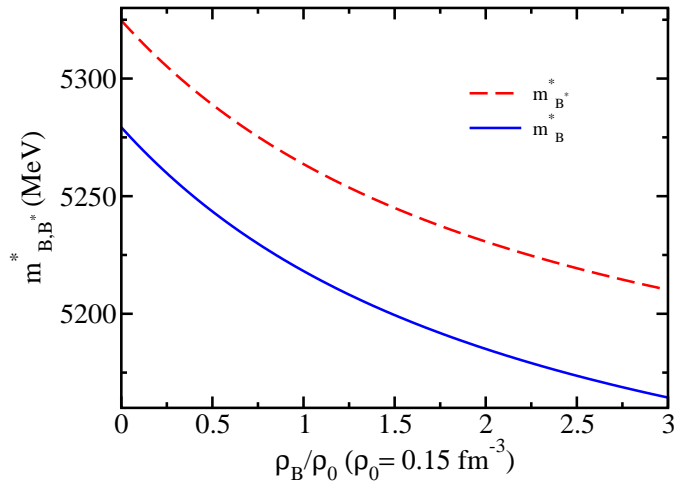}
 \caption{$B$ and $B^{*}$ meson effective Lorentz-scalar masses  
 in symmetric nuclear matter versus baryon density.}
 \label{fig2}
\end{figure}

To determine the in-medium masses of the $B$ and $B^{*}$ mesons by the Eq.~(\ref{mstar})
($h = B,B^{*}$, and $Q = b$), we obtain the following bag parameters for the bag radius of the nucleon in free space with value $R_{N}$ = 0.8 fm as input:
$g^{q}_{\sigma}$ = 5.69, $g^{q}_{\omega}$ = 2.72, $g^{q}_{\rho}$ = 9.33, 
$B^{1/4}_{p}$ = 170 MeV, $Z_{B}$ = -1.136 and $Z_{B^{*}}$ = -1.334, with he vacuum mass values (input)
$m_B$ = 5279 MeV and $m_{B^{*}}$ = 5325 MeV, and the chosen values ($m_{q}$, $m_{b}$) = (5, 4200) MeV for the current quark masses. 
The QMC model predicts a similar amount in the decrease of the in-medium effective 
Lorentz-scalar masses of the $B$ and $B^{*}$ mesons in symmetric nuclear matter 
as shown in Fig.~\ref{fig2}.
At $\rho_0$ the mass shifts of the $B$ and $B^*$ mesons are respectively, 
$(m^*_B - m_B)=-61$ MeV and $(m^*_{B^*}-m_{B^*})=-61$ MeV, the difference 
in their mass shift values appears in the next digit.
To calculate the $\Upsilon$ and $\eta_b$ meson self-energies in symmetric nuclear matter 
by the excited $B$ and $B^*$ meson intermediate states in the loops, 
we use the calculated in-medium masses of them shown in Fig.~\ref{fig2}.

\chapter{\boldmath{$\Upsilon$} mass shift and bound state energies} 

\label{Chapter4} 

\thispagestyle{empty}

\noindent

The mass of the in-medium $\Upsilon$ meson (and of heavy quarkonia in general) is also expected to 
be modified as a result of the interaction between $\Upsilon$ with nuclei. 
But differently from the $B$ and $B^{*}$ mesons considered before in the last chapter, heavy quarkonium has no light quarks, and so, interactions involving light-quark or light-flavored hadron exchanges do not occur in the lowest order.
However, the heavy quarkonium intermediate state hadrons do contain light quarks, what makes the excitation of these intermediate state hadrons a possible mechanism for the heavy quarkonium interaction with the nuclear medium.
The $\Upsilon$ mass shift in nuclear medium is then studied as the contribution of its possible intermediate state mesons to its self-energy by employing effective Lagrangians for the coupling of these mesons to $\Upsilon$~\cite{Zeminiani:2020aho}.
The resultant negative shift in the $\Upsilon$ mass is interpreted as an attractive potential that
can bind the $\Upsilon$ meson to nuclei.

In this chapter, the $\Upsilon$ mass shift in nuclear medium is studied using effective Lagrangians, which are obtained from a unified SU(5) symmetry Lagrangian by minimal substitutions, introducing also an extra (anomalous coupling) Lagrangian for the $BB^{*}$-loop case.
We present here the results for the contributions of the $BB$, $BB^{*}$ and the $B^{*}B^{*}$ meson loops
to the $\Upsilon$ self-energy, where the in-medium masses of the $B$ and $B^{*}$ mesons are obtained based on the QMC model, as explained in Chapter~\ref{Chapter3}.
This chapter also includes a comparison between the total ($BB$ + $BB^{*}$ + $B^{*}B^{*}$)
meson loop contribution and its correspondent ($DD$ + $DD^{*}$ + $D^{*}D^{*}$) in the $J/\Psi$ case.
The $\Upsilon$-nucleus potential is calculated for various nuclei 
($^{4}$He, $^{12}$C, $^{16}$O, $^{40}$Ca, $^{48}$Ca, $^{90}$Zr and $^{208}$Pb)
using a local density approximation, and the bound state energies are obtained by
numerically solving the Klein-Gordon equation (originally the Proca equation) for each nucleus.

\section{Effective Lagrangians}

\thispagestyle{myheadings}

\noindent

The $\Upsilon$ mass shift in medium comes from the modification of $BB$, $BB^{*}$ and $B^{*}B^{*}$ 
meson loop contributions to the $\Upsilon$ self-energy relative to that in free space,
which is calculated, based on a flavor SU(5) symmetry using an effective Lagrangian 
density~\cite{Lin:2000ke} (hereafter we simply call Lagrangian). 
The free Lagrangian for pseudoscalar and vector mesons is given by 

\begin{equation}
{\cal L}_{0}=Tr \left(  \partial _{ \mu }P^{\dagger} \partial ^{ \mu }P \right)
-\frac{1}{2}Tr \left( F_{ \mu  \nu }^{\dagger}F^{ \mu  \nu } \right),
\end{equation}
with \[ F_{ \mu  \nu }= \partial _{ \mu }V_{ \nu }- \partial _{ \nu }V_{ \mu } ,\]
where $P$ and $V$ (suppressing the Lorentz indices for $V$) are, respectively, the $5 \times 5$ pseudoscalar and vector meson matrices in SU(5):

\begin{eqnarray}
&&\hspace{-4ex} P = \frac{1}{\sqrt{2}} \begin{pmatrix} 
\frac{\pi^{0}}{\sqrt{2}} + \frac{\eta}{\sqrt{6}} + \frac{\eta_{c}}{\sqrt{12}} 
+ \frac{\eta_{b}}{\sqrt{20}}  &  \pi^{+}  &  K^{+}  &  \overline{D}^{0}  &  B^{+}\\
\pi^{-}  &  \frac{-\pi^{0}}{\sqrt{2}} + \frac{\eta}{\sqrt{6}} + \frac{\eta_{c}}{\sqrt{12}} 
+ \frac{\eta_{b}}{\sqrt{20}}  &  K^{0}  &  D^{-}  &  B^{0}\\
K^{-}  &  \overline{K}^{0}  &  \frac{-2\eta}{\sqrt{6}} + \frac{\eta_{c}}{\sqrt{12}} 
+ \frac{\eta_{b}}{\sqrt{20}}  &  D_{s}^{-}  &  B_{s}^{0}\\
D^{0}  &  D^{+}  &  D_{s}^{+}  &  \frac{-3\eta_{c}}{\sqrt{12}} 
+ \frac{\eta_{b}}{\sqrt{20}}  &  B_{c}^{+}\\
B^{-}  &  \overline{B^{0}}  &  \overline{B_{s}^{0}}  &  B_{c}^{-}  &  \frac{-2\eta_{b}}{\sqrt{5}}\\
\end{pmatrix}, \nonumber \label{p} \\
\nonumber\\
\nonumber\\
&&\hspace{-4ex} V = \frac{1}{\sqrt{2}} \begin{pmatrix}
\frac{\rho^{0}}{\sqrt{2}} + \frac{\omega}{\sqrt{6}} + \frac{J/\Psi}{\sqrt{12}} 
+ \frac{\Upsilon}{\sqrt{20}}  &  \rho^{+}  &  K^{*+}  &  \overline{D}^{*0}  &  B^{*+}\\
\rho^{-}  &  \frac{-\rho^{0}}{\sqrt{2}} + \frac{\omega}{\sqrt{6}} + \frac{J/\Psi}{\sqrt{12}} 
+ \frac{\Upsilon}{\sqrt{20}}  &  K^{*0}  &  D^{*-}  &  B^{*0}\\
K^{*-}  &  \overline{K}^{*0}  &  \frac{-2\omega}{\sqrt{6}} + \frac{J/\Psi}{\sqrt{12}} 
+ \frac{\Upsilon}{\sqrt{20}}  &  D_{s}^{*-}  &  B_{s}^{*0}\\
D^{*0}  &  D^{*+}  &  D_{s}^{*+}  &  \frac{-3J/\Psi}{\sqrt{12}} + \frac{\Upsilon}{\sqrt{20}}  &  
B_{c}^{*+}\\
B^{*-}  &  \overline{B^{*0}}  &  \overline{B_{s}^{*0}}  &  B_{c}^{*-}  &  
\frac{-2\Upsilon}{\sqrt{5}} \\ 
\end{pmatrix}. \nonumber \label{v}
\end{eqnarray}

The following minimal substitution is introduced to obtain the couplings (interactions) 
between pseudoscalar mesons and vector mesons, in a similar way as the substitution to get the interaction of a charged particle with a electromagnetic field is introduced~\cite{Fayyazuddin:2012qfa}:

\begin{eqnarray}
&&\partial _{ \mu }P \rightarrow  \partial _{ \mu }P-\frac{ig}{2} \left[ V_{ \mu }\text{, P} 
\right],\\
&&F_{ \mu  \nu } \rightarrow  \partial _{ \mu }V_{ \nu }- \partial _{ \nu }V_{ \mu }
 -\frac{ig}{2} \left[ V_{ \mu },~V_{ \nu } \right].
\end{eqnarray}

Then, the effective Lagrangian is obtained as, 

\begin{eqnarray}
\label{efflag}
 {\cal L}&={\cal L}_{0}+igTr \left(  \partial _{ \mu }P \left[ P,~V_{ \mu } \right]  \right) 
 -\frac{g^{2}}{4}Tr \left(  \left[ \text{P, V}_{ \mu } \right] ^{2} \right)\nonumber \\ 
 &+igTr \left(  \partial ^{ \mu }V^{ \nu } \left[ V_{ \mu },~V_{ \nu } \right]  \right) 
 +\frac{g^{2}}{8}Tr \left(  \left[ V_{ \mu },~V_{ \nu } \right] ^{2} \right). 
\end{eqnarray}

Expanding this in terms of the $P$ and $V$ matrices above,
we obtain the following interaction Lagrangians~\cite{Lin:2000ke} 

\begin{eqnarray}
{\cal L}_{\Upsilon BB} 
&=& i g_{\Upsilon BB}\Upsilon^{\mu}
\left[\overline{B} \partial_{\mu}B 
 - \left(\partial_{\mu}\overline{B}\right)B\right],
\label{lag1}
\\
{\cal L}_{\Upsilon B^{*}B^{*}} 
&=& i g_{\Upsilon B^{*}B^{*}}
\left\{ \Upsilon^{\mu}\left[ (\partial_{\mu}\overline{B^{*}}^{\nu}) B^{*}_{\nu} - 
\overline{B^{*}}^{\nu}\partial_{\mu} B^{*}_{\nu}\right]
+ \left[ (\partial_{\mu} \Upsilon^{\nu}) \overline{B^{*}_{\nu}} - \Upsilon^{\nu}
\partial_{\mu} \overline{B^{*}_{\nu}}\right] 
B^{*\mu} \nonumber
\right.
\\ 
&&\hspace{12ex} 
\left.+ \overline{B^{*}}^{\mu} \left[\Upsilon^{\nu} \partial_{\mu} B^{*}_{\nu} - 
(\partial_{\mu}\Upsilon^{\nu}) B^{*}_{\nu}\right]\right\},
\label{lag2}
\end{eqnarray}
where the following convention was adopted
\begin{align*}
B&=\begin{pmatrix}
        B^{+}\\
        B^{0}
       \end{pmatrix}, & \overline{B}=\begin{pmatrix}
       B^{-} & \overline{B^{0}} \end{pmatrix},  
			& &B^{*} =\begin{pmatrix}
        B^{*+}\\
        B^{*0}
       \end{pmatrix}, & &\overline{B^{*}}&=\begin{pmatrix}
       B^{*-} & \overline{B^{*0}} \end{pmatrix}.\\         
\end{align*}
In addition we also include the $\Upsilon BB^*$ 
anomalous-coupling~\cite{Leinweber:2001ac,Eletsky:1982py} interaction Lagrangian, 
similar to the case of $J/\Psi$ that was introduced in the $J/\Psi DD^*$ interaction 
Lagrangian in Refs.~\cite{Oh:2000qr,Krein:2010vp},

\begin{equation}
{\cal L}_{\Upsilon BB^{*}}
= \frac{g_{\Upsilon BB^{*}}}{m_{\Upsilon}}\varepsilon_{\alpha \beta \mu \nu}
\left(\partial^{\alpha}\Upsilon^{\beta}\right)\left[\left(\partial^{\mu} 
\overline{B^{*}}^{\nu}\right)B
+ \overline{B}\left(\partial^{\mu}{B^{*}}^{\nu}\right)\right],
\label{lag3}
\end{equation}
where, we assume $g_{\Upsilon BB^*} = g_{\Upsilon BB} = g_{\Upsilon B^*B^*}$,  
the corresponding relation adopted for the $J/\Psi$ case~\cite{Krein:2010vp}.
\section{Coupling constants}

\noindent

\begin{figure}[htb]
\vspace{4ex}
\centering
 \includegraphics[scale=0.58]{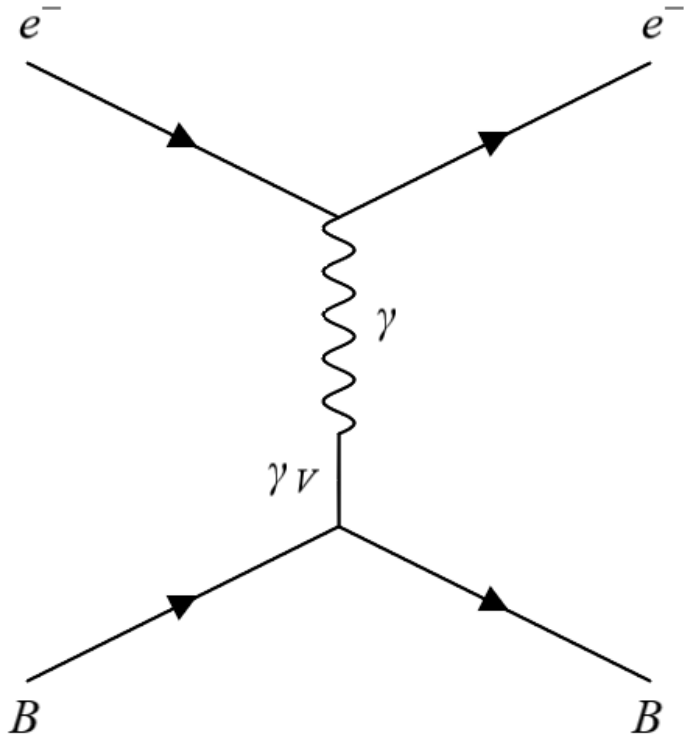}
 \caption{Representation of the process $e^{-} B^{+} \rightarrow e^{-} B^{+}$ in the VMD model.}
 \label{vmd}
\end{figure}

The values of the coupling constants appearing in the Lagrangians of the last section are 
here determined by the Vector Meson Dominance model (VMD)~\cite{Lin:2000ke}.
In this model, represented in Fig.~\ref{vmd}, the virtual photon in the process 
$e^{-} B^{+} \rightarrow e^{-} B^{+}$ is coupled to the vector mesons $\rho$, $\omega$ and $\Upsilon$.
If we consider the case of zero momentum transfer, we can use the following relation:

\begin{equation}
\sum_{V = \rho, \omega, \Upsilon}^{}
\frac{\gamma_{V} g_{V B^{+} B^{-}}}{m^{2}_{V}} = e,
\end{equation}
where photon-vector-meson mixing amplitude $\gamma_V$ is determined from the vector-meson 
decay width

\begin{equation}
\Gamma_{Vee} = \frac{\alpha \gamma^{2}_{V}}{3 m^{3}_{V}},
\end{equation}
with $\alpha = e^{2}/4\pi$ being the fine structure constant, which is a dimensionless measure of the strengths of the electromagnetic interactions. We have then

\begin{equation}
\frac{\gamma_{\Upsilon} g_{\Upsilon BB}}{m^{2}_{\Upsilon}} = \frac{1}{3} e,
\end{equation}
thus obtaining 

\begin{equation}
g_{\Upsilon BB} = \frac{1}{3} \alpha 
\left( \frac{4 \pi m_{\Upsilon}}{3 \Gamma_{\Upsilon ee}} \right)^{1/2}
= 13.228 \approx 13.2.
\end{equation}
%

\section{Self-energy}
\label{Ch5Sc3}

\noindent

The in-medium potential for the $\Upsilon$ meson is given by the difference between the in-medium, 
$m^{*}_{\Upsilon}$, and free space, 
$m_{\Upsilon}$, masses of $\Upsilon$,
\begin{equation}
 V = m_{\Upsilon}^{*} - m_{\Upsilon},
\end{equation}
with the free space physical $\Upsilon$ mass being reproduced first by,   
\begin{equation}
m^{2}_{\Upsilon} = \left(m^{0}_{\Upsilon}\right)^{2} + 
\Sigma_\Upsilon (k^{2}=m^{2}_{\Upsilon}),
\end{equation}
where $m^{0}_{\Upsilon}$ is the bare mass, and the total self-energy 
$\Sigma_\Upsilon$ is calculated by the sum
of the contributions from the free space $BB$, $BB^{*}$ and $B^{*}B^{*}$ meson loops    
in the case we include all the meson loops considered in this study. 
Note that, we ignore the possible width, or the imaginary part in the self-energy in the present  
study. The in-medium mass, $m^{*}_{\Upsilon}$,
is calculated likewise, by the total self-energy in medium using the medium-modified $B$ and 
$B^{*}$ meson masses with the same $m_\Upsilon^0$ value fixed in free space.
We remind that the $m^0_\Upsilon$ value depends on the loops included in the self-energy 
of the $\Upsilon$ in free space.

The S-Matrix is (see Appendix~\ref{apx-smat})
\begin{equation}
S = 1 + iT = T exp \left( i \int d^{4}x {\cal L}_{int}(x) \right) \sim -i {\cal M},
\end{equation}
where the second term containing the time order operator $T$ contains the information on the interactions and ${\cal M}$ is the invariant amplitude, which
is the probability amplitude of transition from the incoming states to the outgoing states in a scattering process.
We then find the self-energy contribution for each interaction lagrangian in equations~(\ref{lag1}), 
(\ref{lag2}) and (\ref{lag3}) by making use of the following relation (see expression~(\ref{ampli}))
\begin{equation}
-i {\cal M} = -i \Sigma.
\end{equation}

We then sum each meson loop contribution for the $\Upsilon$ self-energy as
\begin{equation}
\Sigma_{\Upsilon} = \sum_l \Sigma_\Upsilon^l = \sum_l (- \frac{g^{2}_{\Upsilon l}}{3\pi^{2}})
\int_{0}^{\infty} dq\, \textbf{q}^{2} F_{l}(\textbf{q}^{2}) 
K_{l}(\textbf{q}^{2}),
\end{equation}
where $l = BB, BB^{*}, B^{*}B^{*}$ and $F_{l}\left(\textbf{q}\right)$ is the product of 
vertex form factors (to be discussed later). The $K_{l}$ for each meson loop contribution 
is given, similarly to the $J/\Psi$ case~\cite{Krein:2010vp},
\begin{eqnarray}
&&K_{BB}\left(\textbf{q}^{2}\right) = \frac{1}{\omega_{B}} 
\left(\frac{\textbf{q}^{2}}{\omega^{2}_{B} - m^{2}_{\Upsilon}/4}\right),
\\
&&K_{BB^{*}}\left(\textbf{q}^{2}\right) =\frac{\textbf{q}^{2}\overline{\omega}_{B}} 
{\omega_{B}\omega_{B^{*}}} \frac{1}{\overline{\omega}^{2}_{B} - m^{2}_{\Upsilon}/4},
\\
&&K_{B^{*}B^{*}}\left(\textbf{q}^{2}\right) =\frac{1}{4m_{\Upsilon} \omega_{B^*} }
\left[
\frac{A\left(q^{0} = \omega_{B^*} \right)}{\omega_{B^*} - m_\Upsilon/2} 
- \frac{A\left( q^{0} = \omega_{B^*} + m_{\Upsilon} \right)}{\omega_{B^*} + m_{\Upsilon}/2}
\right],
\end{eqnarray}
where $\omega_{B} = \left(\textbf{q}^{2} + m^{2}_{B}\right)^{1/2}$,
$\omega_{B^{*}} = \left(\textbf{q}^{2} + m^{2}_{B^{*}}\right)^{1/2}$,
$\overline{\omega}_{B} = \left(\omega_{B} + \omega_{B^{*}}\right)$ and

\begin{equation}
A\left(q\right) = \sum^{4}_{i=1} A_{i}\left(q\right),
\end{equation}
with
%
\begin{eqnarray}
A_1 (q) 
&=& -4q^2 \left[
4-\frac{q^2 + (q-k)^2}{m^{2}_{B^*}}
+ \frac{\left[q \cdot (q-k) \right]^2}{m^{4}_{B^*}} 
\right],
\label{EqA1}
\\
A_2 (q) 
&=& 8\left[q^2 - \frac{\left[q \cdot (q-k) \right]^2}{m^2_{B^*}}\right]
\left[2 + \frac{(q^0)^{2}}{m^{2}_{B^{*}}} \right],
\\
A_3 (q ) 
&=& 8\, (2q^0 - m_\Upsilon)
\left[
q^0 - (2q^{0} - m_\Upsilon) \frac{q^2 + q \cdot (q-k)}{m^{2}_{B^*}}
+q^{0}\frac{\left[q \cdot\left(q-k\right)\right]^{2}}{m^{4}_{B^*}}
\right],
\\
A_4 (q) 
&=& -8 \left[ q^0 - (q^0 - m_\Upsilon) \frac{q\cdot(q-k)}{m^2_{B^*}} \right]  
\left[ (q^0 - m_\Upsilon) - q^0\frac{q\cdot(q-k)}{m^2_{B^*}} \right],	
\label{EqA4}
\end{eqnarray}
where $q = \left(q^{0}, \textbf{q}\right)$, and the $\Upsilon$ is taken at rest, 
$k = \left(m_{\Upsilon}, 0\right)$.\par

We use phenomenological form factors to regularize the self-energy
loop integrals following Refs.~\cite{Krein:2010vp,Leinweber:1999ig}, 
\begin{equation}
u_{B,B^{*}}(\textbf{q}^{2}) = \left(\frac{\Lambda^{2}_{B,B^{*}} + m^{2}_{\Upsilon}}
{\Lambda^{2}_{B,B^{*}} + 4\omega^{2}_{B,B^{*}}\left(\textbf{q}^{2}\right)}\right)^{2}.
\label{ffups}
\end{equation}

For the vertices $\Upsilon BB$, $\Upsilon BB^{*}$ and $\Upsilon B^{*}B^{*}$, we use the form 
factors 
$F_{BB}\left(\textbf{q}^{2}\right) = u^{2}_{B}\left(\textbf{q}^{2}\right)$, 
$F_{BB^{*}}\left(\textbf{q}^{2}\right) = 
u_{B}\left(\textbf{q}^{2}\right)u_{B^{*}}\left(\textbf{q}^{2}\right)$, and 
$F_{B^{*}B^{*}}\left(\textbf{q}^{2}\right) = u^{2}_{B^{*}}\left(\textbf{q}^{2}\right)$, 
respectively, with $\Lambda_B$ ($\Lambda_{B^*}$) being the corresponding cutoff 
mass associated with $B$ ($B^*$) meson, and the common value, $\Lambda_{B} = \Lambda_{B^{*}}$,
will be used in this study. 

We have to point out that the choice of the cutoff mass values in the form factors for the 
$\Upsilon BB$, $\Upsilon BB^{*}$ and $\Upsilon B^{*}B^{*}$ vertices 
has nonnegligible impact on the results.
But the form factors are necessary to include the effects of the finite 
sizes of the mesons for the overlapping regions associated with the vertices. 
The cutoff values $\Lambda_{B,B^*}$ may be associated with the energies used to probe 
the internal structure of the mesons or the overlapping regions associated with the vertices. 
When these values get closer to the corresponding meson masses, 
the Compton wavelengths ($\lambda$) associated with the values of $\Lambda_{B,B^*}$ 
are comparable to the sizes of the mesons
(for $\hbar = c = 1$ and $\Lambda_{B,D} \sim m_{\Upsilon, J/\Psi}$, we have
$\lambda_{\Lambda_{B,D}} = 1/\Lambda_{B,D} \sim \lambda_{\Upsilon, J/\Psi} = 1/m_{\Upsilon, J/\Psi}$), 
and the use of the form factors does not make reasonable sense.
Then, in order to have a physical meaning, we may be able to 
constrain the choice for the cutoff mass values,   
in such a way that the form factors reflect the finite size effect of the participating mesons 
reasonably. Later, an analysis on this issue will be made   
taking the $J/\Psi DD$, $J/\Psi DD^*$ and $J/\Psi D^*D^*$  
vertices as examples.

By the heavy quark and heavy meson symmetry in QCD, the charm and bottom quark sectors  
are expected to have mostly similar properties (but quantitatively need to be shown if possible). 
We thus follow this naive expectation and choose the similar cutoff mass values 
as the ones used in the previous work of the $J/\Psi$ mass shift~\cite{Krein:2017usp}, 
varying the $\Lambda_{B,B^*}$ values   
between $2000~\text{MeV} \leq \Lambda_{B,B^*} \leq 6000~\text{MeV}$, 
but with the larger upper values, since the $B$ and $B^*$ masses are 
larger than those of the $D$ and $D^*$ mesons.


\section{Mass shift}

\noindent

\subsection{Results for $\Upsilon$ mass shift}
\label{ures}

In the following we present the results for the in-medium mass shift of $\Upsilon$ meson 
together with each meson loop contribution    
for five different values of the cutoff mass $\Lambda_B (= \Lambda_{B^*})$, 
where we use the in-medium $B$ and $B^{*}$ meson masses shown in Fig.~\ref{fig2}.
The values used for the free space masses of $\Upsilon$, $B$ and $B^{*}$ mesons are, 
respectively, 9460, 5279 and 5325 MeV~\cite{PDG2020}.

The values of $m^{0}_{\Upsilon}$, as well as each loop contributions to $m^{0}_{\Upsilon}$,
in the range of the cutoff masses $\Lambda_B$ are: 9653 to 9802 MeV for the total $BB$ loop contribution, 9836 to 10123 MeV for the ($BB$ + $BB^{*}$) total loop contributions and 
19649 to 24966 MeV for the ($BB$ + $BB^{*}$ + $B^{*}B^{*}$) total loop contributions.
The $BB$ loop contribution to $m^{0}_{\Upsilon}$ ranges from 193 to 342 MeV, the contribution
corresponding to the $BB^{*}$ loop being 186 to 332 MeV, and the one from the $B^{*}B^{*}$
loop ranging from 10003 to 15244 MeV.

\begin{figure}[htb]%
\vspace{4ex}
\centering
\includegraphics[scale=0.34]{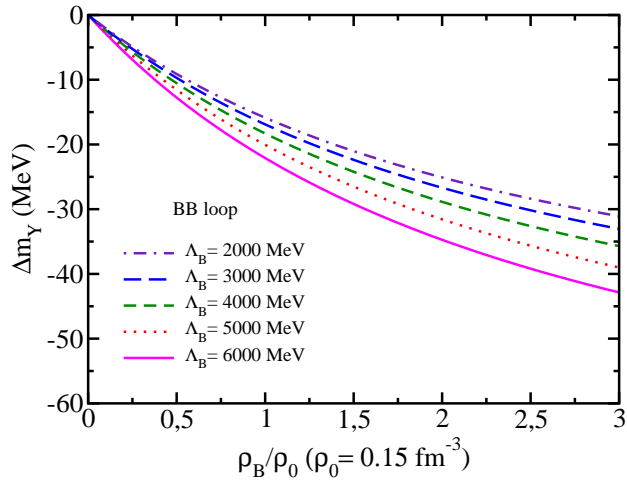}%
\caption{Contribution from the $BB$ meson loop (total is $BB$ meson loop) to 
the $\Upsilon$ mass shift versus nuclear matter density for five different values of 
the cutoff mass $\Lambda_{B}$.}%
\label{fig3}%
\end{figure}

\begin{figure}[htb]%
\vspace{2ex}
\centering
\includegraphics[width=6.5cm]{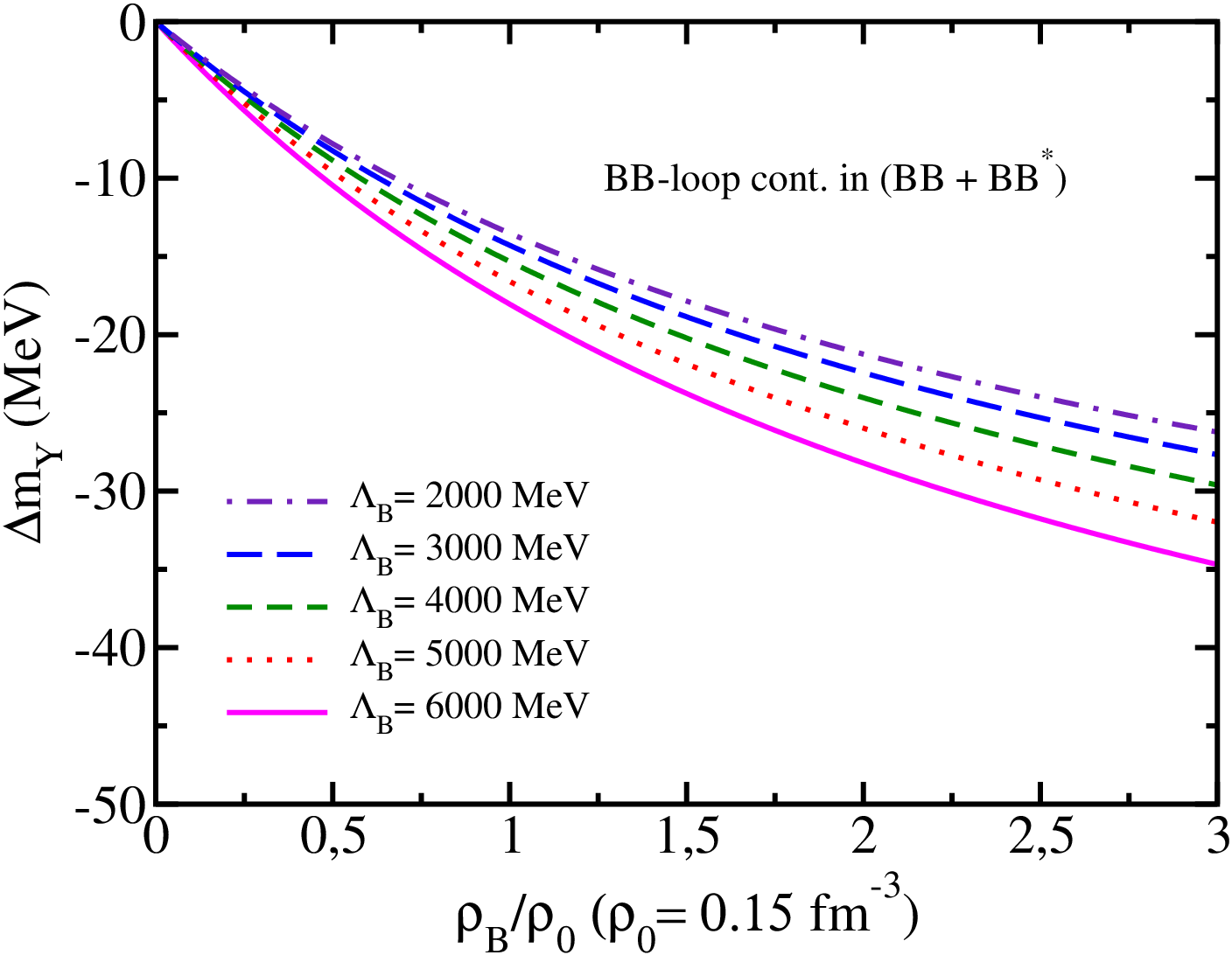}
\hspace{2ex}
\includegraphics[width=6.5cm]{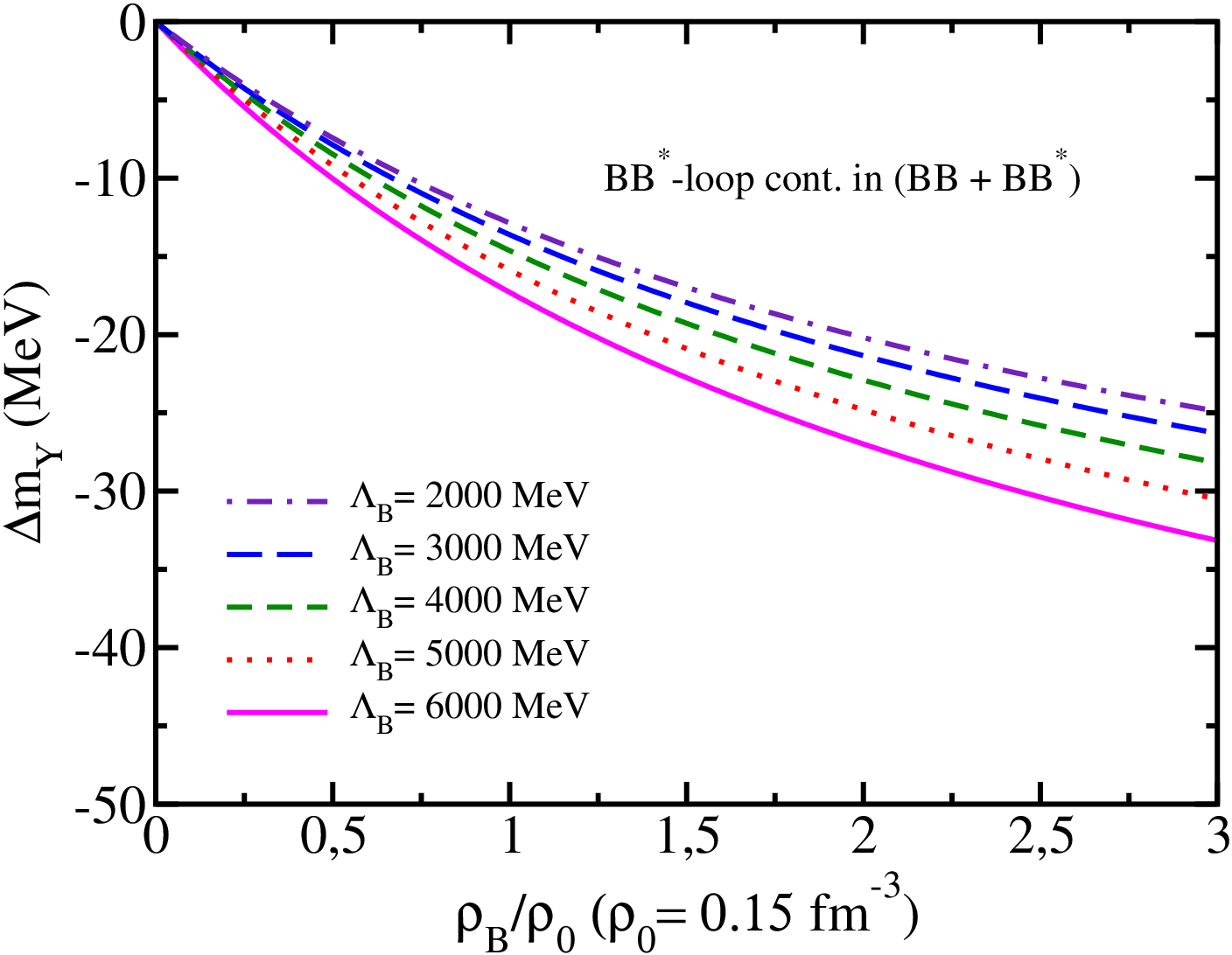}
\\
\vspace{6ex}
\includegraphics[width=6.5cm]{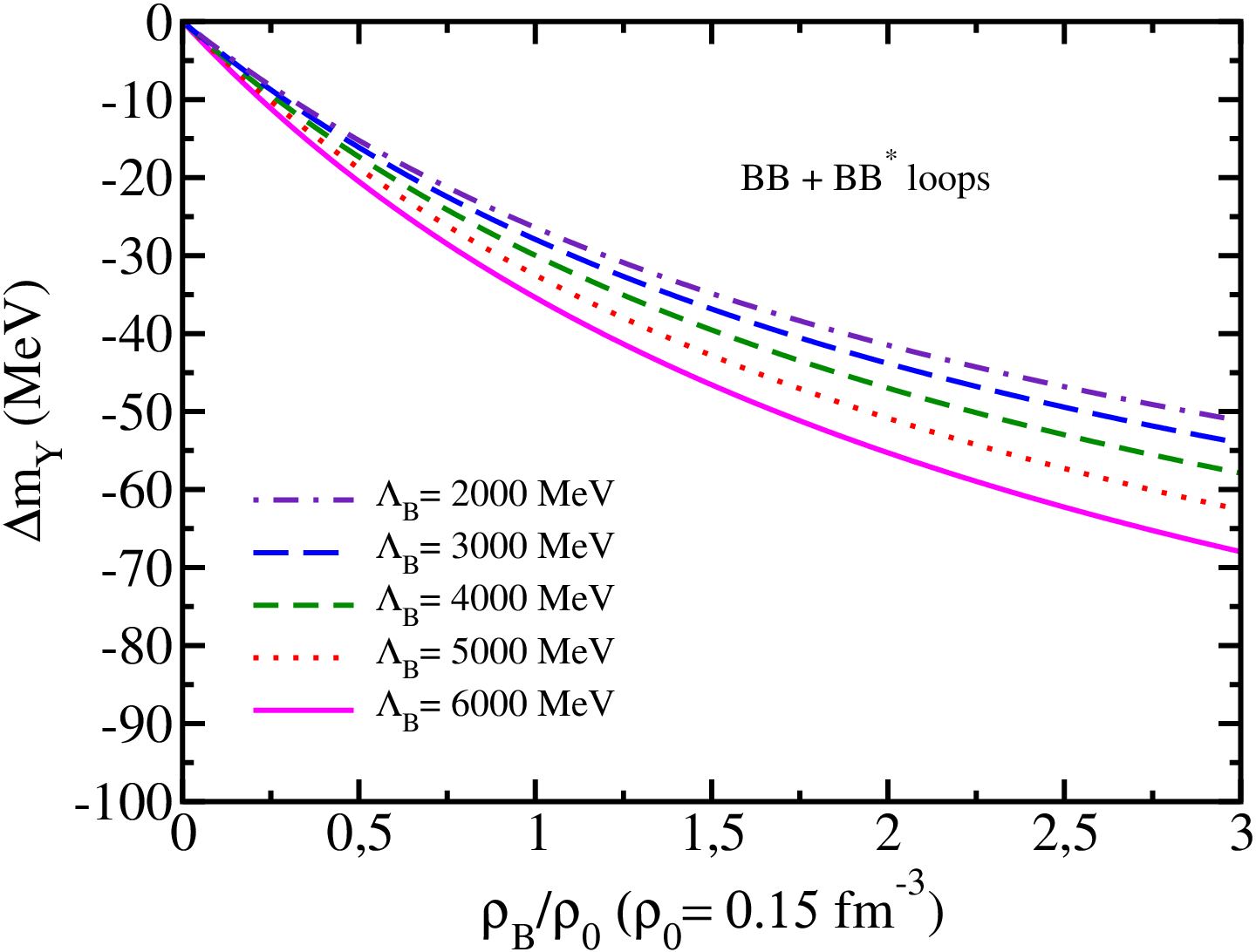}
\caption{$BB$ (top left), $BB^{*}$ (top right) and total (bottom) meson loop contributions to the 
$\Upsilon$ mass shift 
versus nuclear matter density for five different values of the cutoff 
mass $\Lambda_{B} (=\Lambda_{B^*})$.}
\label{fig4}
\end{figure}

\begin{figure}[htb]%
\vspace{2ex}
\centering
\includegraphics[width=6.5cm]{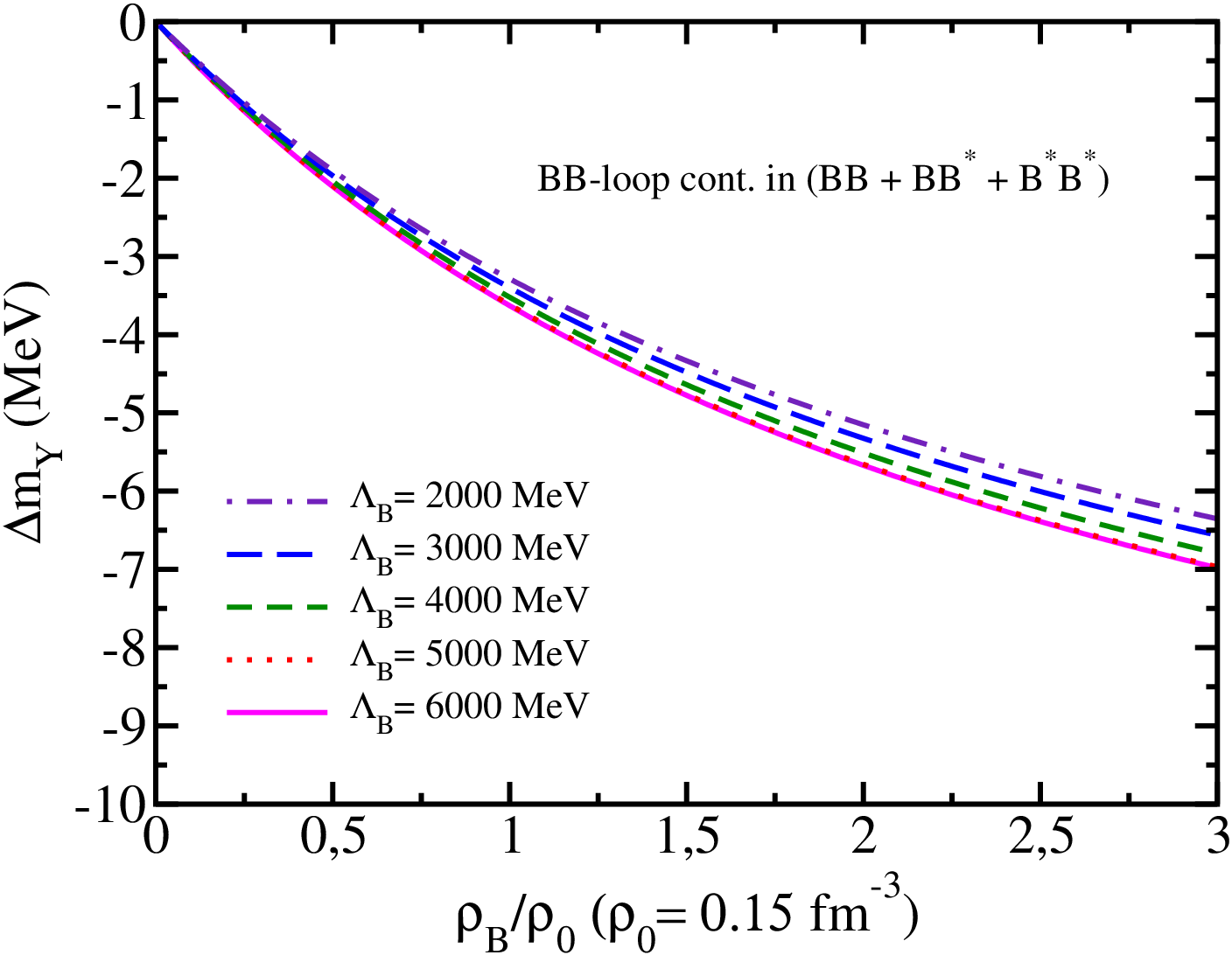}
\hspace{2ex}
\includegraphics[width=6.5cm]{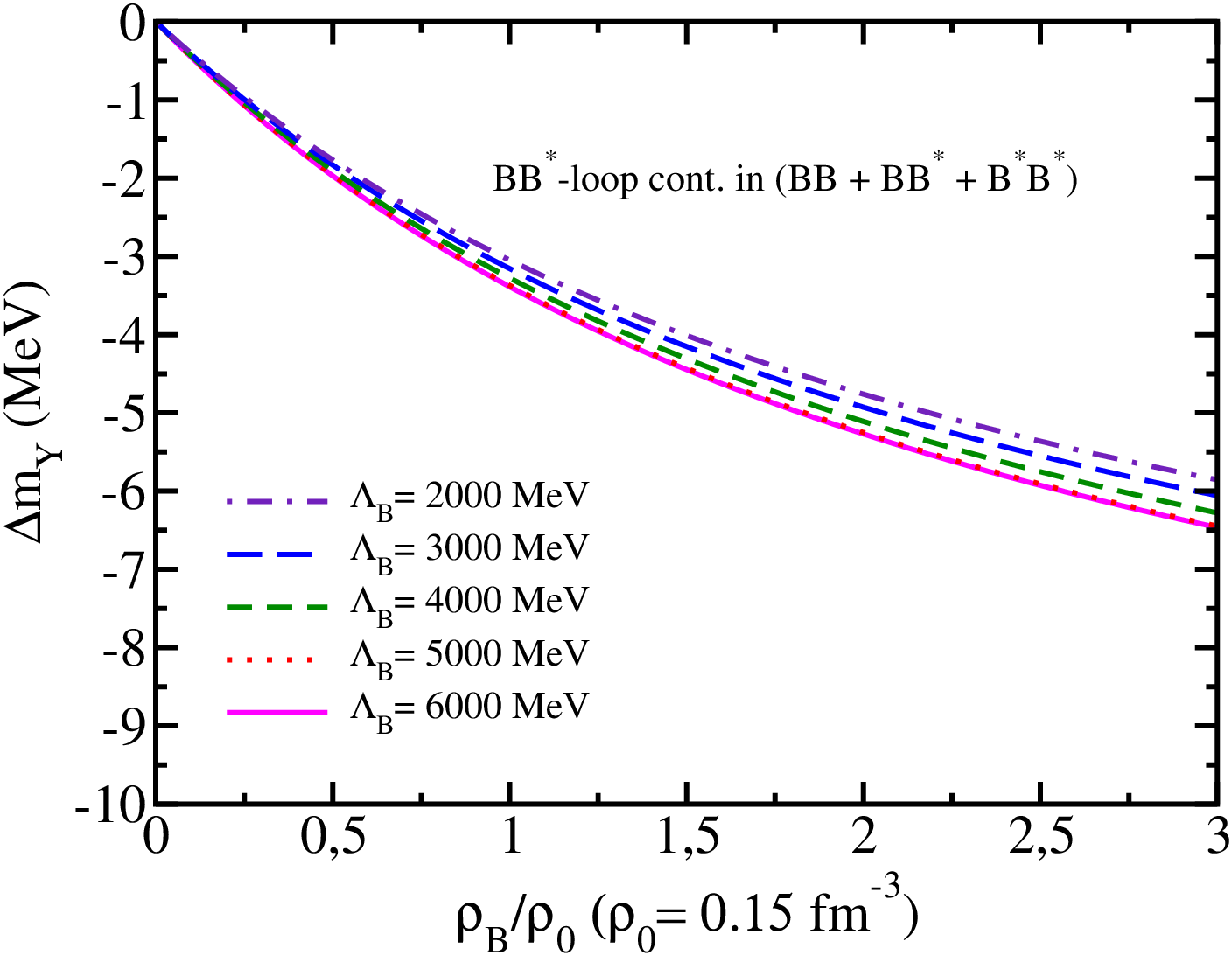}
\\
\vspace{6ex}
\includegraphics[width=6.5cm]{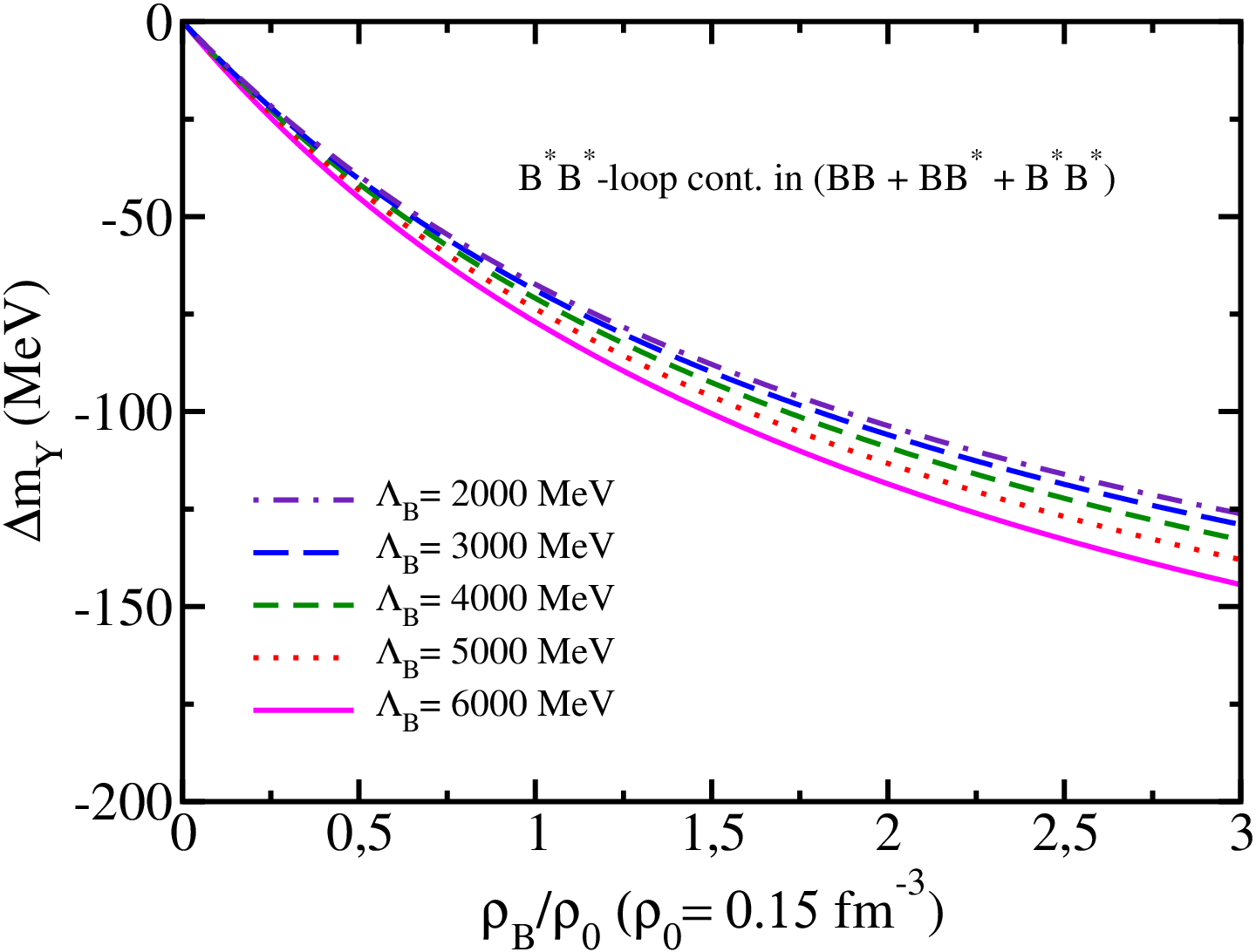}
\hspace{2ex}
\includegraphics[width=6.5cm]{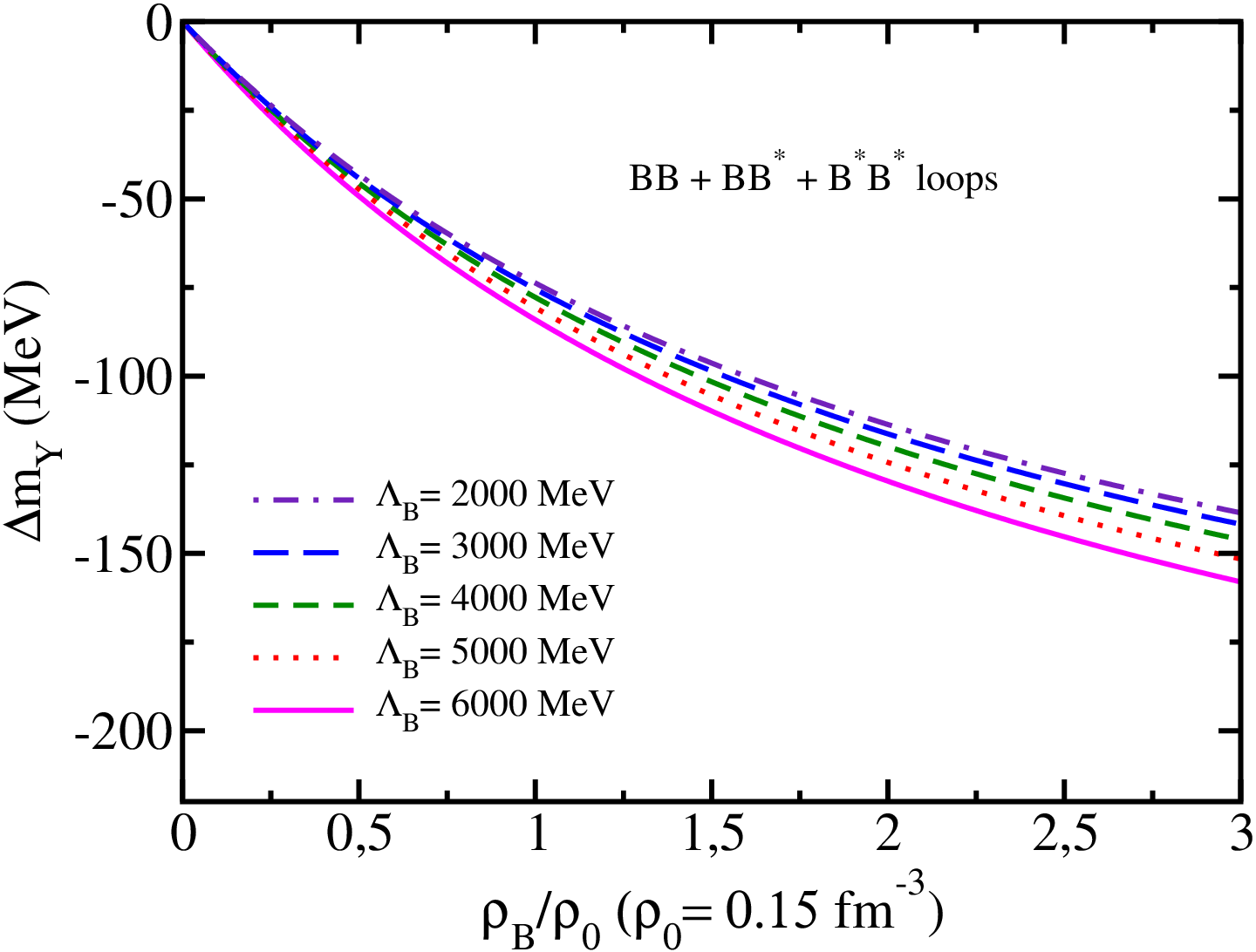}
\caption{$BB$ (top left), $BB^{*}$ (top right), $B^{*}B^{*}$ (bottom left) and
total (bottom right) meson loop contributions to the $\Upsilon$ mass shift 
versus nuclear matter density for five different values of the cutoff mass 
$\Lambda_{B} (=\Lambda_{B^*})$.}%
\label{fig5}%
\end{figure}

In Fig.~\ref{fig3} we show the $\Upsilon$ mass shift, taking the total contribution 
to be the $BB$ meson loop for five values of the cutoff mass $\Lambda_{B}$, 
2000, 3000, 4000, 5000 and 6000 MeV 
(these values will be applied for all the studies in the following with $\Lambda_B = 
\Lambda_{B^*}$). 
As one can see, the effect of the decrease in the $B$ meson in-medium mass    
yields a negative mass shift of the $\Upsilon$. 
The decrease of the $B$ meson mass in (symmetric) nuclear matter enhances the $BB$ meson loop 
contribution, thus the self-energy contribution in the medium becomes larger 
than that in the free space. 
This negative shift of the $\Upsilon$ mass is also dependent on the
value of the cutoff mass $\Lambda_B$, i.e., the amount of the mass shift 
increases as $\Lambda_B$ value increases, ranging 
from -16 to -22 MeV at the symmetric nuclear matter 
saturation density, $\rho_0 = 0.15$ fm$^{-3}$.

Next, in Fig.~\ref{fig4} we show the $\Upsilon$ mass shift taking the total self-energy 
contribution to be the $(BB + BB^*)$ meson loops. 
The contributions are shown for the $BB$ meson loop (top left), 
$BB^{*}$ meson loop (top right), and the total $(BB + BB^*)$ meson loops (bottom). 
The total mass shift at $\rho_0$ ranges from -26 to -35 MeV for 
the same range of the $\Lambda_B (= \Lambda_{B^*})$ values. 

Finally, we show in Fig.~\ref{fig5} the $\Upsilon$ mass shift 
taking the total self-energy contribution to be the $(BB + BB^* + B^*B^*)$ meson loops. 
The contributions are shown for the $BB$ meson loop (top left), 
the $BB^{*}$ meson loop (top right), $B^*B^*$ meson loop (bottom left),  
and the total $(BB + BB^* + B^*B^*)$ meson loops (bottom right).
The total mass shift at $\rho_0$ ranges from -74 to -84 MeV 
for the same range of the $\Lambda_B (= \Lambda_{B^*})$ values.

It is important to note that due to the unexpectedly    
larger contribution from the heavier meson-pair $B^{*}\overline{B^{*}}$   
meson loop ($B^*B^*$ meson loop) to the $\Upsilon$ mass shift than the other 
lighter-meson-pair loops $BB$ and $BB^*$   
presented in Fig.~\ref{fig5}, we regard the form factor used for 
the vertices in the $B^{*}B^{*}$ meson loop may not be appropriate, 
and need to consider either different 
form factors, or adopt an alternative regularization 
method in the future.
\vspace{1ex}

\newpage
\noindent
{\it Summary for the $\Upsilon$ mass shift:}\\ 
The $\Upsilon$ mass shift is shown separately in Figs.~\ref{fig3},~\ref{fig4} 
and ~\ref{fig5}, by the difference in the intermediate states contributing for 
the total $\Upsilon$ self-energy, namely, by the $BB$, 
$\left(BB + BB^{*}\right)$, and $\left(BB + BB^{*} + B^{*}B^{*}\right)$ meson 
loops. The corresponding $\Upsilon$ mass shift at $\rho_0$ ranges, 
(-16 to -22) MeV, (-26 to -35) MeV, and (-74 to -84) MeV, for the adopted range 
of the $\Lambda_B (=\Lambda_{B^*}$) values.
The results indicate that the dependence on the values of the cutoff 
mass $\Lambda_B (=\Lambda_{B^*})$ is rather small compared to that of the $\Lambda_D 
(=\Lambda_{D^*})$ for the $J/\Psi$ case as will be discussed later, 
and this gives smaller ambiguities for our prediction 
originating from the cutoff mass values.

\subsection{Comparison with $J/\Psi$ mass shift}
\label{comp}

\begin{figure}[htb]
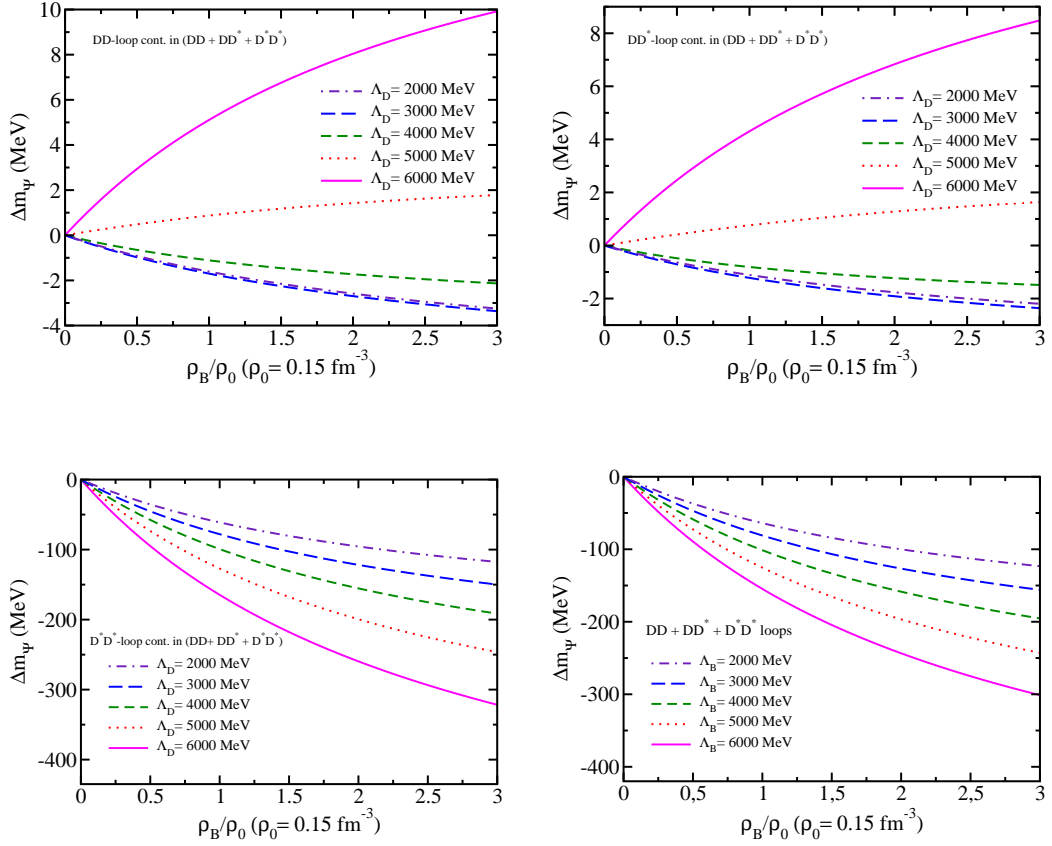
%
\vspace{2ex}
\centering
\includegraphics[width=6.5cm]{3_DD.eps}
\hspace{2ex}
\includegraphics[width=6.5cm]{3_DDs.eps}
\\
\vspace{6ex}
\includegraphics[width=6.5cm]{3_DsDs.eps}
\hspace{2ex}
\includegraphics[width=6.5cm]{3_Jtotal.eps}
\caption{$DD$ (top left), $DD^{*}$ (top right), $D^{*}D^{*}$ (bottom left) and
total (bottom right) meson loop contributions to the $J/\Psi$ mass shift 
versus nuclear matter density for five different values of the cutoff mass 
$\Lambda_D (=\Lambda_{D^*})$.}%
\label{fig6}%
\end{figure}

The issue of the larger contribution from the heavier vector meson loop,  
in the present case $B^{*}B^{*}$ meson loop, was already observed 
in a previous study of the $J/\Psi$ 
mass shift due to the heavier meson $D^{*}D^{*}$ loop contribution, where a 
similar nongauged effective Lagrangian was used and 
no cutoff readjustment was made for the heavier vector meson intermediate 
states~\cite{Krein:2010vp}. 
The cutoff mass value readjustment in a proper manner is important, 
because it controls the fluctuations from the shorter distances.
However, we do not try this in the present exploratory study, 
since we first need to see the bare result without 
readjusting, so that we are able to compare with those of the $J/\Psi$ case, 
focusing on the heavy quark and heavy meson symmetry.

We have calculated the total ($DD + DD^* + D^{*}D^{*}$) meson loop contribution  
for the $J/\Psi$ mass shift as featured in Ref.~\cite{Krein:2017usp}  
using the same effective Lagrangian and cutoff mass values to compare with the  
total ($BB + BB^* + B^*B^*$) meson loop contribution in the $\Upsilon$ mass shift.
The free space masses of the $J/\Psi$, $D$ and $D^{*}$ mesons used are
3097, 1867 and 2009 MeV~\cite{PDG2020}, respectively.

The values of $m^{0}_{J/\Psi}$, as well as each loop contributions to $m^{0}_{J/\Psi}$,
in the range of the cutoff masses $\Lambda_D$ are: 3120 to 3316 MeV for the total $DD$ loop contribution, 3138 to 3499 MeV for the ($DD$ + $DD^{*}$) total loop contributions and 
4462 to 12000 MeV for the ($DD$ + $DD^{*}$ + $D^{*}D^{*}$) total loop contributions.
The $DD$ loop contribution to $m^{0}_{J/\Psi}$ ranges from 23 to 219 MeV, the contribution
corresponding to the $DD^{*}$ loop being 18 to 196 MeV, and the the one from the $D^{*}D^{*}$
loop ranging from 1336 to 8792 MeV.
  
The result is presented in Fig.~\ref{fig6}.  
The $D^{*}D^{*}$ meson loop contribution 
for the $J/\Psi$ mass shift ranges from -61 to -164 MeV at $\rho_0$, 
which is mostly larger than that of the $B^{*}B^{*}$ (-67 to -77 MeV at $\rho_0$) 
for the same range of the cutoff mass values in the corresponding form factors. 
Note that, the larger cutoff mass values, $\Lambda_D (=\Lambda_{D^*}) = 5000$ and $6000$ MeV,  
may not be appropriate as will be discussed in the following.
We can see from Fig.~\ref{fig6} that the  
closer the cutoff mass value gets to the $J/\Psi$ mass, 
less pronounced the negative mass shift becomes, 
until it reaches a transition point (when $\Lambda_{D}$ is larger than 
the $J/\Psi$ free space mass), where the potential starts to become even positive. 
Naively, according to the second order perturbation theory in quantum mechanics, 
they should give the negative contribution, but the positive contributions 
for $\Lambda_D = 5000$ and $6000$ MeV, thus suggest that such larger values of 
the cutoff mass may not be justified for the form factor used.
One can expect a similar behavior in the $BB$ meson loop in the 
total $(BB+BB^*+B^*B^*)$ meson loop contribution when the cutoff mass value gets 
closer to the $\Upsilon$ mass.
Indeed, such behavior is observed for the $BB$ and $BB^*$ meson loop contributions, 
for the cutoff mass values larger than 10000 MeV.
As already commented in Sec.~\ref{Ch5Sc3}, the large cutoff-mass values than 
the corresponding vector meson mass means that the distance for the interaction between 
the vector meson and the intermediate state meson included is shorter than the meson 
overlapping region, and a physical picture as an isolated vector meson 
is lost --- one also needs to consider the quark-quark, quark-antiquark, and antiquark-antiquark  
interactions and/or the corresponding correlations at the quark level  
in such short distances, where the present approach does not have. 

The bad high-energy behavior of the vector meson propagator is well known. 
To evaluate amplitudes in high-energy region that contain vector meson propagators 
in spontaneously broken gauge theory such as the weak interaction in the Standard Model, 
the generalized renormalizable gauge ($R_{\xi}$ gauge) is usually used.
This gauge uses an adjustable parameter to include other existing gauges as special cases, 
such as the $U$, $R$ and 't Hooft-Feynman gauges. 
In the $R_{\xi}$ gauge, the massive-vector-boson propagator is 

\begin{eqnarray}
\Delta_{\mu \nu} (p, \xi) &=& -i \left[ g_{\mu \nu} - \left(1- \frac{1}{\xi}\right) 
\frac{P_{\mu} P_{\nu}}{P^{2} - M^{2}/\xi}\right] \frac{1}{P^{2} - M^{2}} \nonumber \\
&=& -i \left(g_{\mu \nu} - \frac{1}{M^{2}} P_{\mu} P_{\nu} \right) \frac{1}{P^{2} - M^{2}}
- \frac{1}{M^{2}} P_{\mu} P_{\nu} \frac{1}{P^{2} - M^{2}/\xi},
\end{eqnarray}
where $M$ is the vector-boson mass and $\xi$ can vary continuously from 0 to $\infty$.

The $R_{\xi}$ gauge is connected to the other gauges as follows:

\begin{enumerate}
\item $R$ gauge: Is a generalization of the Landau gauge in QED, obtained from the
$R_{\xi}$ gauge by setting $\xi$ = 0.

\item 't Hooft-Feynman gauge: Is obtained for $\xi$ = 1. In this gauge, the vector-boson
propagator is proportional to the metric $g_{\mu \nu}$ (see appendix~\ref{apx-conv} for the metric convention), and the unphysical scalar-boson propagator
has a pole at $P^{2} = M^{2}$.

\item $U$ gauge: There's no unphysical scalar bosons in this formulation, so the vector-boson
propagator is just

\begin{equation}
\Delta_{\mu \nu} (p) = -i \left(g_{\mu \nu} - \frac{P_{\mu} P_{\nu}}{M^{2}} \right)
\frac{1}{P^{2} - M^{2}},
\end{equation}
which is equivalent as taking $\xi \rightarrow 0$ in the $R_{\xi}$ gauge.
\end{enumerate}

The $R_{\xi}$ gauge with $\xi = 1$ ('t Hooft-Feynman gauge) makes the high-energy behavior of the vector meson propagators similar to that of the spin-0 meson 
propagators~\cite{tHooft:1971qjg,tHooft:1971akt,Lee:1971kj,Fujikawa:1972fe}. 
$R_{\xi}$ gauge then removes unphysical degrees of freedom associated 
with the Goldstone bosons. In the present case, we cannot justify to use such vector meson 
propagators, so we need to tame the bad high-energy behavior phenomenologically. 
We can do this by introducing a phenomenological form factor for the $BB$ meson loop case.  
But for the $BB^{*}$ and $B^{*}B^{*}$ meson loops we simply discard their contributions 
in the present study as was practiced in Ref.~\cite{Krein:2017usp}. 
Therefore, our prediction should be regarded based on the minimum contribution  
with respect to the intermediate state meson loops, namely by only the 
$BB$ meson loop contribution as in Ref.~\cite{Krein:2017usp}, 
which took only the $DD$ meson loop contribution for estimating the $J/\Psi$ mass shift.
Regarding the form factors, another choice of form factors is possible 
to moderate the high-energy behavior~\cite{Gryniuk:2020mlh,Tsushima:1994rj,Lin:1999ad}, 
and an initial study of using a different form factor will be performed in Sec.~\ref{Ch6Sc2}.

Furthermore, although we have chosen the same coupling constants for 
$\Upsilon BB$, $\Upsilon BB^{*}$, and 
$\Upsilon B^{*}B^{*}$, it is certainly possible to use the  
different values for the coupling constants. 
Some studies of SU(4) flavor symmetry breaking couplings in charm sector offer 
alternative ways for the calculation of these coupling constants. 
This can be extended to include SU(5) symmetry 
breaking couplings. But for the flavor SU(5) sector, the breaking 
effect is expected to be even larger than that for the SU(4) sector, since bottom quark mass is 
much heavier than the charm quark, and the SU(5) symmetry breaking 
is expected to be larger. 
There are some studies focused on the SU(4) symmetry 
breaking of the coupling 
constants, although the results are not conclusive.  
A recent calculation~\cite{Lucha:2015dda} used dispersion formulation of the relativistic 
constituent quark model, where the couplings were 
obtained as residues at the poles of suitable form factors. 
Two other studies are made by the Schwinger-Dyson-equation-based  
approaches for QCD~\cite{ElBennich:2011py,El-Bennich:2016bno,El-Bennich:2021ldv}. 
In the both approaches, the obtained results for the SU(4) symmetry 
breaking are considerably larger than those obtained using QCD sum-rule approach. 
We plan to do more dedicated studies on the issues in the future.
In the present study, the coupling constant $g_{\Upsilon BB}=13.2$ contains 
SU(5) symmetry breaking effect with respect to that of the corresponding 
charm sector, $g_{J/\Psi DD}=7.64$, where both of them are determined 
using the VMD model with experimental data.

We emphasize again that, the prediction for the $J/\Psi$ mass shift   
made solely by the $DD$ meson loop, gives -3.0 to -6.5 MeV based on 
Refs.~\cite{Krein:2010vp,Krein:2017usp} (-5 to -21 MeV for the same range 
of the cutoff $\Lambda_D$ value, 2000 to 6000 MeV), while for the $\Upsilon$ mass shift, 
taking only the contribution from the $BB$ meson loop, gives -16 to -22 MeV.
In Sec.~\ref{Ch6Sc1} we will make some study for the $\Upsilon$ and $\eta_b$ 
mass shifts focusing on the SU(5) symmetric coupling constant   
between the charm and bottom sectors, as well as a coupling constant 
in a broken SU(5) symmetry scheme between the $\Upsilon$ and $\eta_b$.

One might question further, as to why the $\Upsilon$ ($\eta_b$) mass shift is larger than 
that of the $J/\Psi$ ($\eta_c$), although we have already commented 
the main reason by the larger coupling constant obtained by 
the VMD model with the experimental data. 
(The other way, why the bottom sector coupling constant is larger than that of the charm sector, 
or the corresponding experimental data in free space to determine the coupling constant is larger.)
Of course, the heavier $B$ and $B^*$ meson masses 
than the corresponding $D$ and $D^*$ meson masses also 
influence the $\Upsilon$ and $J/\Psi$ mass shift difference, 
although the heavier $B$ and $B^*$ meson masses counteract to reduce 
the $\Upsilon$ mass shift, since the heavier particles are more difficult 
to be excited in the intermediate states of the $\Upsilon$ self-energy meson loops.
To understand better, let us consider the systems of the bottom and charm sectors, 
$(D,D^*)$ and $(B,B^*)$ meson systems. 
For these two sets of systems, 
we can estimate the difference in the (heavy quark)-(light quark) interactions 
by $(m_D-m_c, m_{D^*}-m_c)$ and $(m_B-m_b, m_{B^*}-m_b)$, 
since the existence of the light quark and its interaction with the heavy quark 
in each system gives the total mass of each meson. 
Using the values (all in MeV in the following), $m_c=1270, m_D=1870, m_{D^*}=2010$, 
$m_b=4180, m_B=5279$, and $m_{B^*}=5325$, 
we get $(m_D-m_c, m_{D^*}-mc)=(600,740)$ 
and $(m_B-m_b, m_{B^*}-m_b)=(1099,1145)$.
These results indicate that the $b$-(light quark) interaction 
is more attractive than that of the $c$-(light quark),
since the larger mass differences for the $b$-quark sector mesons without light quarks 
than those corresponding for the $c$-quark sector mesons, are diminished more than those for the 
corresponding $c$-quark sector mesons as the experimentally observed masses --- the consequence of 
more attractive $b$-(light quark) interaction.
This implies that the bottomonium-nucleon (bottomonium-(nuclear matter)) interaction 
is more attractive than that of the charmonium-nucleon 
(charmonium-(nuclear matter)). 
In this way, we may be able to understand the larger mass shift of  
the $\Upsilon$ ($\eta_b$) than that of the $J/\Psi$ ($\eta_c$)  
due to the interaction with the nuclear medium --- composed 
of infinite number of light quarks.


\section{Nuclear potentials}

\noindent
 
For a $\Upsilon$-meson produced inside a nucleus A with baryon density distribution
$\rho^{A}_{B} (r)$, the $\Upsilon$-meson potential within nucleus A for a distance $r$
from the center of the nucleus is given by

\begin{equation}
\label{eqn:Vsu}
V_{\Upsilon A} (r) = \Delta m_{\Upsilon} \left( \rho^{A}_{B} (r) \right),
\end{equation}
where a local density approximation was used, meaning that the potential is calculated for a given baryon density obtained for a given point $r$ inside the nucleus.
Details are given in Appendix~\ref{apx-nucp}.
Also the nuclear density distributions were calculated within the QMC model (see Sec.~\ref{Ch3Sc2-5}),
with the exception of the $^4$He nucleus, for which the parametrization was obtained 
in Ref.~\cite{Saito:1997ae}.

Are considered in this study the following nuclei: 
$^{4}$He, $^{12}$C, $^{16}$O, $^{40}$Ca, $^{48}$Ca, $^{90}$Zr and $^{208}$Pb.
The $\Upsilon$-nucleus potential is calculated for each of the above mentioned
nuclei, varying accordingly the cutoff parameter $\Lambda_{B}$. 
In Figs.~\ref{nuclpot1} and~\ref{nuclpot2} the attractive potentials are presented separately for each nucleus. As one can see, the potential depth depends on the value of the cutoff mass, where the
larger $\Lambda_{B}$ gets, deeper is the potential.

\vspace{2ex}
\begin{figure}[htb]%
\centering
\includegraphics[width=6.5cm]{Upsipot_He4.eps}
\hspace{2ex}
\includegraphics[width=6.5cm]{Upsipot_C12.eps}
\\
\vspace{6ex}
\includegraphics[width=6.5cm]{Upsipot_O16.eps}
\caption{$\Upsilon$-nucleus potentials for various nuclei and values of the cutoff mass $\Lambda_{B}$.}
\label{nuclpot1}
\end{figure}

\begin{figure}[htb]
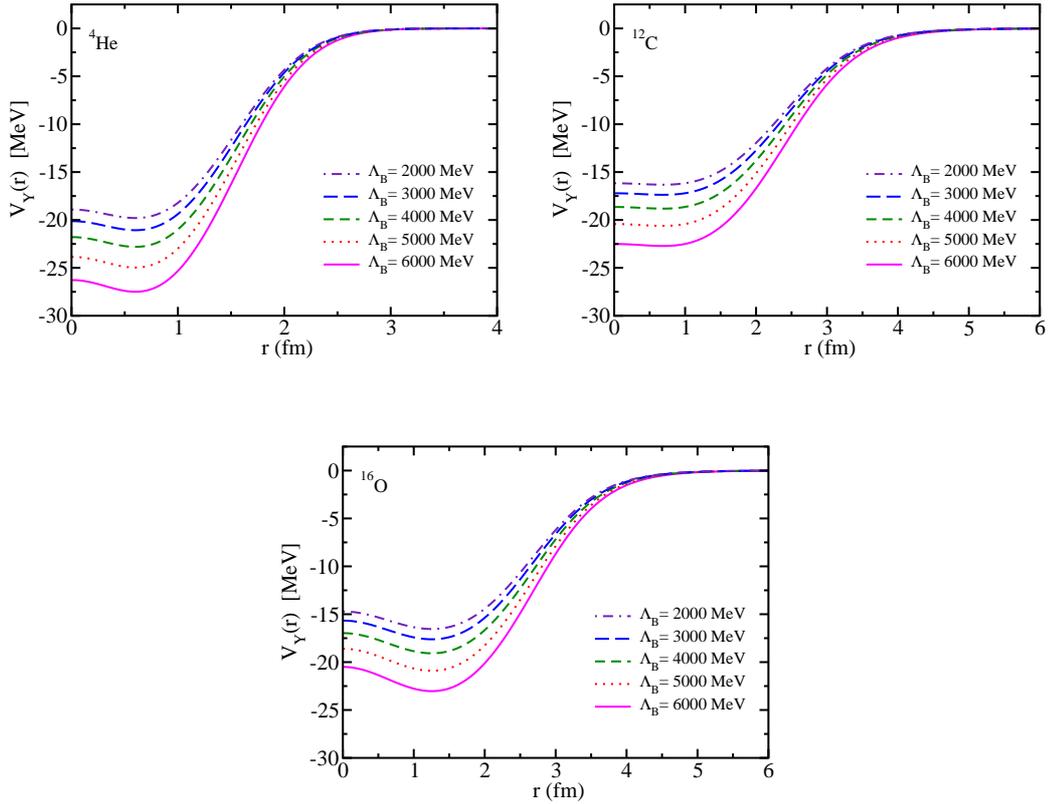
%
\centering
\includegraphics[width=6.5cm]{Upsipot_Ca40.eps}
\hspace{2ex}
\includegraphics[width=6.5cm]{Upsipot_Ca48.eps}
\\
\vspace{6ex}
\includegraphics[width=6.5cm]{Upsipot_Zr90.eps}
\hspace{2ex}
\includegraphics[width=6.5cm]{Upsipot_Pb208.eps}
\caption{$\Upsilon$-nucleus potentials for various nuclei and values of the cutoff mass $\Lambda_{B}$.}
\label{nuclpot2}%
\end{figure}


\section{$\Upsilon$-nucleus bound state energies}
\label{Ch4Sc6}

\noindent

In order to obtain the $\Upsilon$--nucleus single particle energies, we
solve the Klein-Gordon equation (Since $\Upsilon$ is a spin-1 particle, we make an approximation where 
the transverse and longitudinal components in the Proca equation are expected to be very similar for $\Upsilon$ at rest, hence it is reduced to only one component, which corresponds to the Klein-Gordon equation) numerically. 
The following calculation was done in collaboration with Prof.~J.J.~Cobos-Mart\'{\i}nez from
University of Sonora, Mexico, who developed the computational code used in the calculation and also provided the results presented in this section.
The author has been studying to handle this code, check it's results and to extend it's use 
for future studies.

The Klein-Gordon equation is obtained by treating the hadron-nucleus potential as an 
scalar and adding it to the mass term: 
\begin{equation}
    \label{eqn:kge}
    [\mathbf{p}^{\,2}+(m+V_{N})^{2}]\psi=\mathcal{E}^{2}\psi.
\end{equation}
\noindent where $V_N(\mathbf{r})$ is a scalar (nuclear) potential (see Eq.~(\ref{eqn:Vsu})), and 
$m$ is the reduced mass of the $\Upsilon$--nucleus system 
$m=m_{\Upsilon}m_{A}/(m_{\Upsilon}+m_{A})$, with $m_{\Upsilon}$ $(m_{A}$) 
being the vacuum mass of the $\Upsilon$ meson (nucleus).

We solve the \eqn{eqn:kge} with the potential given in
\eqn{eqn:Vsu} for various nuclei, using momentum space methods.
Here, the Klein-Gordon equation is first converted to a momentum space representation
via Fourier transform, followed by a partial wave decomposition of the 
Fourier-transformed potential. 

In momentum space and after a partial wave decomposition, \eqn{eqn:kge} is given by 
\begin{multline}
\label{eqn:kge_pwd}
(p^{\,2}+m^{2})\phi_{nl}(p)+2m\int_{0}^{\infty}\dx{p'}p'^{2}V_{l}(p,p')
\phi_{nl}(p')
+ \int_{0}^{\infty}\dx{p'}p'^{2}\left[\int_{0}^{\infty}\dx{p''}p''^{2}V_{l}(p,p'')
V_{l}(p'',p')\right]\phi_{nl}(p') \\ = \mathcal{E}_{nl}^{2}\phi_{nl}(p)
\end{multline}
where $\mathcal{E}_{nl}$ is the energy eigenvalue and $V_{l}(p,p')$ is the partial wave decomposition of the momentum space, $V(q)$, interaction potential  
\begin{equation}
\label{eqn:v_pwd}
V_{l}(p,p')=2\pi\int_{-1}^{1}V(q)\, P_{l}(\cos\theta)
    d\cos\theta
\end{equation}
where $q=(p^{2}+p'^{2}-2pp'\cos\theta)^{1/2}$ and $P_{l}(\cos\theta)$ are the Legendre
polynomials.

For a given value of orbital angular momentum $l$, the eigenvalues $\mathcal{E}_{nl}$
of the resulting equation are found by the inverse iteration eigenvalue algorithm.
This algorithm is used to solve equations of the type

\begin{equation}
H \ket{\psi_{n}} = \mathcal{E}_{n} \ket{\psi_{n}}.
\end{equation}

For that, an operator is introduced as

\begin{equation}
B_{n} = (H - \epsilon_{n})^{-1},
\end{equation}
where $\epsilon_{n}$ are reasonable approximations for the eigenvalues. 
Also, an arbitrary state $\ket{X^{0}_{n}}$ is introduced, being described by the expansion

\begin{equation}
\ket{X^{0}_{n}} = \sum_{n'} C_{n'} \ket{\psi_{n'}}.
\end{equation}

Operating $N$ times in this state with $B_{n}$, we have

\begin{equation}
\ket{X^{N}_{n}} = (B_{n})^{N} \ket{X^{0}_{n}} = \sum_{n'} C_{n'}
(\mathcal{E}_{n'} - \epsilon_{n})^{-N} \ket{\psi_{n'}}.
\end{equation}

If $\epsilon_{n}$ is a good guess approximation, $n' = n$ will rapidly take over the sum, 
so for a large enough N

\begin{eqnarray}
\ket{X^{N}_{n}} &\approx& C_{n} (\mathcal{E}_{n} - \epsilon_{n})^{-N} \ket{\psi_{n}} \nonumber \\
\ket{X^{N+1}_{n}} &\approx& (\mathcal{E}_{n} - \epsilon_{n})^{-1} \ket{X^{N}_{n}}.
\end{eqnarray}

After enough iterations, $\mathcal{E}_{n}$ is obtained for each $l$.

The hadron-nucleus bound state energies, given by $E_{nl}=\mathcal{E}_{nl}-m$,
are listed in Tables~\ref{tab:upsilon-He4-kge}-\ref{tab:upsilon-Pb208-kge}.
As can be seen, the $\Upsilon$ is expected to form bound state with all of the nuclei studied.
However, the bound state energies are dependent on the value of the cutoff mass
$\Lambda_B$ used.

\begin{table}[h]
  \caption{\label{tab:upsilon-He4-kge} $^{4}_{\Upsilon}\text{He}$ bound state
  energies. When $|E| < 10^{-1}$ MeV we consider there is no bound state,
  which we denote with ``n''. All dimensioned quantities are in MeV.}
\begin{center}
\scalebox{0.9}{
\begin{tabular}{ll|r|r|r|r|r}
  \hline \hline
  & & \multicolumn{5}{c}{Bound state energies} \\
  \hline
& $n\ell$ & $\Lambda_{B}=2000$ & $\Lambda_{B}=3000$ & $\Lambda_{B}= 4000$ &
$\Lambda_{B}= 5000$ & $\Lambda_{B}= 6000$ \\
\hline
$^{4}_{\Upsilon}\text{He}$
& 1s &  -5.6 &  -6.4 & -7.5 & -9.0 & -10.8\\
\hline
\end{tabular}
}
\end{center}
\end{table}
\newpage

\begin{table}[h]
  \caption{\label{tab:upsilon-C12-kge} $^{12}_{\Upsilon}\text{C}$ bound state
  energies. When $|E| < 10^{-1}$ MeV we consider there is no bound state,
  which we denote with ``n''. All dimensioned quantities are in MeV.}
\begin{center}
\scalebox{0.9}{
\begin{tabular}{ll|r|r|r|r|r}
  \hline \hline
  & & \multicolumn{5}{c}{Bound state energies} \\
  \hline
& $n\ell$ & $\Lambda_{B}=2000$ & $\Lambda_{B}=3000$ & $\Lambda_{B}= 4000$ &
$\Lambda_{B}= 5000$ & $\Lambda_{B}= 6000$ \\
\hline
$^{12}_{\Upsilon}\text{C}$
& 1s & -10.6 & -11.6 & -12.8 & -14.4 & -16.3 \\
& 1p & -6.1 & -6.8 & -7.9 & -9.3 & -10.9 \\
& 1d & -1.5 & -2.1 & -2.9 & -4.0 & -5.4 \\
& 2s & -1.6 & -2.1 & -2.8 & -3.8 & -5.1 \\
& 2p & n & n & n & -0.1 & -0.7 \\
\hline
\end{tabular}
}
\end{center}
\end{table}

\begin{table}[h]
  \caption{\label{tab:upsilon-O16-kge} $^{16}_{\Upsilon}\text{O}$ bound state
  energies. When $|E| < 10^{-1}$ MeV we consider there is no bound state,
  which we denote with ``n''. All dimensioned quantities are in MeV.}
\begin{center}
\scalebox{0.9}{
\begin{tabular}{ll|r|r|r|r|r}
  \hline \hline
  & & \multicolumn{5}{c}{Bound state energies} \\
  \hline
& $n\ell$ & $\Lambda_{B}=2000$ & $\Lambda_{B}=3000$ & $\Lambda_{B}= 4000$ &
$\Lambda_{B}= 5000$ & $\Lambda_{B}= 6000$ \\
\hline
$^{16}_{\Upsilon}\text{O}$
& 1s & -11.9 & -12.9 & -14.2 & -15.8 & -17.8 \\
& 1p & -8.3 & -9.2 & -10.4 & -11.9 & -13.7 \\
& 1d & -4.4 & -5.1 & -6.2 & -7.5 & -9.2 \\
& 2s & -3.7 & -4.4 & -5.4 & -6.7 & -8.3 \\
& 1f & n & -0.9 & -1.8 & -2.9 & -4.3 \\
& 2p & -0.3 & -0.7 & -1.4 & -2.3 & -3.5 \\
& 3s & n & n & n & n & -0.1 \\
\hline
\end{tabular}
}
\end{center}
\end{table}
\newpage

\begin{table}[h]
  \caption{\label{tab:upsilon-Ca40-kge} $^{40}_{\Upsilon}\text{Ca}$ bound state
  energies. When $|E| < 10^{-1}$ MeV we consider there is no bound state,
  which we denote with ``n''. All dimensioned quantities are in MeV.}
\begin{center}
\scalebox{0.9}{
\begin{tabular}{ll|r|r|r|r|r}
  \hline \hline
  & & \multicolumn{5}{c}{Bound state energies} \\
  \hline
& $n\ell$ & $\Lambda_{B}=2000$ & $\Lambda_{B}=3000$ & $\Lambda_{B}= 4000$ &
$\Lambda_{B}= 5000$ & $\Lambda_{B}= 6000$ \\
\hline
$^{40}_{\Upsilon}\text{Ca}$
& 1s & -15.5 & -16.6 & -18.2 & -20.0 & -22.3 \\
& 1p & -13.3 & -14.4 & -15.9 & -17.7 & -19.8 \\
& 1d & -10.8 & -11.9 & -13.3 & -15.0 & -17.1 \\
& 2s & -10.3 & -11.3 & -12.7 & -14.4 & -16.4 \\
& 1f & -8.1 & -9.1 & -10.4 & -12.1 & -14.0 \\
& 2p & -7.4 & -8.3 & -9.6 & -11.2 & -13.1 \\
& 1g & -5.2 & -6.1 & -7.4 & -8.9 & -10.8 \\
& 2d & -4.5 & -5.4 & -6.5 & -8.0 & -9.8 \\
& 3s & -4.3 & -5.1 & -6.2 & -7.7 & -9.4 \\
& 1h & -2.3 & -3.1 & -4.2 & -5.6 & -7.3 \\
& 2f & -1.8 & -2.5 & -3.5 & -4.8 & -6.4 \\
& 3p & -1.6 & -2.3 & -3.2 & -4.4 & -6.0 \\
& 1i & n & n & -0.9 & -2.2 & -3.8 \\
& 2g & n & n & -0.6 & -1.7 & -3.0 \\
& 3d & n & n & -0.7 & -1.6 & -2.8 \\
& 4s & n & -0.3 & -0.7 & -1.6 & -2.7 \\
& 4p & n & n & n & n & -0.4 \\
& 3f & n & n & n & n & -0.1 \\
\hline
\end{tabular}
}
\end{center}
\end{table}
\newpage

\begin{table}[h]
  \caption{\label{tab:upsilon-Ca48-kge} $^{48}_{\Upsilon}\text{Ca}$ bound state
  energies. When $|E| < 10^{-1}$ MeV we consider there is no bound state,  
  which we denote with ``n''. All dimensioned quantities are in MeV.}
\begin{center}
\scalebox{0.9}{
\begin{tabular}{ll|r|r|r|r|r}
  \hline \hline
  & & \multicolumn{5}{c}{Bound state energies} \\
  \hline
& $n\ell$ & $\Lambda_{B}=2000$ & $\Lambda_{B}=3000$ & $\Lambda_{B}= 4000$ &
$\Lambda_{B}= 5000$ & $\Lambda_{B}= 6000$ \\
\hline
$^{48}_{\Upsilon}\text{Ca}$
& 1s & -15.3 & -16.4 & -17.9 & -19.7 & -21.8 \\
& 1p & -13.5 & -14.6 & -16.0 & -17.8 & -19.9 \\
& 1d & -11.4 & -12.4 & -13.8 & -15.6 & -17.6 \\
& 2s & -10.8 & -11.8 & -13.2 & -14.9 & -16.9 \\
& 1f & -9.1 & -10.1 & -11.4 & -13.1 & -15.0 \\
& 2p & -8.3 & -9.2 & -10.5 & -12.2 & -14.1 \\
& 1g & -6.6 & -7.5 & -8.8 & -10.3 & -12.2 \\
& 2d & -5.7 & -6.6 & -7.8 & -9.3 & -11.1 \\
& 3s & -5.4 & -6.2 & -7.4 & -8.9 & -10.7 \\
& 1h & -3.9 & -4.8 & -6.0 & -7.5 & -9.2 \\
& 2f & -3.1 & -3.8 & -4.9 & -6.3 & -8.0 \\
& 3p & -2.7 & -3.5 & -4.5 & -5.8 & -7.5 \\
& 1i & -1.2 & -1.9 & -3.0 & -4.4 & -6.1 \\
& 2g & -0.6 & -1.2 & -2.1 & -3.4 & -4.9 \\
& 3d & -0.5 & -1.0 & -1.8 & -3.0 & -4.4 \\
& 4s & -0.5 & -1.0 & -1.8 & -2.8 & -4.2 \\
& 1k & n & n & n & -1.3 & -2.8 \\
& 2h & n & n & n & -0.5 & -1.9 \\
& 3f & n & n & n & -0.4 & -1.5 \\
& 4p & n & n & n & -0.5 & -1.5 \\
\hline
\end{tabular}
}
\end{center}
\end{table}
\newpage

\begin{center}
\begin{longtable}{ll|r|r|r|r|r}
  \caption{\label{tab:upsilon-Zr90-kge} $^{90}_{\Upsilon}\text{Zr}$ bound state
  energies. When $|E| < 10^{-1}$ MeV we consider there is no bound state,
  which we denote with ``n''. All dimensioned quantities are in MeV.} \\
  \hline \hline
  & & \multicolumn{5}{c}{Bound state energies} \\
  \hline
& $n\ell$ & $\Lambda_{D}=2000$ & $\Lambda_{D}=3000$ & $\Lambda_{D}= 4000$ &
  $\Lambda_{D}= 5000$ & $\Lambda_{D}= 6000$ \\
\hline
\endfirsthead
\multicolumn{7}{c}%
{\tablename\ \thetable\ -- \textit{Continued from previous page}} \\
\hline
& $n\ell$ & $\Lambda_{B}=2000$ & $\Lambda_{B}=3000$ & $\Lambda_{B}= 4000$ &
$\Lambda_{B}= 5000$ & $\Lambda_{B}= 6000$ \\
\hline
\endhead
\hline \multicolumn{6}{r}{\textit{Continued on next page}} \\
\endfoot
\hline \hline
\endlastfoot
$^{90}_{\Upsilon}\text{Zr}$
& 1s & -15.5 & -16.6 & -18.1 & -19.9 & -22.0 \\
& 1p & -14.5 & -15.5 & -17.0 & -18.8 & -20.9 \\
& 1d & -13.2 & -14.2 & -15.7 & -17.4 & -19.5 \\
& 2s & -12.7 & -13.8 & -15.2 & -16.9 & -19.0 \\
& 1f & -11.7 & -12.7 & -14.1 & -15.9 & -17.9 \\
& 2p & -11.0 & -12.0 & -13.4 & -15.1 & -17.1 \\
& 1g & -10.1 & -11.1 & -12.4 & -14.1 & -16.1 \\
& 2d & -9.2 & -10.2 & -11.5 & -13.2 & -15.1 \\
& 3s & -8.8 & -9.8 & -11.1 & -12.8 & -14.7 \\
& 1h & -8.3 & -9.3 & -10.6 & -12.2 & -14.2 \\
& 2f & -7.2 & -8.2 & -9.4 & -11.1 & -13.0 \\
& 3p & -6.7 & -7.7 & -8.9 & -10.5 & -12.4 \\
& 1i & -6.4 & -7.3 & -8.6 & -10.2 & -12.1 \\
& 2g & -5.2 & -6.1 & -7.3 & -8.9 & -10.7 \\
& 3d & -4.6 & -5.5 & -6.7 & -8.2 & -10.0 \\
& 1k & -4.3 & -5.2 & -6.4 & -8.0 & -9.8 \\
& 4s & -4.4 & -5.2 & -6.4 & -7.9 & -9.7 \\
& 2h & -3.1 & -4.0 & -5.1 & -6.6 & -8.4 \\
& 3f & -2.5 & -3.3 & -4.4 & -5.8 & -7.5 \\
& 1l & -2.2 & -3.0 & -4.2 & -5.7 & -7.5 \\
& 4p & -2.3 & -3.0 & -4.1 & -5.4 & -7.1 \\
& 2i & -1.0 & -1.8 & -2.8 & -4.2 & -5.9 \\
& 3g & -0.6 & -1.2 & -2.2 & -3.4 & -5.0 \\
& 1m & n & -0.7 & -1.8 & -3.2 & -5.0 \\
& 4d & -0.5 & -1.0 & -1.9 & -3.0 & -4.5 \\
& 5s & -0.5 & -0.9 & -1.8 & -2.9 & -4.4 \\
& 2k & n & n & -0.6 & -1.8 & -3.4 \\
& 3h & n & n & n & -1.2 & -2.6 \\
& 1n & n & n & n & -0.7 & -2.4 \\
& 4f & n & n & n & -0.9 & -2.2 \\
& 5p & n & n & -0.1 & -0.8 & -2.0 \\
& 2l & n & n & n & n & -0.9 \\
& 3i & n & n & n & n & -0.3 \\
& 6s & n & n & n & n & -0.2 \\
& 5d & n & n & n & n & -0.2 \\
& 4g & n & n & n & n & -0.1 \\
\hline
\end{longtable}
\end{center}
\newpage

\begin{center}
\begin{longtable}{ll|r|r|r|r|r}
  \caption{\label{tab:upsilon-Pb208-kge} $^{208}_{\Upsilon}\text{Pb}$ bound state
  energies. When $|E| < 10^{-1}$ MeV we consider there is no bound state,
  which we denote with ``n''. All dimensioned quantities are in MeV.} \\
  \hline \hline
  & & \multicolumn{5}{c}{Bound state energies} \\
  \hline
& $n\ell$ & $\Lambda_{D}=2000$ & $\Lambda_{D}=3000$ & $\Lambda_{D}= 4000$ &
  $\Lambda_{D}= 5000$ & $\Lambda_{D}= 6000$ \\
\hline
\endfirsthead
\multicolumn{7}{c}%
{\tablename\ \thetable\ -- \textit{Continued from previous page}} \\
\hline
& $n\ell$ & $\Lambda_{B}=2000$ & $\Lambda_{B}=3000$ & $\Lambda_{B}= 4000$ &
$\Lambda_{B}= 5000$ & $\Lambda_{B}= 6000$ \\
\hline
\endhead
\hline \multicolumn{6}{r}{\textit{Continued on next page}} \\
\endfoot
\hline \hline
\endlastfoot
$^{208}_{\Upsilon}\text{Pb}$
& 1s & -15.7 & -16.8 & -18.2 & -20.0 & -22.1 \\
& 1p & -15.2 & -16.2 & -17.7 & -19.4 & -21.5 \\
& 1d & -14.5 & -15.5 & -16.9 & -18.7 & -20.8 \\
& 2s & -14.1 & -15.2 & -16.6 & -18.3 & -20.4 \\
& 1f & -13.6 & -14.7 & -16.1 & -17.8 & -19.9 \\
& 2p & -13.2 & -14.2 & -15.6 & -17.3 & -19.4 \\
& 1g & -12.7 & -13.7 & -15.1 & -16.8 & -18.9 \\
& 2d & -12.1 & -13.1 & -14.5 & -16.2 & -18.2 \\
& 3s & -11.8 & -12.8 & -14.2 & -15.9 & -17.9 \\
& 1h & -11.6 & -12.7 & -14.0 & -15.7 & -17.8 \\
& 2f & -10.9 & -11.9 & -13.2 & -14.9 & -16.9 \\
& 1i & -10.5 & -11.5 & -12.9 & -14.6 & -16.6 \\
& 3p & -10.5 & -11.5 & -12.9 & -14.6 & -16.5 \\
& 2g & -9.6 & -10.6 & -11.9 & -13.6 & -15.6 \\
& 1k & -9.3 & -10.3 & -11.6 & -13.3 & -15.2 \\
& 3d & -9.1 & -10.1 & -11.4 & -13.1 & -15.1 \\
& 4s & -8.8 & -9.8 & -11.1 & -12.8 & -14.7 \\
& 2h & -8.3 & -9.2 & -10.5 & -12.2 & -13.9 \\
& 1l & -8.0 & -8.9 & -10.2 & -11.9 & -13.8 \\
& 3f & -7.7 & -8.6 & -9.9 & -11.5 & -13.5 \\
& 4p & -7.4 & -8.3 & -9.6 & -11.2 & -13.1 \\
& 2i & -6.9 & -7.8 & -9.1 & -10.7 & -12.6 \\
& 1m & -6.6 & -7.5 & -8.8 & -10.4 & -12.4 \\
& 3g & -6.2 & -7.1 & -8.4 & -10.0 & -11.9 \\
& 4d & -5.8 & -6.7 & -8.0 & -9.5 & -11.4 \\
& 5s & -5.6 & -6.5 & -7.8 & -9.4 & -11.3 \\
& 2k & -5.4 & -6.3 & -7.6 & -9.2 & -11.0 \\
& 1n & -5.1 & -6.0 & -7.3 & -8.9 & -10.8 \\
& 3h & -4.7 & -5.5 & -6.8 & -8.3 & -10.2 \\
& 4f & -4.2 & -5.1 & -6.3 & -7.8 & -9.6 \\
& 5p & -4.0 & -4.9 & -6.1 & -7.6 & -9.3 \\
& 2l & -3.9 & -4.7 & -5.9 & -7.5 & -9.3 \\
& 1o & -3.6 & -4.5 & -5.7 & -7.3 & -9.1 \\
& 3i & -3.1 & -3.9 & -5.1 & -6.6 & -8.3 \\
& 4g & -2.6 & -3.4 & -4.5 & -6.0 & -7.7 \\
& 2m & -2.3 & -3.1 & -4.3 & -5.8 & -7.6 \\
& 1q & -2.0 & -2.8 & -4.1 & -5.6 & -7.4 \\
& 5d & -2.4 & -3.1 & -4.2 & -5.6 & -7.4 \\
& 6s & -2.3 & -3.0 & -4.1 & -5.5 & -7.3 \\
& 3k & -1.5 & -2.3 & -3.4 & -4.9 & -6.8 \\
& 6p & -0.8 & -1.5 & -2.6 & -4.2 & -6.4 \\
& 4h & -1.1 & -1.8 & -2.9 & -4.5 & -6.4 \\
& 5f & -0.9 & -1.6 & -2.7 & -4.3 & -6.4 \\
& 2n & -0.7 & -1.5 & -2.7 & -4.2 & -5.9 \\
& 1r & -0.3 & -1.1 & -2.3 & -3.8 & -5.6 \\
& 7s & n & -0.4 & -1.4 & -3.0 & -5.3 \\
& 6d & n & -0.4 & -1.4 & -2.9 & -5.2 \\
& 3l & n & -0.8 & -1.9 & -3.3 & -5.1 \\
& 5g & n & -0.4 & -1.4 & -2.9 & -5.0 \\
& 4i & n & -0.5 & -1.5 & -3.0 & -4.9 \\
& 2o & n & n & -0.9 & -2.3 & -4.0 \\
& 1t & n & n & -0.5 & -2.0 & -3.7 \\
& 7p & n & n & -0.2 & -1.2 & -3.7 \\
& 6f & n & n & n & -1.1 & -3.6 \\
& 5h & n & n & n & -0.9 & -3.4 \\
& 3m & n & n & -0.2 & -1.4 & -2.9 \\
& 6g & n & n & n & -0.3 & -2.9 \\
& 7d & n & n & n & -0.5 & -2.8 \\
& 8s & n & n & n & -0.5 & -2.8 \\
& 4k & n & n & n & -0.9 & -2.8 \\
& 5i & n & n & n & n & -2.8 \\
& 4l & n & n & n & n & -2.2 \\
& 2q & n & n & n & -0.4 & -2.0 \\
& 5k & n & n & n & n & -2.0 \\
& 4m & n & n & n & n & -1.8 \\
& 1u & n & n & n & n & -1.8 \\
& 3n & n & n & n & n & -1.6 \\
& 7f & n & n & n & n & -1.3 \\
& 8p & n & n & n & n & -1.2 \\
& 3o & n & n & n & n & -0.8 \\
& 2r & n & n & n & n & -0.5 \\
\end{longtable}
\end{center}
\vspace{-1cm}
As can be seen by the above tables, the absolute values of the bound state energies increases
according to the increase in the value of $\Lambda_B$.
This behavior comes from the $\Upsilon$-nucleus potentials, which gets deeper as the value 
of the cutoff mass increases. One can also note that the strength of the biding is higher
for heavier nuclei.
Despite the fact that the particular values for the bound state energies
are clearly dependent on the cutoff parameter $\Lambda_B$, the overall results, namely the expectation
that $\Upsilon$ may form bound states with all of the nuclei studied, do not depend on $\Lambda_B$.
Furthermore, the few MeV difference between the bound state energies would make it difficult to 
experimentally distinguish between the states. And since we have no evidence to support the idea
that the effect of the widths would be suppressed in medium, the inclusion of the imaginary
part of the potentials (or equivalently, the in-medium width of the $\Upsilon$ meson) 
in the calculations will be done in the near future, so we can test how much it will 
affect the predicted bound states.

\chapter{\boldmath{$\eta_b$} mass shift and bound state energies} 

\label{Chapter5} 

\thispagestyle{empty}

\noindent


\section{Mass shift}
\label{Ch5Sc1}

\thispagestyle{myheadings}

\noindent

Based on the discussion and analysis made for the $\Upsilon$ mass shift 
so far, we proceed to study the $\eta_b$ mass shift.
By the same philosophy as adopted for the $\Upsilon$ mass shift, 
we take only the $BB^*$ meson loop contribution for the $\eta_b$ self-energy 
as our prediction, 
namely, participants in the self-energy diagram of the $\eta_b$ meson are, one 
vector meson $B^*$, and two pseudoscalar mesons $\eta_b$ and $B$.

Before going into the details of the $\eta_b$ mass shift, 
we comment on an issue discussed in the pionic-atom study, 
the Ericson-Ericson-Lorentz-Lorenz (EELL) double (multiple) scattering 
correction~\cite{Ericson:1966fm,Ericson,Brown:1990wyp}. 
Since the mean field potentials become constant and the coupling constants 
are determined within the Hartree approximation (local) in the present QMC model treatment,   
the EELL double scattering correction (nonlocal effect), which was also considered 
for the $\eta$ and $\eta'$ meson mass shifts in nuclear matter~\cite{Bass:2005hn}, 
may be regarded as effectively included in our calculation.
In fact, based on this argument with some discussions, the EELL effect was not 
included explicitly in the study of the $\eta_c$ mass shift~\cite{Cobos-Martinez:2020ynh}. 
We simply follow Ref.~\cite{Cobos-Martinez:2020ynh} on the issue of the EELL effect 
in the present study.
Aside from this, we mention that there is a lack of useful information in the literature 
on the $\eta_b$-nucleon scattering length, even if one wants to estimate  
the EELL effect. 

The effective Lagrangian for the $\eta_b BB^*$ interaction is obtained from Eq.~(\ref{efflag}) 
in the same way as those for the $\Upsilon$, and we get,  
\begin{eqnarray}
{\cal L}_{\eta_b BB^*} 
&=& i g_{\eta_b BB^*}
\left\{ (\partial^\mu \eta_b) 
\left( \overline{B^*}_\mu B - \overline{B} B^*_\mu \right)
- \eta_b 
\left[ \overline{B^*}_\mu (\partial^\mu B) - (\partial^\mu \overline{B}) B^*_\mu \right]
\right\}, 
\label{Letab1}
\end{eqnarray}
where the coupling constant in the SU(5) scheme is used 
for $g_{\eta_b BB^*}$: 
\begin{equation}
g_{\eta_b BB^*} = g_{\Upsilon BB} = g_{\Upsilon B^*B^*} = \frac{5g}{4\sqrt{10}}.
\end{equation}

We also study the anomalous coupling $\eta_bB^*B^*$ contribution,
\begin{equation}
{\cal L}_{\eta_b B^*B^*} 
= \frac{g_{\eta_b B^*B^*}}{m_{\eta_b}}\varepsilon_{\alpha \beta \mu \nu}
\eta_b \left[ (\partial^{\alpha} \overline{B^*}^\beta) (\partial^{\mu}B^{* \nu}) \right], 
\label{Letab2} 
\end{equation}
assuming 
$g_{\eta_b B^*B^*} \left(\frac{m_{\eta_b}}{m_\Upsilon}\right) = g_{\eta_b BB^*}
(= g_{\Upsilon B^{*}B^{*}})$.
If we rely on the heavy quark symmetry and/or heavy meson (spin) symmetry, 
the above relation, $g_{\eta_b BB^*} = g_{\Upsilon BB}$, which is used for our  
prediction of the $\eta_b$ mass shift, may be justified.

The $\eta_b$ self-energy is expressed by~\cite{Cobos-Martinez:2020ynh} 
\bea
\hspace{-40ex}\Sigma_{\eta_b} = \Sigma_{\eta_b}^{BB^*} &+& \Sigma_{\etab}^{B^*B^*},   
\eea
with
\bea
\Sigma_{\eta_b}^{BB^*} 
&=& \frac{8 g_{\etab B B^*}^{2}}{\pi^{2}}\int_{0}^{\infty}
    \dx{q} \tbf{q}^{2} \tilde{F}_{BB^*}(\tbf{q}^2) I_{BB^*}(\tbf{q}^{2}),  
\label{SigetabBBs}
\\
\Sigma_{\etab}^{B^*B^*}(\tbf{q}^{2})
&=& \frac{2g_{\etab B^* B^{*}}^{2}}{\pi^{2}}\int_{0}^{\infty}
    \dx{q} \tbf{q}^{4} \tilde{F}_{B^*B^*}(\tbf{q}^2) I_{B^*B^*}(\tbf{q}^{2}),  
\label{SigetabBsBs}
\eea
and for the $\eta_b$ at rest, 
\bea
I_{BB^*}(\tbf{q}^{2})
&=& \left. \frac{m_{\etab}^{2}(-1+(q^0)^2/m_{B^{*}}^{2})}
{(q^{0}+\omega_{B^{*}})(q^{0}-\omega_{B^*}) 
(q^{0}-m_{\etab}-\omega_{B})}\right|_{q^{0}=m_{\etab}- \omega_B} 
\nonumber \\
&&\hspace{5ex} + \left. \frac{m_{\etab}^{2}(-1+(q^0)^2/m_{B^{*}}^{2})}
{(q^{0}-\omega_{B^{*}})(q^{0}-m_{\etab}+\omega_{B}) 
(q^{0}-m_{\etab}-\omega_{B})}\right|_{q^{0}=-\omega_{B^*}},  
\label{IBBs}
\\
\\
I_{B^*B^*}(\tbf{q}^{2})
&=& \left. \frac{1}{(q^0 + \omega_{B^*})(q^0 -\omega_{B^*}) 
(q^{0}-m_{\etab} -\omega_{B^*})} \right|_{q^0=m_{\etab}-\omega_{B^*}}  
\nonumber \\
&&\hspace{5ex} + \left. \frac{1}{(q^0-\omega_{B^*})(q^0-m_{\etab} + \omega_{B^*}) 
(q^0-m_{\etab} - \omega_{B^*})} \right|_{q^0=-\omega_{B^*}},
\label{IBsBs}
\eea
and $\omega_{B,B^*}=(\tbf{q}^{2}+m_{B,B^*}^{2})^{1/2}$.
For the $\etab BB^*$ and $\etab B^*B^*$ vertices, we use the similar form factors
as for the $\Upsilon$ case, 
$\tilde{F}_{BB^*}(\tbf{q}^2) = \tilde{u}_B(\tbf{q}^2) \tilde{u}_{B^*}(\tbf{q}^2)$ 
and $\tilde{F}_{B^*B^*}(\tbf{q}^2) = \tilde{u}_{B^*}^2(\tbf{q}^2)$, 
respectively, with
\begin{equation}
\label{eqn:FF}
\tilde{u}_{B,B^*}(\tbf{q}^{2})=
\left(
\frac{\Lambda_{B,B^*}^{2} + m_{\etab}^{2}}{\Lambda_{B,B^*}^{2}
    +4\omega_{B,B^*}^{2}(\tbf{q}^{2})}
\right)^{2}.
\end{equation}

Note that, in Eq.~(\ref{IBsBs}) no terms arise  
originating from $\propto (q^\lambda q^\sigma/m_{B^*}^2)$ in the $B^*$ propagators, 
due to the (two multiplication of) totally antisymmetric $\varepsilon$ tensor in the amplitude.
Thus, in contrast to the $B^*B^*$ loop contribution for the $\Upsilon$ self-energy, 
the $B^*B^*$ loop contribution for the $\eta_b$ self-energy is expected to be small, 
and to give a less divergent high energy behavior than 
that in the $\Upsilon$ self-energy.

\subsection{Results for $\eta_b$ mass shift}

\noindent

\begin{figure}[htb]
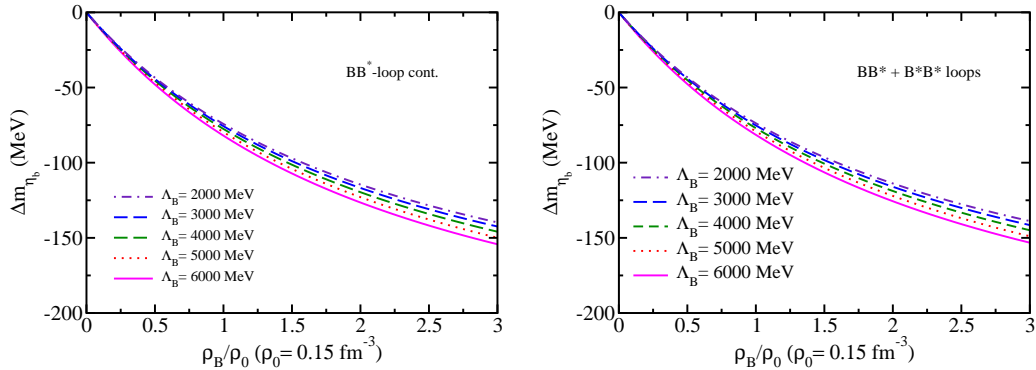
%
\vspace{4ex}
\centering
\includegraphics[width=6.5cm]{etab_BBs.eps}%
\hspace{2ex}
\includegraphics[width=6.5cm]{etab_total.eps}
\caption{$\eta_b$ mass shift from the $BB^*$ 
(left panel) and $BB^*+B^*B^*$ (right panel) meson loops  
versus nuclear matter density for five different values of the cutoff mass $\Lambda_{B} 
(=\Lambda_{B^*})$.}%
\label{fig7}%
\end{figure}

To be consistent, we show in Fig.~\ref{fig7} the calculated $\eta_b$ mass shift for including only 
the $BB^*$ loop --- our prediction (left panel), and that including the 
($BB^*+B^*B^*$) loops (right panel) for five different cutoff mass values, 
the same as those applied for the $\Upsilon$.
We have used $m_{\eta_b} = 9399$ MeV~\cite{PDG2020} for the free space value, with the values
of $m^{0}_{\eta_b}$ in the range of the cutoff masses $\Lambda_B$ being 14962 to 17726 MeV for the 
sole $BB^{*}$ loop contribution, and 14800 to 17475 MeV for the ($BB^{*}$ + $B^{*}B^{*}$) total loop contributions.
The $BB^{*}$ loop contribution to $m^{0}_{\eta_b}$ ranges from 5563 to 8327 MeV, and the one from the $B^{*}B^{*}$ loop contribution being -261 to -483 MeV.

The $\eta_b$ mass shift at $\rho_0$ with the $BB^*$ loop only (left panel) 
ranges from -75 to -82 MeV, while that with the ($BB^*+B^*B^*$) loops (right panel) 
ranges from -74 to -81 MeV. The two results show very similar mass shifts.
In the latter case, the $B^*B^*$ loop contribution ranges from +3 to +5 MeV at $\rho_0$.
The reason for the smaller contribution compared to that for the $\Upsilon$ self-energy 
is explained already in Sec.~\ref{Ch5Sc1}.
By the above difference and the fact that the smallness of the $B^*B^*$ loop contribution for 
the $\eta_b$ self-energy, we can conclude that the large $B^*B^*$ loop contribution for 
the $\Upsilon$ self energy arises due to (summation of) the $\Upsilon$ polarization vector, 
correlating with the momentum dependent part in numerators of the two $B^*$ propagators, 
$\propto (q^\mu q^\nu / m_{B^*}^2)$, which also often gives a divergent high energy behavior 
with including the integral measure $\int d^4 q$ in the $B^*$ meson propagator. 

Note that, similar to the $\Upsilon$ mass shift, dependence of the $\eta_b$ mass shift 
on the cutoff mass value $\Lambda_B (=\Lambda_{B^*})$ is again small, 
and it gives less ambiguity for the prediction originating from the cutoff mass value.
Unexpectedly, the $\eta_b$ mass shift is much larger than the  
predicted $\Upsilon$ mass shift due to only the $BB$ meson loop contribution, 
although the same lowest order $BB^*$ meson loop contribution 
(one vector and two pseudoscalar mesons) is included in the self-energy with 
the same range of the cutoff mass $\Lambda_B 
(=\Lambda_{B^*})$ values. 
One of the main reasons lies in the Lagrangian  
Eq.~(\ref{Letab1}). By the explicit calculation one can show that  
the large number of the interaction terms in the Lagrangian   
contributes to the self-energy, results to make 
the total contribution large. This is reflected in the coefficient in 
Eq.~(\ref{SigetabBBs}), and in contrast to the case of the $BB$ meson loop contribution 
in the $\Upsilon$ self-energy.
The similar, larger mass shift of the $\eta_c$ than that of the $J/\Psi$ 
was also observed in Ref.~\cite{Cobos-Martinez:2020ynh}, 
using the corresponding Lagrangians in the SU(4) sector.


\section{Nuclear potentials}

\noindent

\begin{figure}[!htb]
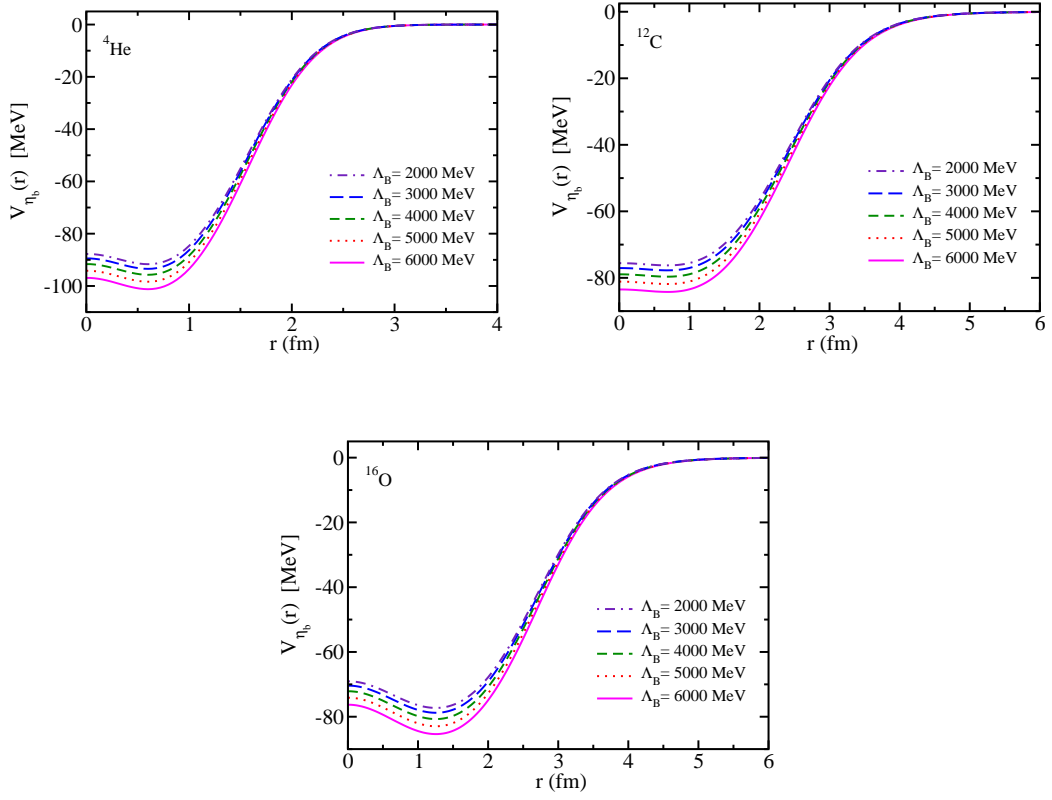
%
\centering
\includegraphics[width=6.5cm]{etabpot_He4.eps}
\hspace{2ex}
\includegraphics[width=6.5cm]{etabpot_C12.eps}
\\
\vspace{6ex}
\includegraphics[width=6.5cm]{etabpot_O16.eps}
\caption{$\eta_b$-nucleus potentials for various nuclei and values of the cutoff mass $\Lambda_{B}$.}
\label{enuclpot1}
\end{figure}

\begin{figure}[!htb]
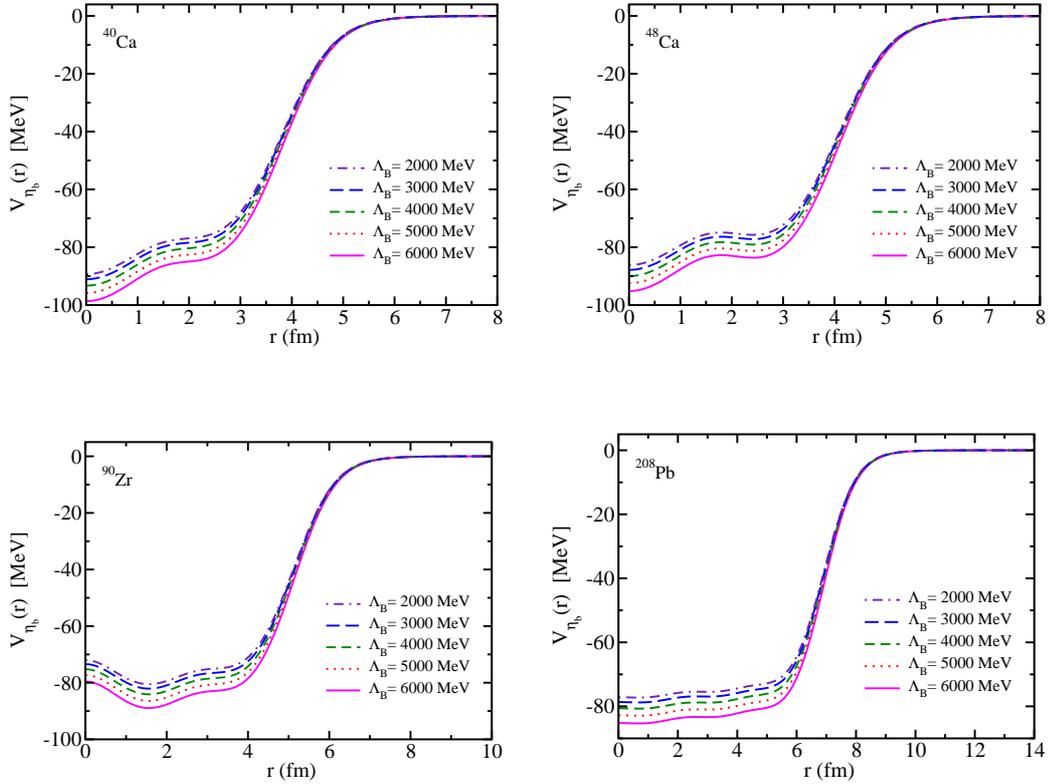
%
\centering
\vspace{2.5ex}
\includegraphics[width=6.5cm]{etabpot_Ca40.eps}
\hspace{2ex}
\includegraphics[width=6.5cm]{etabpot_Ca48.eps}
\\
\vspace{6ex}
\includegraphics[width=6.5cm]{etabpot_Zr90.eps}
\hspace{2ex}
\includegraphics[width=6.5cm]{etabpot_Pb208.eps}
\caption{$\eta_b$-nucleus potentials for various nuclei and values of the cutoff mass $\Lambda_{B}$.}
\label{enuclpot2}%
\end{figure}

The $\eta_b$-nucleus scalar potential is calculated in the local density approximation
from the $\eta_b$ mass shift in nuclear matter, in the same way as for $\Upsilon$ and for the same set of nuclei, together with the baryon density profile $\rho_{B}^{A}(r)$ of the given 
nucleus, also calculated in the QMC model, as

\begin{equation}
\label{eqn:Vs}
V_{\eta_b A}(r)= \Delta m_{\eta_b}(\rho_{B}^{A}(r)),
\end{equation}
\noindent where $r$ is the distance from the center of the nucleus, and 
$\Delta m_{\eta_b}$ is the $\eta_b$ mass shift in the nucleus $A$.
The attractive potentials are presented for each nucleus in Figs.~\ref{enuclpot1}
and~\ref{enuclpot2}.

\vspace{2ex}

\section{$\eta_b$-nucleus bound state energies}

\noindent

We obtain the $\eta_b$-nucleus bound state energies by solving \eqn{eqn:kge} with the potential 
given in \eqn{eqn:Vs}. In the same way that was done for the $\Upsilon$ case, the calculation
was done with the code provided by Prof.~J.J.~Cobos-Mart\'{\i}nez.
Again, the author is studying the code in order to check the results and to apply it to
further extensions of this study. 

\newpage
We present the results for the binding energies for each nucleus in 
Tables~\ref{tab:etab-He4-kge}-\ref{tab:etab-Pb208-kge}. By the results obtained, 
the $\eta_b$ meson is expected to be bound to all of the studied nuclei.

\begin{table}[!h]
  \caption{\label{tab:etab-He4-kge} $^{4}_{\eta_b}\text{He}$ bound state
  energies. When $|E| < 10^{-1}$ MeV we consider there is no bound state,
  which we denote with ``n''. All dimensioned quantities are in MeV.}
\begin{center}
\scalebox{0.9}{
\begin{tabular}{ll|r|r|r|r|r}
  \hline \hline
  & & \multicolumn{5}{c}{Bound state energies} \\
  \hline
& $n\ell$ & $\Lambda_{B}=2000$ & $\Lambda_{B}=3000$ & $\Lambda_{B}= 4000$ &
$\Lambda_{B}= 5000$ & $\Lambda_{B}= 6000$ \\
\hline
$^{4}_{\eta_b}\text{He}$
& 1s & -63.1 & -64.7 & -66.7 & -69.0 & -71.5 \\ 
& 1p & -40.6 & -42.0 & -43.7 & -45.8 & -48.0 \\ 
& 1d & -17.2 & -18.3 & -19.7 & -21.4 & -23.2 \\ 
& 2s & -15.6 & -16.6 & -17.9 & -19.4 & -21.1 \\ 
& 2p & n & n & -0.3 & -0.9 & -1.7 \\
\hline
\end{tabular}
}
\end{center}
\end{table}
\newpage

\begin{table}[!h]
  \caption{\label{tab:etab-C12-kge} $^{12}_{\eta_b}\text{C}$ bound state
  energies. When $|E| < 10^{-1}$ MeV we consider there is no bound state,
  which we denote with ``n''. All dimensioned quantities are in MeV.}
\begin{center}
\scalebox{0.9}{
\begin{tabular}{ll|r|r|r|r|r}
  \hline \hline
  & & \multicolumn{5}{c}{Bound state energies} \\
  \hline
& $n\ell$ & $\Lambda_{B}=2000$ & $\Lambda_{B}=3000$ & $\Lambda_{B}= 4000$ &
$\Lambda_{B}= 5000$ & $\Lambda_{B}= 6000$ \\
\hline
$^{12}_{\eta_b}\text{C}$
& 1s & -65.8 & -67.2 & -69.0 & -71.1 & -73.4 \\
& 1p & -57.0 & -58.4 & -60.1 & -62.1 & -64.3 \\
& 1d & -47.5 & -48.8 & -50.4 & -52.3 & -54.4 \\
& 2s & -46.3 & -47.5 & -49.1 & -51.0 & -53.0 \\
& 1f & -37.5 & -38.7 & -40.2 & -42.0 & -43.9 \\
& 2p & -36.0 & -37.1 & -38.6 & -40.3 & -42.2 \\
& 1g & -27.1 & -28.2 & -29.6 & -31.3 & -33.1 \\
& 2d & -25.7 & -26.7 & -28.1 & -29.7 & -31.4 \\
& 3s & -25.1 & -26.2 & -27.5 & -29.1 & -30.8 \\
& 1h & -16.6 & -17.7 & -18.9 & -20.4 & -22.0 \\
& 2f & -15.7 & -16.6 & -17.8 & -19.2 & -20.8 \\
& 3p & -15.4 & -16.3 & -17.4 & -18.8 & -20.3 \\
& 1i & -6.1 & -6.9 & -8.1 & -9.4 & -10.9 \\
& 4s & -6.9 & -7.6 & -8.5 & -9.5 & -10.7 \\
& 3d & -6.7 & -7.4 & -8.3 & -9.4 & -10.6 \\
& 2g & -6.3 & -7.1 & -8.1 & -9.3 & -10.6 \\
& 4p & -1.0 & -1.3 & -1.8 & -2.5 & -3.2 \\
& 3f & n & -0.3 & -0.9 & -1.6 & -2.4 \\
& 2h & n & n & n & -0.2 & -1.2 \\
\hline
\end{tabular}
}
\end{center}
\end{table}
\newpage

\begin{table}[!h]
  \caption{\label{tab:etab-O16-kge} $^{16}_{\eta_b}\text{O}$ bound state
  energies. When $|E| < 10^{-1}$ MeV we consider there is no bound state,
  which we denote with ``n''. All dimensioned quantities are in MeV.}
\begin{center}
\scalebox{0.9}{
\begin{tabular}{ll|r|r|r|r|r}
  \hline \hline
  & & \multicolumn{5}{c}{Bound state energies} \\
  \hline
& $n\ell$ & $\Lambda_{B}=2000$ & $\Lambda_{B}=3000$ & $\Lambda_{B}= 4000$ &
$\Lambda_{B}= 5000$ & $\Lambda_{B}= 6000$ \\
\hline
$^{16}_{\eta_b}\text{O}$
& 1s & -67.8 & -69.2 & -71.0 & -73.1 & -75.4 \\
& 1p & -61.8 & -63.2 & -64.9 & -67.0 & -69.2 \\
& 1d & -54.9 & -56.2 & -57.9 & -59.9 & -62.0 \\
& 2s & -53.2 & -54.6 & -56.3 & -58.2 & -60.3 \\
& 1f & -47.3 & -48.6 & -50.2 & -52.1 & -54.2 \\
& 2p & -45.0 & -46.2 & -47.9 & -49.7 & -51.8 \\
& 1g & -39.1 & -40.3 & -41.9 & -43.7 & -45.7 \\
& 2d & -36.4 & -37.6 & -39.1 & -40.9 & -42.9 \\
& 3s & -35.3 & -36.5 & -38.0 & -39.8 & -41.7 \\
& 1h & -30.5 & -31.6 & -33.1 & -34.9 & -36.8 \\
& 2f & -27.6 & -28.7 & -30.1 & -31.8 & -33.7 \\
& 3p & -26.2 & -27.3 & -28.7 & -30.3 & -32.1 \\
& 1i & -21.4 & -22.5 & -23.9 & -25.6 & -27.4 \\
& 2g & -18.7 & -19.7 & -21.0 & -22.6 & -24.3 \\
& 3d & -17.3 & -18.3 & -19.5 & -21.0 & -22.7 \\
& 4s & -16.7 & -17.7 & -19.0 & -20.4 & -22.0 \\
& 1k & -12.1 & -13.1 & -14.4 & -16.0 & -17.6 \\
& 2h & -9.8 & -10.8 & -12.0 & -13.4 & -14.9 \\
& 3f & -8.8 & -9.7 & -10.8 & -12.1 & -13.6 \\
& 4p & -8.5 & -9.3 & -10.4 & -11.6 & -13.0 \\
& 1l & -2.6 & -3.5 & -4.7 & -6.2 & -7.7 \\
& 2i & -1.3 & -2.1 & -3.2 & -4.4 & -5.8 \\
& 3g & -1.3 & -2.0 & -2.9 & -3.9 & -5.1 \\
& 5s & -2.1 & -2.6 & -3.2 & -4.1 & -5.1 \\
& 4d & -1.8 & -2.3 & -3.1 & -4.0 & -5.1 \\
& 5p & n & n & n & n & -0.2 \\
\hline
\end{tabular}
}
\end{center}
\end{table}
\newpage

\begin{longtable}{ll|r|r|r|r|r}
  \caption{\label{tab:etab-Ca40-kge} $^{40}_{\eta_{b}}\text{Ca}$ bound
    state energies. When $|E| < 10^{-1}$ MeV we consider there is no bound
    state, which we denote with ``n''. All dimensioned quantities are in
    MeV.} \\
  \hline \hline
  & & \multicolumn{5}{c}{Bound state energies} \\
  \hline
& $n\ell$ & $\Lambda_{B}=2000$ & $\Lambda_{B}=3000$ & $\Lambda_{B}= 4000$ &
  $\Lambda_{B}= 5000$ & $\Lambda_{B}= 6000$ \\
\hline
\endfirsthead
\multicolumn{7}{c}%
{\tablename\ \thetable\ -- \textit{Continued from previous page}} \\
\hline
& $n\ell$ & $\Lambda_{B}=2000$ & $\Lambda_{B}=3000$ & $\Lambda_{B}= 4000$ &
  $\Lambda_{B}= 5000$ & $\Lambda_{B}= 6000$ \\
\hline
\endhead
\hline \multicolumn{6}{r}{\textit{Continued on next page}} \\
\endfoot
\hline \hline
\endlastfoot
$^{40}_{\eta_b}\text{Ca}$
& 1s & -79.0 & -80.6 & -82.6 & -85.0 & -87.5 \\
& 1p & -75.4 & -77.0 & -79.0 & -81.4 & -83.9 \\
& 1d & -71.4 & -73.0 & -74.9 & -77.2 & -79.7 \\
& 2s & -70.5 & -72.0 & -74.0 & -76.3 & -78.8 \\
& 1f & -67.0 & -68.5 & -70.4 & -72.7 & -75.1 \\
& 2p & -65.6 & -67.1 & -69.0 & -71.3 & -73.7 \\
& 1g & -62.2 & -63.6 & -65.6 & -67.8 & -70.2 \\
& 2d & -60.4 & -61.9 & -63.8 & -66.0 & -68.4 \\
& 3s & -59.7 & -61.2 & -63.1 & -65.3 & -67.6 \\
& 1h & -57.0 & -58.5 & -60.4 & -62.5 & -64.9 \\
& 2f & -55.0 & -56.4 & -58.3 & -60.4 & -62.8 \\
& 3p & -53.9 & -55.4 & -57.2 & -59.4 & -61.7 \\
& 1i & -51.6 & -53.0 & -54.8 & -57.0 & -59.3 \\
& 2g & -49.3 & -50.8 & -52.6 & -54.7 & -56.9 \\
& 3d & -48.0 & -49.4 & -51.2 & -53.3 & -55.5 \\
& 4s & -47.4 & -48.8 & -50.6 & -52.7 & -54.9 \\
& 1k & -45.9 & -47.3 & -49.2 & -51.1 & -53.4 \\ 
& 2h & -43.5 & -44.9 & -46.6 & -48.7 & -50.9 \\
& 3f & -42.0 & -43.3 & -45.0 & -47.1 & -49.2 \\
& 4p & -41.1 & -42.5 & -44.2 & -46.2 & -48.4 \\
& 1l & -40.0 & -41.3 & -43.0 & -45.1 & -47.2 \\
& 2i & -37.5 & -38.8 & -40.5 & -42.5 & -44.6 \\
& 3g & -35.8 & -37.1 & -38.8 & -40.7 & -42.8 \\
& 4d & -34.8 & -36.1 & -37.8 & -39.7 & -41.8 \\
& 5s & -34.4 & -35.7 & -37.3 & -39.2 & -41.3 \\
& 1m & -33.8 & -35.1 & -36.8 & -38.8 & -40.9 \\
& 2k & -31.3 & -32.6 & -34.2 & -36.1 & -38.2 \\
& 3h & -29.6 & -30.8 & -32.4 & -34.3 & -36.3 \\
& 4f & -28.5 & -29.7 & -31.3 & -33.2 & -35.2 \\
& 5p & -28.0 & -29.2 & -30.7 & -32.5 & -34.5 \\
& 1n & -27.5 & -28.7 & -30.4 & -32.3 & -34.3 \\
& 2l & -25.0 & -26.3 & -27.8 & -29.7 & -31.7 \\
& 3i & -23.4 & -24.5 & -26.1 & -27.9 & -29.8 \\
& 4g & -22.3 & -23.4 & -24.9 & -26.7 & -28.6 \\
& 5d & -21.6 & -22.8 & -24.2 & -26.0 & -27.8 \\
& 1o & -21.0 & -22.2 & -23.8 & -25.6 & -27.7 \\
& 6s & -21.4 & -22.5 & -24.0 & -25.7 & -27.5 \\
& 2m & -18.7 & -19.9 & -21.4 & -23.1 & -25.0 \\ 
& 3k & -17.1 & -18.3 & -19.7 & -21.4 & -23.2 \\ 
& 4h & -16.2 & -17.2 & -18.6 & -20.3 & -22.1 \\
& 5f & -15.6 & -16.6 & -18.0 & -19.6 & -21.3 \\
& 6p & -15.3 & -16.3 & -17.6 & -19.2 & -20.9 \\
& 1q & -14.3 & -15.5 & -17.0 & -18.8 & -20.7 \\
& 2n & -12.3 & -13.4 & -14.8 & -16.5 & -18.3 \\
& 3l & -11.0 & -12.0 & -13.4 & -15.0 & -16.7 \\
& 4i & -10.2 & -11.2 & -12.5 & -14.0 & -15.7 \\
& 5g & -9.8 & -10.8 & -12.0 & -13.4 & -15.0 \\
& 6d & -9.6 & -10.5 & -11.7 & -13.1 & -14.7 \\
& 7s & -9.5 & -10.4 & -11.6 & -13.0 & -14.5 \\
& 1r & -7.6 & -8.7 & -10.1 & -11.8 & -13.6 \\
& 2o & -5.9 & -6.9 & -8.3 & -9.9 & -11.6 \\
& 3m & -5.0 & -6.0 & -7.2 & -8.7 & -10.3 \\
& 4k & -4.6 & -5.5 & -6.6 & -8.0 & -9.5 \\
& 5h & -4.6 & -5.3 & -6.4 & -7.7 & -9.1 \\
& 6f & -4.6 & -5.4 & -6.4 & -7.6 & -8.9 \\
& 7p & -4.7 & -5.4 & -6.4 & -7.5 & -8.8 \\
& 1t & -0.7 & -1.8 & -3.1 & -4.8 & -6.5 \\
& 2q & n & -0.5 & -1.8 & -3.2 & -4.8 \\
& 8s & -1.2 & -1.6 & -2.2 & -3.0 & -4.0 \\
& 3n & n & n & -1.2 & -2.5 & -4.0 \\
& 7d & -1.0 & -1.4 & -2.1 & -2.9 & -4.0 \\
& 6g & -0.6 & -1.1 & -1.8 & -2.7 & -3.8 \\
& 4l & n & -0.2 & -1.2 & -2.3 & -3.6 \\
& 5i & n & -0.6 & -1.4 & -2.5 & -3.6 \\
& 8p & n & n & n & -0.2 & -0.7 \\
\end{longtable}

\newpage

\begin{longtable}{ll|r|r|r|r|r}
  \caption{\label{tab:etab-Ca48-kge} $^{48}_{\eta_{b}}\text{Ca}$ bound
    state energies. When $|E| < 10^{-1}$ MeV we consider there is no bound
    state, which we denote with ``n''. All dimensioned quantities are in
    MeV.} \\
  \hline \hline
  & & \multicolumn{5}{c}{Bound state energies} \\
  \hline
& $n\ell$ & $\Lambda_{B}=2000$ & $\Lambda_{B}=3000$ & $\Lambda_{B}= 4000$ &
  $\Lambda_{B}= 5000$ & $\Lambda_{B}= 6000$ \\
\hline
\endfirsthead
\multicolumn{7}{c}%
{\tablename\ \thetable\ -- \textit{Continued from previous page}} \\
\hline
& $n\ell$ & $\Lambda_{B}=2000$ & $\Lambda_{B}=3000$ & $\Lambda_{B}= 4000$ &
  $\Lambda_{B}= 5000$ & $\Lambda_{B}= 6000$ \\
\hline
\endhead
\hline \multicolumn{6}{r}{\textit{Continued on next page}} \\
\endfoot
\hline \hline
\endlastfoot
$^{48}_{\eta_b}\text{Ca}$

& 1s & -76.7 & -78.2 & -80.2 & -82.5 & -85.0 \\
& 1p & -74.0 & -75.5 & -77.4 & -79.7 & -82.1 \\
& 1d & -70.8 & -72.3 & -74.2 & -76.4 & -78.8 \\
& 2s & -69.9 & -71.4 & -73.3 & -75.5 & -77.9 \\
& 1f & -67.2 & -68.6 & -70.6 & -72.8 & -75.1 \\
& 2p & -65.8 & -67.3 & -69.2 & -71.4 & -73.8 \\
& 1g & -63.2 & -64.7 & -66.6 & -68.7 & -71.1 \\
& 2d & -61.5 & -63.0 & -64.8 & -67.0 & -69.3 \\
& 3s & -60.8 & -62.2 & -64.1 & -66.3 & -68.6 \\
& 1h & -59.0 & -60.4 & -62.3 & -64.4 & -66.7 \\
& 2f & -56.9 & -58.4 & -60.2 & -62.3 & -64.6 \\
& 3p & -55.9 & -57.3 & -59.1 & -61.2 & -63.5 \\
& 1i & -54.4 & -55.9 & -57.7 & -59.8 & -62.1 \\
& 2g & -52.1 & -53.5 & -55.3 & -57.4 & -59.7 \\
& 3d & -50.8 & -52.1 & -53.9 & -56.0 & -58.2 \\
& 4s & -50.2 & -51.5 & -53.3 & -55.4 & -57.6 \\
& 1k & -49.6 & -51.0 & -52.8 & -54.9 & -57.1 \\
& 2h & -47.1 & -48.5 & -50.2 & -52.3 & -54.5 \\
& 3f & -45.5 & -46.8 & -48.6 & -50.6 & -52.8 \\
& 1l & -44.6 & -45.9 & -47.7 & -49.7 & -51.9 \\
& 4p & -44.6 & -46.0 & -47.7 & -49.7 & -51.9 \\
& 2i & -41.9 & -43.2 & -44.9 & -46.9 & -49.1 \\
& 3g & -40.0 & -41.4 & -43.1 & -45.0 & -47.2 \\
& 1m & -39.3 & -40.6 & -42.5 & -44.3 & -46.5 \\
& 4d & -39.0 & -40.3 & -41.9 & -43.9 & -46.0 \\
& 5s & -38.5 & -39.8 & -41.5 & -43.4 & -45.5 \\
& 2k & -36.5 & -37.8 & -39.4 & -41.4 & -43.5 \\
& 3h & -34.5 & -35.8 & -37.4 & -39.3 & -41.4 \\
& 1n & -33.8 & -35.1 & -36.8 & -38.7 & -40.8 \\
& 4f & -33.2 & -34.5 & -36.1 & -38.0 & -40.1 \\
& 5p & -32.6 & -33.8 & -35.5 & -37.3 & -39.4 \\
& 2l & -30.9 & -32.2 & -33.8 & -35.7 & -37.7 \\
& 3i & -28.9 & -30.1 & -31.7 & -33.5 & -35.6 \\
& 1o & -28.2 & -29.4 & -31.0 & -32.9 & -35.0 \\
& 4g & -27.5 & -28.7 & -30.3 & -32.1 & -34.1 \\
& 5d & -26.7 & -27.9 & -29.4 & -31.2 & -33.2 \\ 
& 6s & -26.3 & -27.5 & -29.1 & -30.9 & -32.8 \\
& 2m & -25.2 & -26.4 & -28.0 & -29.9 & -31.9 \\
& 3k & -23.2 & -24.4 & -25.9 & -27.7 & -29.6 \\
& 1q & -22.3 & -23.5 & -25.1 & -26.9 & -28.9 \\
& 4h & -21.8 & -22.9 & -24.4 & -26.2 & -28.1 \\
& 5f & -20.9 & -22.0 & -23.5 & -25.2 & -27.1 \\
& 6p & -20.5 & -21.6 & -23.0 & -24.7 & -26.6 \\
& 2n & -19.4 & -20.6 & -22.1 & -23.9 & -25.8 \\
& 3l & -17.4 & -18.6 & -20.0 & -21.7 & -23.6 \\
& 1r & -16.3 & -17.5 & -19.0 & -20.8 & -22.7 \\
& 4i & -16.1 & -17.2 & -18.6 & -20.3 & -22.1 \\
& 5g & -15.3 & -16.3 & -17.7 & -19.3 & -21.1 \\
& 6d & -14.8 & -15.8 & -17.2 & -18.7 & -20.5 \\
& 7s & -14.6 & -15.6 & -16.9 & -18.5 & -20.2 \\
& 2o & -13.6 & -14.7 & -16.2 & -17.9 & -19.7 \\
& 3m & -11.7 & -12.8 & -14.2 & -15.8 & -17.6 \\
& 1t & -10.2 & -11.3 & -12.8 & -14.5 & -16.4 \\
& 4k & -10.5 & -11.6 & -12.9 & -14.4 & -16.1 \\
& 5h & -9.8 & -10.8 & -12.0 & -13.5 & -15.2 \\
& 6f & -9.4 & -10.4 & -11.6 & -13.0 & -14.6 \\
& 7p & -9.2 & -10.2 & -11.3 & -12.8 & -14.3 \\
& 2q & -7.7 & -8.7 & -10.1 & -11.7 & -13.5 \\
& 3n & -6.1 & -7.1 & -8.4 & -9.9 & -11.6 \\
& 4l & -5.2 & -6.1 & -7.3 & -8.7 & -10.3 \\
& 1u & -3.9 & -5.0 & -6.4 & -8.1 & -9.9 \\
& 5i & -4.7 & -5.6 & -6.7 & -8.0 & -9.5 \\ 
& 6g & -4.6 & -5.3 & -6.4 & -7.7 & -9.1 \\
& 7d & -4.5 & -5.3 & -6.3 & -7.5 & -8.9 \\
& 8s & -4.5 & -5.2 & -6.2 & -7.4 & -8.7 \\
& 2r & -1.7 & -2.7 & -4.0 & -5.6 & -7.3 \\
& 3o & -0.5 & -1.4 & -2.6 & -4.0 & -5.6 \\
& 4m & n & -0.9 & -1.9 & -3.2 & -4.6 \\
& 5k & -0.1 & -0.8 & -1.8 & -2.9 & -4.2 \\
& 8p & -1.0 & -1.5 & -2.2 & -3.0 & -4.1 \\
& 7f & -0.8 & -1.3 & -2.0 & -3.0 & -4.0 \\
& 6h & -0.5 & -1.0 & -1.8 & -2.9 & -4.0 \\
& 1v & n & n & n & -1.5 & -3.3 \\
& 2t & n & n & n & n & -1.0 \\
& 9s & n & n & n & -0.3 & -0.8 \\
& 8d & n & n & n & -0.1 & -0.6 \\
& 7g & n & n & n & n & -0.3 \\
\end{longtable}

\newpage

\begin{longtable}{ll|r|r|r|r|r}
  \caption{\label{tab:etab-Zr90-kge} $^{90}_{\eta_{b}}\text{Zr}$ bound
    state energies. When $|E| < 10^{-1}$ MeV we consider there is no bound
    state, which we denote with ``n''. All dimensioned quantities are in
    MeV.} \\
  \hline \hline
  & & \multicolumn{5}{c}{Bound state energies} \\
  \hline
& $n\ell$ & $\Lambda_{B}=2000$ & $\Lambda_{B}=3000$ & $\Lambda_{B}= 4000$ &
  $\Lambda_{B}= 5000$ & $\Lambda_{B}= 6000$ \\
\hline
\endfirsthead
\multicolumn{7}{c}%
{\tablename\ \thetable\ -- \textit{Continued from previous page}} \\
\hline
& $n\ell$ & $\Lambda_{B}=2000$ & $\Lambda_{B}=3000$ & $\Lambda_{B}= 4000$ &
  $\Lambda_{B}= 5000$ & $\Lambda_{B}= 6000$ \\
\hline
\endhead
\hline \multicolumn{6}{r}{\textit{Continued on next page}} \\
\endfoot
\hline \hline
\endlastfoot
$^{90}_{\eta_b}\text{Zr}$
& 1s & -75.5 & -77.0 & -78.9 & -81.1 & -83.5 \\
& 1p & -74.1 & -75.6 & -77.5 & -79.7 & -82.1 \\ 
& 1d & -72.3 & -73.8 & -75.7 & -77.9 & -80.2 \\ 
& 2s & -71.6 & -73.0 & -74.9 & -77.1 & -79.5 \\ 
& 1f & -70.2 & -71.7 & -73.6 & -75.8 & -78.1 \\ 
& 2p & -69.2 & -70.7 & -72.6 & -74.8 & -77.1 \\ 
& 1g & -67.9 & -69.4 & -71.2 & -73.4 & -75.8 \\ 
& 2d & -66.6 & -68.1 & -69.9 & -72.1 & -74.4 \\ 
& 3s & -66.0 & -67.4 & -69.3 & -71.5 & -73.8 \\ 
& 1h & -65.4 & -66.8 & -68.7 & -70.8 & -73.2 \\ 
& 2f & -63.8 & -65.2 & -67.0 & -69.2 & -71.5 \\ 
& 3p & -62.9 & -64.4 & -66.2 & -68.4 & -70.7 \\ 
& 1i & -62.6 & -64.0 & -65.9 & -68.0 & -70.3 \\ 
& 2g & -60.7 & -62.1 & -64.0 & -66.1 & -68.4 \\ 
& 1k & -59.7 & -61.1 & -62.9 & -65.0 & -67.3 \\ 
& 3d & -59.6 & -61.0 & -62.9 & -65.0 & -67.3 \\ 
& 4s & -59.1 & -60.5 & -62.3 & -64.4 & -66.7 \\ 
& 2h & -57.5 & -58.9 & -60.7 & -62.8 & -65.1 \\ 
& 1l & -56.5 & -57.9 & -59.7 & -61.8 & -64.1 \\ 
& 3f & -56.1 & -57.5 & -59.4 & -61.5 & -63.7 \\ 
& 4p & -55.4 & -56.8 & -58.6 & -60.7 & -63.0 \\ 
& 2i & -54.1 & -55.5 & -57.3 & -59.4 & -61.6 \\ 
& 1m & -53.2 & -54.6 & -56.4 & -58.5 & -60.7 \\ 
& 3g & -52.5 & -53.9 & -55.7 & -57.8 & -60.0 \\ 
& 4d & -51.5 & -52.9 & -54.7 & -56.8 & -59.0 \\ 
& 5s & -51.1 & -52.4 & -54.2 & -56.3 & -58.5 \\ 
& 2k & -50.5 & -51.9 & -53.7 & -55.8 & -58.0 \\ 
& 1n & -49.7 & -51.1 & -52.8 & -54.9 & -57.1 \\ 
& 3h & -48.7 & -50.1 & -51.9 & -53.9 & -56.1 \\ 
& 4f & -47.6 & -48.9 & -50.7 & -52.7 & -54.9 \\ 
& 5p & -46.9 & -48.3 & -50.1 & -52.1 & -54.3 \\ 
& 2l & -46.8 & -48.2 & -50.0 & -52.0 & -54.2 \\ 
& 1o & -46.0 & -47.4 & -49.1 & -51.2 & -53.4 \\ 
& 3i & -44.8 & -46.2 & -47.9 & -49.9 & -52.1 \\
& 4g & -43.5 & -44.8 & -46.5 & -48.5 & -50.7 \\ 
& 2m & -43.0 & -44.3 & -46.1 & -48.1 & -50.2 \\ 
& 5d & -42.6 & -44.0 & -45.7 & -47.7 & -49.9 \\ 
& 1q & -42.2 & -43.5 & -45.2 & -47.3 & -49.4 \\ 
& 6s & -42.2 & -43.5 & -45.2 & -47.2 & -49.4 \\ 
& 3k & -40.8 & -42.1 & -43.8 & -45.8 & -48.0 \\ 
& 4h & -39.2 & -40.6 & -42.3 & -44.2 & -46.4 \\ 
& 2n & -39.0 & -40.3 & -42.0 & -44.0 & -46.1 \\ 
& 1r & -38.2 & -39.5 & -41.2 & -43.2 & -45.3 \\ 
& 5f & -38.2 & -39.6 & -41.2 & -43.2 & -45.3 \\ 
& 6p & -37.7 & -39.0 & -40.7 & -42.7 & -44.8 \\ 
& 3l & -36.6 & -37.9 & -39.6 & -41.6 & -43.7 \\ 
& 4i & -35.0 & -36.2 & -37.9 & -39.9 & -42.0 \\ 
& 2o & -34.8 & -36.2 & -37.8 & -39.8 & -41.9 \\ 
& 1t & -34.1 & -35.4 & -37.0 & -39.0 & -41.1 \\ 
& 5g & -33.8 & -35.1 & -36.7 & -38.7 & -40.7 \\ 
& 6d & -33.1 & -34.4 & -36.0 & -37.9 & -40.0 \\ 
& 7s & -32.7 & -34.0 & -35.6 & -37.5 & -39.6 \\ 
& 3m & -32.4 & -33.7 & -35.3 & -37.2 & -39.3 \\ 
& 2q & -30.6 & -31.9 & -33.5 & -35.5 & -37.5 \\ 
& 4k & -30.6 & -31.8 & -33.5 & -35.4 & -37.4 \\ 
& 1u & -29.8 & -31.0 & -32.7 & -34.6 & -36.7 \\ 
& 5h & -29.3 & -30.5 & -32.2 & -34.0 & -36.1 \\ 
& 6f & -28.5 & -29.7 & -31.3 & -33.2 & -35.2 \\ 
& 3n & -28.0 & -29.3 & -30.9 & -32.8 & -34.8 \\ 
& 7p & -28.0 & -29.2 & -30.8 & -32.7 & -34.7 \\ 
& 2r & -26.2 & -27.5 & -29.1 & -31.0 & -33.1 \\ 
& 4l & -26.1 & -27.4 & -29.0 & -30.8 & -32.8 \\ 
& 1v & -25.3 & -26.6 & -28.3 & -30.1 & -32.2 \\ 
& 5i & -24.8 & -26.0 & -27.5 & -29.4 & -31.4 \\ 
& 6g & -23.8 & -25.0 & -26.6 & -28.4 & -30.3 \\ 
& 3o & -23.6 & -24.8 & -26.4 & -28.3 & -30.3 \\ 
& 7d & -23.3 & -24.4 & -26.0 & -27.8 & -29.7 \\ 
& 8s & -22.9 & -24.1 & -25.6 & -27.4 & -29.4 \\ 
& 2t & -21.8 & -23.0 & -24.6 & -26.4 & -28.5 \\ 
& 4m & -21.6 & -22.8 & -24.4 & -26.2 & -28.2 \\ 
& 1w & -20.8 & -22.0 & -23.6 & -25.5 & -27.5 \\ 
& 5k & -20.2 & -21.4 & -22.9 & -24.7 & -26.6 \\ 
& 3q & -19.1 & -20.3 & -21.8 & -23.6 & -25.6 \\ 
& 6h & -19.2 & -20.4 & -21.9 & -23.6 & -25.5 \\ 
& 7f & -18.6 & -19.7 & -21.2 & -22.9 & -24.8 \\ 
& 8p & -18.2 & -19.4 & -20.8 & -22.5 & -24.4 \\ 
& 2u & -17.2 & -18.4 & -20.0 & -21.8 & -23.7 \\ 
& 4n & -17.1 & -18.3 & -19.8 & -21.5 & -23.4 \\ 
& 1x & -16.1 & -17.3 & -18.9 & -20.8 & -22.8 \\ 
& 5l & -15.7 & -16.8 & -18.3 & -20.0 & -21.8 \\ 
& 3r & -14.5 & -15.7 & -17.2 & -18.9 & -20.8 \\ 
& 6i & -14.6 & -15.7 & -17.2 & -18.8 & -20.7 \\ 
& 7g & -14.0 & -15.0 & -16.4 & -18.1 & -19.9 \\ 
& 8d & -13.6 & -14.6 & -16.0 & -17.6 & -19.4 \\ 
& 9s & -13.3 & -14.4 & -15.8 & -17.4 & -19.1 \\ 
& 2v & -12.5 & -13.7 & -15.2 & -17.0 & -18.9 \\ 
& 4o & -12.5 & -13.7 & -15.1 & -16.8 & -18.7 \\ 
& 1y & -11.3 & -12.5 & -14.1 & -15.9 & -17.9 \\ 
& 5m & -11.1 & -12.2 & -13.6 & -15.3 & -17.1 \\ 
& 3t & -9.9 & -11.0 & -12.5 & -14.2 & -16.0 \\ 
& 6k & -10.2 & -11.2 & -12.6 & -14.1 & -15.9 \\ 
& 7h & -9.5 & -10.5 & -11.8 & -13.4 & -15.1 \\ 
& 8f & -9.1 & -10.1 & -11.4 & -12.9 & -14.6 \\ 
& 9p & -8.9 & -9.9 & -11.2 & -12.7 & -14.3 \\ 
& 2w & -7.8 & -8.9 & -10.4 & -12.1 & -14.0 \\ 
& 4q & -8.0 & -9.1 & -10.5 & -12.1 & -13.9 \\ 
& 1z & -6.4 & -7.6 & -9.1 & -10.9 & -12.8 \\ 
& 5n & -6.7 & -7.7 & -9.0 & -10.6 & -12.3 \\ 
& 6l & -5.8 & -6.8 & -8.1 & -9.5 & -11.2 \\ 
& 3u & -5.2 & -6.3 & -7.7 & -9.4 & -11.2 \\ 
& 7i & -5.3 & -6.2 & -7.4 & -8.9 & -10.4 \\
& 8g & -5.0 & -5.9 & -7.0 & -8.4 & -9.9 \\ 
& 9d & -4.9 & -5.7 & -6.8 & -8.2 & -9.7 \\ 
& 10s & -4.8 & -5.6 & -6.7 & -8.0 & -9.5 \\ 
& 4r & -3.5 & -4.5 & -5.8 & -7.4 & -9.1 \\ 
& 2x & -3.0 & -4.1 & -5.5 & -7.2 & -9.0 \\ 
& 5o & -2.4 & -3.3 & -4.5 & -6.0 & -7.6 \\ 
& 6m & -1.8 & -2.6 & -3.7 & -5.1 & -6.6 \\ 
& 3v & -0.6 & -1.6 & -3.0 & -4.5 & -6.3 \\ 
& 7k & -1.5 & -2.3 & -3.3 & -4.6 & -6.0 \\ 
& 8h & -1.4 & -2.1 & -3.1 & -4.3 & -5.6 \\ 
& 9f & -1.5 & -2.1 & -3.0 & -4.1 & -5.4 \\ 
& 10p & -1.6 & -2.2 & -3.0 & -4.1 & -5.3 \\ 
& 4t & n & n & -1.2 & -2.7 & -4.3 \\ 
& 2y & n & n & -0.6 & -2.2 & -4.0 \\ 
& 5q & n & n & -0.2 & -1.5 & -3.0 \\ 
& 6n & n & n & n & -0.9 & -2.2 \\ 
& 7l & n & n & n & -0.6 & -1.9 \\ 
& 10d & n & n & -0.3 & -0.9 & -1.8 \\ 
& 11s & n & n & -0.4 & -1.0 & -1.8 \\ 
& 9g & n & n & n & -0.8 & -1.7 \\ 
& 8i & n & n & n & -0.7 & -1.7 \\ 
& 3w & n & n & n & n & -1.4 \\
\end{longtable}

\newpage

\begin{longtable}{ll|r|r|r|r|r}
  \caption{\label{tab:etab-Pb208-kge} $^{208}_{\eta_{b}}\text{Pb}$ bound
    state energies. When $|E| < 10^{-1}$ MeV we consider there is no bound
    state, which we denote with ``n''. All dimensioned quantities are in
    MeV.} \\
  \hline \hline
  & & \multicolumn{5}{c}{Bound state energies} \\
  \hline
& $n\ell$ & $\Lambda_{B}=2000$ & $\Lambda_{B}=3000$ & $\Lambda_{B}= 4000$ &
  $\Lambda_{B}= 5000$ & $\Lambda_{B}= 6000$ \\
\hline
\endfirsthead
\multicolumn{7}{c}%
{\tablename\ \thetable\ -- \textit{Continued from previous page}} \\
\hline
& $n\ell$ & $\Lambda_{B}=2000$ & $\Lambda_{B}=3000$ & $\Lambda_{B}= 4000$ &
  $\Lambda_{B}= 5000$ & $\Lambda_{B}= 6000$ \\
\hline
\endhead
\hline \multicolumn{6}{r}{\textit{Continued on next page}} \\
\endfoot
\hline \hline
\endlastfoot
$^{208}_{\eta_b}\text{Pb}$
& 1s & -74.7 & -76.2 & -78.1 & -80.3 & -82.6 \\ 
& 1p & -74.2 & -75.7 & -77.5 & -79.7 & -82.1 \\ 
& 1d & -73.2 & -74.7 & -76.6 & -78.8 & -81.1 \\ 
& 2s & -72.7 & -74.1 & -76.0 & -78.2 & -80.5 \\ 
& 1f & -72.1 & -73.6 & -75.5 & -77.6 & -80.0 \\ 
& 2p & -71.5 & -73.0 & -74.9 & -77.1 & -79.4 \\ 
& 1g & -70.9 & -72.3 & -74.2 & -76.4 & -78.7 \\ 
& 2d & -70.1 & -71.5 & -73.4 & -75.6 & -77.9 \\ 
& 3s & -69.6 & -71.0 & -72.9 & -75.0 & -77.3 \\ 
& 1h & -69.5 & -71.0 & -72.8 & -75.0 & -77.3 \\ 
& 2f & -68.5 & -70.0 & -71.8 & -74.0 & -76.3 \\ 
& 1i & -68.0 & -69.5 & -71.3 & -73.5 & -75.8 \\ 
& 3p & -68.0 & -69.5 & -71.3 & -73.5 & -75.8 \\ 
& 2g & -66.8 & -68.2 & -70.1 & -72.2 & -74.6 \\ 
& 1k & -66.4 & -67.8 & -69.7 & -71.8 & -74.1 \\ 
& 3d & -66.1 & -67.6 & -69.4 & -71.5 & -73.9 \\ 
& 4s & -65.6 & -67.1 & -68.9 & -71.0 & -73.3 \\ 
& 2h & -65.0 & -66.4 & -68.3 & -70.4 & -72.7 \\ 
& 1l & -64.6 & -66.1 & -67.9 & -70.0 & -72.3 \\ 
& 3f & -64.1 & -65.5 & -67.4 & -69.5 & -71.8 \\ 
& 4p & -63.7 & -65.1 & -66.9 & -69.0 & -71.4 \\ 
& 2i & -63.0 & -64.5 & -66.3 & -68.4 & -70.7 \\ 
& 1m & -62.8 & -64.2 & -66.0 & -68.2 & -70.4 \\ 
& 3g & -62.0 & -63.4 & -65.3 & -67.4 & -69.7 \\ 
& 4d & -61.4 & -62.8 & -64.6 & -66.8 & -69.0 \\ 
& 2k & -61.0 & -62.4 & -64.3 & -66.4 & -68.7 \\ 
& 5s & -60.9 & -62.3 & -64.2 & -66.3 & -68.6 \\ 
& 1n & -60.8 & -62.2 & -64.0 & -66.2 & -68.4 \\ 
& 3h & -59.8 & -61.2 & -63.0 & -65.1 & -67.4 \\ 
& 4f & -59.0 & -60.4 & -62.3 & -64.4 & -66.6 \\ 
& 2l & -58.9 & -60.3 & -62.1 & -64.2 & -66.5 \\ 
& 1o & -58.7 & -60.1 & -61.9 & -64.0 & -66.3 \\ 
& 5p & -58.6 & -60.0 & -61.8 & -63.9 & -66.2 \\ 
& 3i & -57.5 & -58.9 & -60.7 & -62.8 & -65.1 \\ 
& 2m & -56.6 & -58.0 & -59.8 & -61.9 & -64.2 \\ 
& 4g & -56.6 & -58.0 & -59.8 & -61.9 & -64.1 \\ 
& 1q & -56.5 & -57.9 & -59.7 & -61.8 & -64.1 \\ 
& 5d & -56.0 & -57.4 & -59.2 & -61.3 & -63.5 \\ 
& 6s & -55.6 & -57.0 & -58.8 & -60.8 & -63.1 \\ 
& 3k & -55.1 & -56.5 & -58.3 & -60.4 & -62.6 \\ 
& 2n & -54.3 & -55.7 & -57.5 & -59.5 & -61.8 \\ 
& 1r & -54.2 & -55.6 & -57.4 & -59.5 & -61.8 \\ 
& 4h & -54.0 & -55.4 & -57.2 & -59.3 & -61.5 \\ 
& 5f & -53.3 & -54.7 & -56.5 & -58.6 & -60.8 \\ 
& 6p & -52.9 & -54.3 & -56.1 & -58.2 & -60.4 \\ 
& 3l & -52.6 & -54.0 & -55.8 & -57.8 & -60.1 \\ 
& 1t & -51.9 & -53.2 & -55.0 & -57.1 & -59.3 \\ 
& 2o & -51.8 & -53.2 & -55.0 & -57.1 & -59.3 \\ 
& 4i & -51.4 & -52.8 & -54.6 & -56.6 & -58.8 \\ 
& 5g & -50.5 & -51.9 & -53.7 & -55.8 & -58.0 \\ 
& 3m & -50.0 & -51.4 & -53.2 & -55.2 & -57.5 \\ 
& 6d & -50.0 & -51.4 & -53.2 & -55.2 & -57.4 \\ 
& 7s & -49.6 & -51.0 & -52.7 & -54.8 & -57.0 \\ 
& 1u & -49.4 & -50.8 & -52.5 & -54.6 & -56.8 \\ 
& 2q & -49.3 & -50.7 & -52.5 & -54.5 & -56.7 \\ 
& 4k & -48.7 & -50.0 & -51.8 & -53.9 & -56.1 \\ 
& 5h & -47.7 & -49.1 & -50.8 & -52.9 & -55.1 \\ 
& 3n & -47.4 & -48.7 & -50.5 & -52.5 & -54.7 \\ 
& 6f & -47.0 & -48.4 & -50.2 & -52.2 & -54.4 \\ 
& 1v & -46.8 & -48.2 & -49.9 & -52.0 & -54.2 \\ 
& 2r & -46.7 & -48.0 & -49.8 & -51.8 & -54.0 \\ 
& 7p & -46.7 & -48.1 & -49.8 & -51.8 & -54.0 \\ 
& 4l & -45.9 & -47.2 & -49.0 & -51.0 & -53.2 \\ 
& 5i & -44.8 & -46.1 & -47.9 & -49.9 & -52.1 \\ 
& 3o & -44.6 & -46.0 & -47.7 & -49.7 & -51.9 \\ 
& 1w & -44.1 & -45.5 & -47.2 & -49.3 & -51.5 \\ 
& 2t & -44.0 & -45.3 & -47.1 & -49.1 & -51.3 \\ 
& 6g & -44.0 & -45.4 & -47.1 & -49.1 & -51.3 \\ 
& 7d & -43.5 & -44.9 & -46.6 & -48.6 & -50.8 \\ 
& 8s & -43.2 & -44.5 & -46.2 & -48.2 & -50.4 \\ 
& 4m & -43.0 & -44.4 & -46.1 & -48.1 & -50.3 \\ 
& 3q & -41.8 & -43.1 & -44.9 & -46.9 & -49.0 \\ 
& 5k & -41.8 & -43.1 & -44.9 & -46.9 & -49.0 \\ 
& 1x & -41.4 & -42.7 & -44.5 & -46.5 & -48.7 \\ 
& 2u & -41.2 & -42.5 & -44.3 & -46.3 & -48.4 \\ 
& 6h & -40.9 & -42.2 & -44.0 & -46.0 & -48.1 \\ 
& 7f & -40.3 & -41.7 & -43.4 & -45.4 & -47.5 \\ 
& 4n & -40.1 & -41.4 & -43.1 & -45.1 & -47.3 \\ 
& 8p & -40.0 & -41.3 & -43.0 & -45.0 & -47.2 \\ 
& 3r & -38.9 & -40.2 & -41.9 & -43.9 & -46.1 \\ 
& 5l & -38.7 & -40.1 & -41.8 & -43.8 & -45.9 \\ 
& 1y & -38.5 & -39.9 & -41.6 & -43.6 & -45.8 \\ 
& 2v & -38.3 & -39.6 & -41.4 & -43.4 & -45.5 \\ 
& 6i & -37.8 & -39.1 & -40.8 & -42.7 & -44.9 \\ 
& 4o & -37.1 & -38.4 & -40.1 & -42.1 & -44.2 \\ 
& 7g & -37.1 & -38.4 & -40.1 & -42.0 & -44.2 \\ 
& 8d & -36.6 & -37.9 & -39.6 & -41.6 & -43.7 \\ 
& 9s & -36.2 & -37.5 & -39.2 & -41.2 & -43.3 \\ 
& 3t & -35.9 & -37.2 & -38.9 & -40.9 & -43.0 \\ 
& 1z & -35.6 & -36.9 & -38.6 & -40.6 & -42.8 \\ 
& 5m & -35.6 & -37.0 & -38.6 & -40.6 & -42.7 \\ 
& 2w & -35.4 & -36.7 & -38.4 & -40.4 & -42.5 \\ 
& 6k & -34.6 & -35.9 & -37.5 & -39.5 & -41.6 \\ 
& 4q & -34.0 & -35.3 & -37.0 & -38.9 & -41.0 \\ 
& 7h & -33.8 & -35.1 & -36.7 & -38.7 & -40.8 \\ 
& 8f & -33.2 & -34.5 & -36.2 & -38.1 & -40.2 \\ 
& 3u & -32.8 & -34.2 & -35.8 & -37.8 & -39.9 \\ 
& 9p & -32.9 & -34.2 & -35.9 & -37.8 & -39.9 \\ 
& 5n & -32.5 & -33.8 & -35.4 & -37.4 & -39.5 \\ 
& 2x & -32.3 & -33.6 & -35.3 & -37.3 & -39.4 \\ 
& 6l & -31.3 & -32.6 & -34.2 & -36.2 & -38.2 \\ 
& 4r & -30.8 & -32.1 & -33.8 & -35.7 & -37.8 \\ 
& 7i & -30.4 & -31.7 & -33.3 & -35.2 & -37.3 \\ 
& 3v & -29.7 & -31.0 & -32.7 & -34.6 & -36.7 \\ 
& 8g & -29.8 & -31.1 & -32.7 & -34.6 & -36.7 \\ 
& 9d & -29.4 & -30.7 & -32.3 & -34.2 & -36.2 \\ 
& 2y & -29.2 & -30.5 & -32.2 & -34.1 & -36.2 \\ 
& 5o & -29.3 & -30.5 & -32.2 & -34.1 & -36.2 \\ 
& 10s & -29.1 & -30.3 & -32.0 & -33.8 & -35.9 \\
& 6m & -28.0 & -29.3 & -30.9 & -32.8 & -34.8 \\ 
& 4t & -27.7 & -28.9 & -30.6 & -32.5 & -34.5 \\ 
& 7k & -27.0 & -28.3 & -29.9 & -31.8 & -33.8 \\ 
& 3w & -26.6 & -27.8 & -29.5 & -31.4 & -33.4 \\ 
& 8h & -26.3 & -27.6 & -29.2 & -31.1 & -33.1 \\ 
& 2z & -26.0 & -27.3 & -29.0 & -30.9 & -33.0 \\ 
& 5q & -26.0 & -27.2 & -28.9 & -30.8 & -32.8 \\ 
& 9f & -25.9 & -27.1 & -28.7 & -30.6 & -32.6 \\ 
& 10p & -25.6 & -26.9 & -28.4 & -30.3 & -32.3 \\ 
& 6n & -24.7 & -25.9 & -27.5 & -29.4 & -31.4 \\ 
& 4u & -24.4 & -25.7 & -27.3 & -29.2 & -31.2 \\ 
& 7l & -23.7 & -24.9 & -26.5 & -28.3 & -30.3 \\ 
& 3x & -23.3 & -24.6 & -26.2 & -28.1 & -30.1 \\ 
& 8i & -22.9 & -24.1 & -25.7 & -27.5 & -29.5 \\ 
& 5r & -22.7 & -23.9 & -25.5 & -27.4 & -29.4 \\ 
& 9g & -22.3 & -23.6 & -25.1 & -26.9 & -28.9 \\ 
& 10d & -22.0 & -23.2 & -24.8 & -26.6 & -28.5 \\ 
& 11s & -21.7 & -22.9 & -24.4 & -26.2 & -28.2 \\ 
& 6o & -21.3 & -22.5 & -24.1 & -25.9 & -27.9 \\ 
& 4v & -21.1 & -22.3 & -23.9 & -25.8 & -27.8 \\ 
& 7m & -20.2 & -21.4 & -23.0 & -24.8 & -26.8 \\ 
& 3y & -20.0 & -21.2 & -22.9 & -24.7 & -26.7 \\ 
& 5t & -19.3 & -20.6 & -22.1 & -23.9 & -25.9 \\ 
& 8k & -19.4 & -20.6 & -22.1 & -23.9 & -25.9 \\ 
& 9h & -18.8 & -20.0 & -21.5 & -23.3 & -25.2 \\ 
& 10f & -18.4 & -19.6 & -21.1 & -22.9 & -24.8 \\ 
& 11p & -18.2 & -19.4 & -20.9 & -22.6 & -24.6 \\ 
& 6q & -17.9 & -19.1 & -20.7 & -22.5 & -24.4 \\ 
& 4w & -17.8 & -19.0 & -20.5 & -22.4 & -24.4 \\ 
& 3z & -16.6 & -17.9 & -19.5 & -21.3 & -23.3 \\ 
& 7n & -16.8 & -18.0 & -19.5 & -21.3 & -23.2 \\ 
& 5u & -16.0 & -17.2 & -18.7 & -20.5 & -22.4 \\ 
& 8l & -16.0 & -17.1 & -18.6 & -20.4 & -22.3 \\ 
& 9i & -15.4 & -16.5 & -18.0 & -19.7 & -21.6 \\ 
& 10g & -14.9 & -16.0 & -17.5 & -19.2 & -21.1 \\ 
& 6r & -14.5 & -15.7 & -17.2 & -19.0 & -20.9 \\ 
& 4x & -14.4 & -15.6 & -17.1 & -18.9 & -20.9 \\ 
& 11d & -14.6 & -15.8 & -17.2 & -18.9 & -20.8 \\
& 12s & -14.4 & -15.5 & -16.9 & -18.6 & -20.4 \\
& 7o & -13.4 & -14.6 & -16.0 & -17.8 & -19.6 \\ 
& 5v & -12.6 & -13.7 & -15.2 & -17.0 & -18.9 \\ 
& 8m & -12.6 & -13.7 & -15.1 & -16.8 & -18.7 \\ 
& 9k & -11.9 & -13.0 & -14.4 & -16.1 & -17.9 \\ 
& 10h & -11.5 & -12.5 & -13.9 & -15.6 & -17.4 \\ 
& 4y & -11.0 & -12.1 & -13.6 & -15.4 & -17.3 \\ 
& 6t & -11.1 & -12.3 & -13.7 & -15.4 & -17.3 \\ 
& 11f & -11.2 & -12.2 & -13.6 & -15.2 & -17.0 \\
& 12p & -11.0 & -12.1 & -13.4 & -15.0 & -16.8 \\ 
& 7q & -10.0 & -11.1 & -12.6 & -14.2 & -16.0 \\ 
& 5w & -9.2 & -10.3 & -11.8 & -13.5 & -15.3 \\ 
& 8n & -9.2 & -10.3 & -11.6 & -13.3 & -15.1 \\ 
& 9l & -8.6 & -9.6 & -11.0 & -12.6 & -14.3 \\ 
& 10i & -8.1 & -9.1 & -10.5 & -12.0 & -13.7 \\ 
& 6u & -7.8 & -8.8 & -10.3 & -11.9 & -13.7 \\ 
& 4z & -7.5 & -8.6 & -10.1 & -11.8 & -13.7 \\ 
& 11g & -7.8 & -8.8 & -10.1 & -11.7 & -13.4 \\ 
& 12d & -7.7 & -8.6 & -9.9 & -11.4 & -13.1 \\ 
& 13s & -7.5 & -8.4 & -9.7 & -11.2 & -12.9 \\ 
& 7r & -6.7 & -7.7 & -9.1 & -10.7 & -12.5 \\ 
& 5x & -5.7 & -6.8 & -8.3 & -9.9 & -11.7 \\ 
& 8o & -5.9 & -6.9 & -8.2 & -9.8 & -11.5 \\ 
& 9m & -5.3 & -6.3 & -7.6 & -9.1 & -10.8 \\ 
& 10k & -5.0 & -5.9 & -7.1 & -8.6 & -10.2 \\ 
& 6v & -4.4 & -5.4 & -6.8 & -8.4 & -10.2 \\
& 11h & -4.7 & -5.6 & -6.8 & -8.2 & -9.8 \\
& 12f & -4.6 & -5.4 & -6.6 & -8.0 & -9.6 \\ 
& 13p & -4.5 & -5.3 & -6.5 & -7.9 & -9.4 \\ 
& 7t & -3.4 & -4.4 & -5.7 & -7.2 & -8.9 \\ 
& 5y & -2.3 & -3.4 & -4.8 & -6.4 & -8.1 \\ 
& 8q & -2.7 & -3.6 & -4.9 & -6.4 & -8.0 \\ 
& 9n & -2.3 & -3.1 & -4.3 & -5.7 & -7.3 \\ 
& 10l & -2.0 & -2.8 & -3.9 & -5.3 & -6.8 \\ 
& 6w & -1.1 & -2.1 & -3.4 & -4.9 & -6.6 \\ 
& 11i & -1.9 & -2.6 & -3.7 & -5.0 & -6.4 \\ 
& 12g & -1.8 & -2.6 & -3.6 & -4.8 & -6.2 \\ 
& 13d & -1.8 & -2.5 & -3.5 & -4.7 & -6.1 \\ 
& 14s & -1.7 & -2.4 & -3.3 & -4.5 & -5.9 \\ 
& 7u & -0.2 & -1.1 & -2.3 & -3.8 & -5.4 \\ 
& 8r & n & -0.5 & -1.7 & -3.0 & -4.6 \\ 
& 5z & n & n & -1.3 & -2.8 & -4.5 \\ 
& 9o & n & -0.2 & -1.2 & -2.5 & -3.9 \\
& 10m & n & n & -1.0 & -2.2 & -3.5 \\
& 11k & n & -0.1 & -0.9 & -2.0 & -3.3 \\
& 12h & n & -0.2 & -0.9 & -1.9 & -3.1 \\ 
& 6x & n & n & n & -1.4 & -3.1 \\ 
& 13f & n & -0.3 & -1.0 & -1.9 & -3.1 \\
& 14p & n & -0.4 & -1.0 & -1.9 & -3.0 \\
& 7v & n & n & n & n & -2.0 \\ 
& 8t & n & n & n & n & -1.2 \\ 
& 9q & n & n & n & n & -0.8 \\ 
& 14d & n & n & n & n & -0.7 \\
& 15s & n & n & n & n & -0.7 \\ 
& 13g & n & n & n & n & -0.6 \\ 
& 10n & n & n & n & n & -0.6 \\ 
& 12i & n & n & n & n & -0.5 \\ 
& 11l & n & n & n & n & -0.5
\end{longtable}

In the same way as for the $\Upsilon$ case, the values of the $\eta_b$-nucleus bound state energies 
depends on the cutoff parameter, because the depth of the $\eta_b$-nucleus potential 
gets higher with increasing $\Lambda_B$. And in this case, the biding is also stronger for heavier nuclei.
The general results suggests that $\eta_b$ must form bound states with all of the studied nuclei, and this is independent of the value of $\Lambda_B$. Even though, as explained in Sec.~\ref{Ch4Sc6}, the absence of the width in the calculations may prove it difficult to resolve the individual states, we plan to in the near future include the effects of the widths in the calculations and see how much it will impact our results. 

\chapter{Coupling and Form factor influence on the mass shift} 

\label{Chapter6} 

\thispagestyle{empty}

\noindent


\section{Heavy meson symmetry limit}
\label{Ch6Sc1}

\thispagestyle{myheadings}

\noindent

In the following, we consider a heavy quark (heavy meson) symmetry 
limit, by treating the $\Upsilon$ and $J/\Psi$, as well as the $\eta_b$ and 
$\eta_c$ mesons on the same footing, namely, to assign the 
same coupling constant value in the corresponding interaction vertices 
with $g_{J/\Psi DD}=g_{\eta_c DD^*}=7.64$  
used in Refs.~\cite{Krein:2010vp,Cobos-Martinez:2020ynh}. 
Furthermore, to compare with the $\eta_c$ mass shift given in Ref.~\cite{Cobos-Martinez:2020ynh} 
calculated by considering an SU(4) symmetry breaking by   
$g_{\eta_c DD^*}=(0.6/\sqrt{2})\,g_{J/\Psi DD} \simeq 0.424\,g_{J/\Psi DD}$, 
we also study the same case for the $\eta_b$ mass shift.

\begin{figure}[htb]
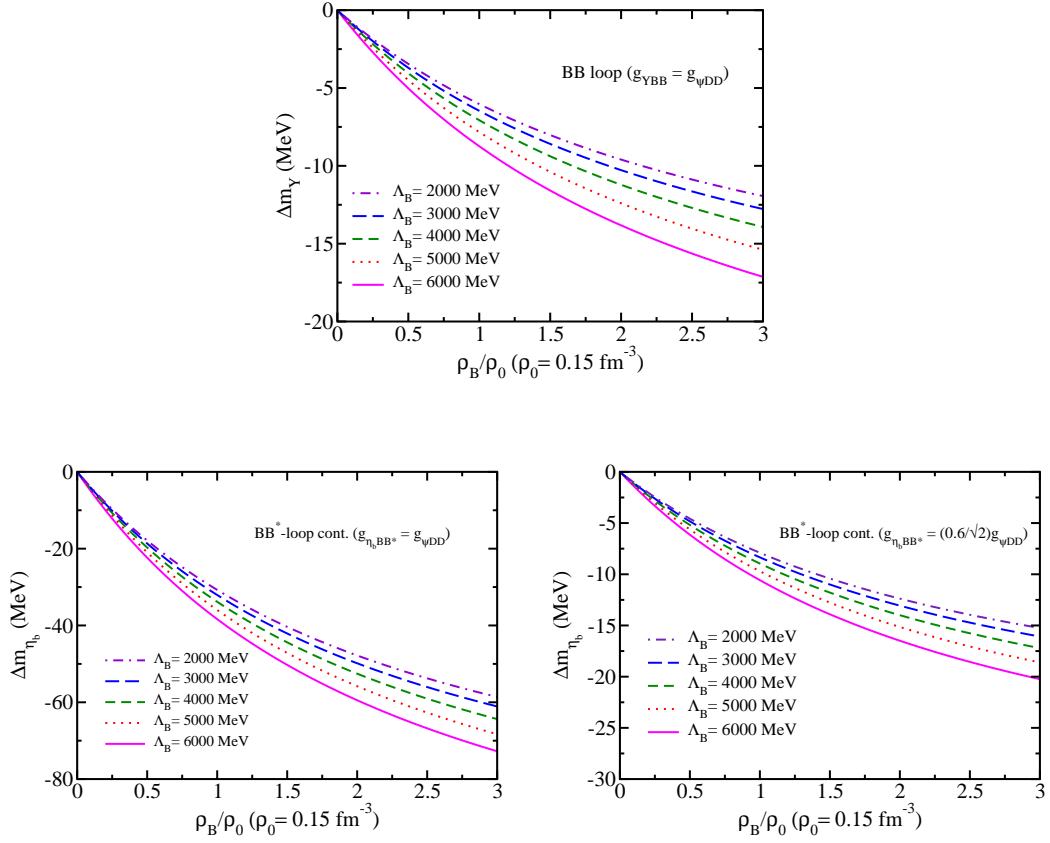
%
\centering
\includegraphics[width=6.5cm]{BB_gJpsi.eps}
\vspace{6ex}
\\
\includegraphics[width=6.5cm]{BBs_gJpsi.eps}
\hspace{2ex}
\includegraphics[width=6.5cm]{BBs_gJpsi_06sqr.eps}
\caption{$\Upsilon$ (top) and $\eta_b$ (bottom left) mass shifts 
calculated using the coupling constant relation 
$g_{\Upsilon BB}=g_{\eta_b BB^*}=g_{\eta_c DD^*}=g_{J/\Psi DD}$, and 
that of the $\eta_b$ calculated by a broken SU(5) symmetry, 
$g_{\eta_b BB^*}=(0.6/\sqrt{2})\,g_{\eta_c DD^*}=(0.6/\sqrt{2})\,g_{J/\Psi DD}$ (bottom right), 
with $g_{J/\Psi DD}=7.64$~\cite{Krein:2010vp,Cobos-Martinez:2020ynh}.
}
\label{figHQsym}
\end{figure}

In Fig.~\ref{figHQsym} we show the mass shifts calculated in the heavy quark (heavy meson) 
symmetry limit, and also the broken SU(5) symmetry in this limit 
for the $\eta_b$, namely,  
(i) $\Upsilon$ mass shift calculated by the coupling constant appearing in the 
self-energy by $g_{\Upsilon BB}=g_{J/\Psi DD}$ (top), 
(ii) $\eta_b$ mass shift by $g_{\eta_b BB^*}=g_{\eta_c D^*}=g_{J/\Psi DD}$ (bottom left), 
and (iii) $\eta_b$ mass shift with a broken SU(5) symmetry by  
$g_{\eta_b BB^*}=(0.6/\sqrt{2})\,g_{\eta_c DD^*}=(0.6/\sqrt{2})\,g_{J/\Psi DD}$ (bottom right), 
where we use $g_{J/\Psi DD}=7.64$~\cite{Krein:2010vp,Cobos-Martinez:2020ynh}.

Although the coupling constant value used for the bottom sector is now the same as that for 
the charm sector, since the relevant mesons in the bottom sector are heavier than 
those of the corresponding charm sector in free space as well as in medium, the 
dependence of the mass shifts on the cutoff mass value becomes more insensitive than 
that of the charm sector for the same range of 
the cutoff mass values. Thus, the ranges of the mass shifts of $\Upsilon$ and $\eta_b$ 
by the values of the cutoff mass becomes smaller than those of the corresponding 
$J/\Psi$ and $\eta_c$. 
In this limit, the obtained mass shifts range at $\rho_0$ corresponding to the cases 
stated above are,  
(i) -6 to -9 MeV (-5 to -21 MeV) for $\Upsilon$ ($J/\Psi$),  
(ii) -31 to -38 MeV (-49 to -87 MeV) for $\eta_b$ ($\eta_c$), 
and (iii) -8 to -11 MeV (-17 to -51 MeV) for $\eta_b$ ($\eta_c$), 
with the same range of the cutoff mass values $\Lambda_B$ in 2000 to 6000 MeV 
($\Lambda_D$ in 2000 to 6000 MeV).
These results indicate, as one can expect, the amounts of mass shifts for the 
$\Upsilon$ and $\eta_b$ become smaller than those of the corresponding $J/\Psi$ and $\eta_c$.
This fact confirms that the larger mass shifts of the $\Upsilon$ and $\eta_b$ 
than those of the $J/\Psi$ and $\eta_c$ obtained in previous sections are 
due to the larger coupling constant $g_{\Upsilon BB} = 13.2$ than   
$g_{J/\Psi DD} = 7.64$, where both values are obtained by the VMD model 
with the corresponding experimental data.
If this heavy quark (heavy meson) symmetry limit is more closely realized in nature, 
we expect to obtain smaller mass shifts for the $\Upsilon$ and $\eta_b$ than those of the 
corresponding $J/\Psi$ and $\eta_c$.

\section{Initial study of using a different form factor}
\label{Ch6Sc2}

\noindent

To see the effects of the form factor on the $\Upsilon$ and $\eta_b$ mass shifts, 
we calculate their mass shifts using a different form factor for  
the lowest order contributions, $BB$ and $BB^*$ loops, respectively (our predictions), 
as an initial study.
(We plan to perform an elaborate study for the effects of different form factors  
on the $\Upsilon$ and $\eta_b$ mass shifts.)
For this purpose we use the form factor~\cite{Tsushima:1991fe,Lin:2000ke,Lin:1999ad}, 
\begin{equation}
u_{B,B^{*}}(\textbf{q}^{2}) = \left(\frac{\Lambda^{2}_{\,\,B,B^{*}}}
{\Lambda^{2}_{\,\,B,B^{*}} + \textbf{q}^2}\right)^{2}, 
\label{newff}
\end{equation}
where the above $u_B$ and $u_{B^*}$ are applied in the same way as those already applied
for the corresponding vertices with $\Lambda_B=\Lambda_{B^*}$. 
The Fourier transform of the function $\Lambda_{\,B,B^*}^2/(\Lambda_{\,B,B^*}^2+{\bf q}^2)$  
in the form factor Eq.~(\ref{newff}), gives the Yukawa-potential type function, 
$\propto \exp(-\Lambda_{B,B^*}\,r)/r$, where $r$ is the distance of $B$ or $B^*$ meson 
from the $\Upsilon$ or $\eta_b$ meson for the corresponding vertices.
Derivative of the integrand $\Lambda_{\,B,B^*}^2/(\Lambda_{\,B,B^*}^2+{\bf q}^2)$ 
in the Fourier transform with respect to $\Lambda_{B,B^*}$ gives the 
dipole form Eq.~(\ref{newff}) aside the irrelevant constant, 
and the function $\propto \exp(-\Lambda_{B,B^*}\,r)$ structure  
remains and keeps controlling the same interaction range.
As is known, for $r > 1/\Lambda_B$, $\exp(-\Lambda_{B,B^*}\,r)$ 
suppresses effectively the interactions between the $\Upsilon$-$B$ and $\eta_b$-$B(B^*)$,  
where the $\Lambda_{\,B,B^*}$ dependence is expected to be more sensitive than 
the form factors in Eqs.~(\ref{ffups}) and~(\ref{eqn:FF}). 
Since the masses of $B$ and $B^*$ mesons are respectively $m_B = 5279 \simeq m_{B^*} = 5325$, 
we expect that the cutoff mass value $\Lambda_{B,B^*} \simeq 5300$ MeV may be a reasonable 
value for the form factor Eq.~(\ref{newff}). 
Thus, for this form factor, we take the cutoff mass central value $\Lambda_B=5300$ MeV, and 
calculate the $\Upsilon$ and $\eta_b$ mass shifts for the cutoff-mass value range, 
4900 MeV $\le \Lambda_B \le$ 5700 MeV.

\begin{figure}[htb]
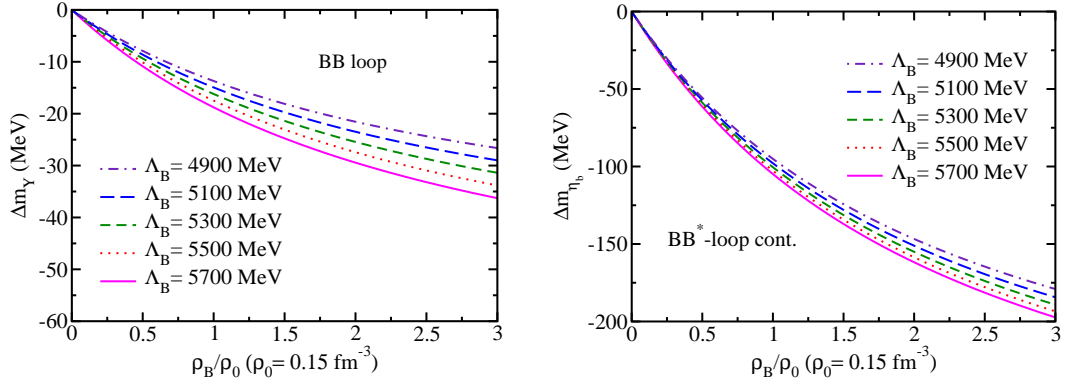
%
\vspace{1ex}
\centering
\includegraphics[width=6.5cm]{BB_dif_ff.eps}
\hspace{2ex}
\includegraphics[width=6.7cm]{BBs_dif_ff.eps}
\caption{$\Upsilon$ (left) and $\eta_b$ (right) mass shifts 
using a different form factor, 
$[\Lambda_{\,B,B^*}^2/(\Lambda_{\,B,B^*}^2+{\bf q}^2)]^2$ 
($\Lambda_B = \Lambda_{B^*}$),
with the cutoff-mass value range, 4900 MeV $\le \Lambda_B \le$ 5700 MeV.
}%
\label{figff}%
\end{figure}

The calculated mass shifts are shown in Fig.~\ref{figff}, 
for the $\Upsilon$ (left panel) and $\eta_b$ (right panel).
The results for the $\Upsilon$ and $\eta_b$ in Fig.~\ref{figff} may be 
compared with the corresponding results shown in Fig.~\ref{fig3} for 
the $\Upsilon$, and Fig.~\ref{fig7} the $\eta_b$, respectively.

The mass shifts at $\rho_0$ with the form factor Eq.~(\ref{newff}) are respectively, 
from -14 to -19 MeV for the $\Upsilon$, and from -95 to -104 MeV 
for the $\eta_b$ for the $\Lambda_B$ range, 4900 MeV $\le \Lambda_B \le$ 5700 MeV. 
The corresponding mass shifts at $\rho_0$ for the $\Upsilon$ with 
the form factor Eq.~(\ref{ffups}) and $\eta_b$ with the form factor Eq.~(\ref{eqn:FF}) are 
respectively, from -16 to -22 MeV, and -75 to -82 MeV for 
the cutoff mass range 2000 MeV $\le \Lambda_B \le$ 6000 MeV.
The different form factor given by Eq.~(\ref{newff}), which is more sensitive 
to the cutoff mass value, gives the similar mass shifts with those  
regarded as our predictions. 
The use of the form factors Eq.~(\ref{newff}) may give a better physical picture 
for the form factor.
The results shown in Fig.~\ref{figff}, especially for $\Upsilon$, 
have turned out to give very similar values 
with those corresponding ones obtained with the form factors  
Eqs.~(\ref{ffups}) but a wider range of the $\Lambda_B$, 
where the form factor Eq.~(\ref{newff}) can provide a better physical picture 
on the interaction range of the corresponding mesons.
As mentioned in Subsec.~\ref{ures}, 
we need, and plan to study further the effects of the other form factors, 
and/or the other regularization methods on the $\Upsilon$ and $\eta_b$ mass shifts, 
as well as on the $J/\Psi$ and $\eta_c$ mass shifts.


\chapter{Summary and Conclusions}
\label{Chapter7}
\thispagestyle{empty} 

\noindent

We have estimated for the first time the $B^*$, $\Upsilon$ and $\eta_b$ mass shifts 
in symmetric nuclear matter, as well as the $\Upsilon$-nucleus and $\eta_b$-nucleus
bound state energies, neglecting any possible widths of the mesons.

For the $\Upsilon$, we have studied the $BB$, $BB^{*}$, and 
$B^{*}B^{*}$ meson loop contributions 
using effective SU(5) symmetry-based Lagrangians and the anomalous coupling one, 
with coupling constants calculated from the experimental data using 
the vector meson dominance model. 
The in-medium $B$ and $B^*$ meson masses necessary to evaluate the 
$\Upsilon$ and $\eta_b$ self-energies in symmetric nuclear matter, are calculated by 
the quark-meson coupling model.
In considering the unexpectedly larger contribution 
from the heavier vector meson $B^*B^*$ meson loop contribution, 
and the similar fact for the $J/\Psi$ mass shift due to the $D^*D^*$ meson loop, 
we regard our prediction for the $\Upsilon$ mass shift 
as taking the minimum meson loop contribution, 
namely, that is estimated by taking only the $BB$ meson loop contribution,  
as was practiced similarly for the $J/\Psi$ mass shift taking  
only the $DD$ meson loop contribution.
Our prediction by this only $BB$-loop, gives the in-medium 
$\Upsilon$ mass shift 
that varies from -16 MeV to -22 MeV at the symmetric nuclear matter saturation density 
($\rho_0 = 0.15$ fm$^{-3}$) for the cutoff mass values in the range from 2000 MeV to 6000 MeV.

A detailed analysis is also made for the $\Upsilon$ self-energy meson loops 
due to the total ($BB+BB^*+B^*B^*$) contribution and the decomposition 
by comparing with the corresponding ($DD+DD^*+D^*D^*$) contribution and the decomposition  
for the $J/\Psi$ mass shift,  
focusing on the form factors in the interaction vertices   
using the correspondence between ($\Upsilon$ and $J/\Psi$), 
($B$ and $D$), and ($B^*$ and $D^*$) mesons. 
We have confirmed that, in the both cases of the $\Upsilon$ and $J/\Psi$ mass shifts, 
the heavier $B^*B^*$ and $D^*D^*$ 
meson loop contributions for the respective self-energies are larger than 
those of the corresponding lighter mesons, ($BB$ and $BB^*$) 
and ($DD$ and $DD^*$) meson loops, respectively. 
This fact suggests that our treatment for the vertices involving $B^*B^*$ mesons 
for the $\Upsilon$ self-energy, as well as the $D^*D^*$ mesons for the $J/\Psi$ self-energy,  
should be improved in treating the short distance fluctuations better.
Furthermore, we have chosen the same coupling constant value for 
$\Upsilon BB$, $\Upsilon BB^{*}$ and $\Upsilon B^{*}B^{*}$.
A more dedicated study on this will be carried out in the near future.

For the $\Upsilon$ meson produced in a large nucleus with a sufficiently low relative momentum 
to the nucleus, the mass shift obtained is enough to form the $\Upsilon$-nucleus bound states with 
the only-$BB$-loop-based mass shift (potential). 
The $\Upsilon$-nucleus bound state energies have been obtained for various nuclei by solving the
Klein-Gordon equation (originally the Proca equation), with the $\Upsilon$ nuclear potentials obtained using a local density approximation, and the the nuclear density distributions calculated within the 
quark-meson coupling (QMC) model.

Based on the detailed analysis on the $\Upsilon$ mass shift, 
we have also studied the $\eta_b$ mass shift on the same footing 
as that for the $\Upsilon$, based on an SU(5) effective Lagrangian. 
By this we have included only the $BB^*$ meson loop contribution for the $\eta_b$ 
self-energy as our prediction.
The obtained $\eta_b$ mass shift at symmetric nuclear matter saturation density 
ranges from -75 to -82 MeV for the same ranges of the cutoff mass values  
used for the $\Upsilon$ mass shift, from 2000 MeV to 6000 MeV.
For the $\eta_b B B^*$ coupling constant, we have used the SU(5) 
universal coupling constant determined by the $\Upsilon BB$ coupling constant   
by the vector meson dominance model with the experimental data.
The $\eta_b$-nucleus bound state energies were calculated in the same 
way as that for $\Upsilon$, with the results indicating that $\eta_b$ 
should form bound states with all of the nuclei studied.
The mass shifts and nuclear binding energies for the $\eta_b$ are larger than those of 
the $\Upsilon$, in our predictions of the present study.

We have also studied the $\Upsilon$ and $\eta_b$ mass shifts in a 
heavy quark (heavy meson) symmetry limit by calculating 
their mass shifts using the same coupling constant value   
with that for the corresponding $J/\Psi$ and $\eta_c$ mass shifts. 
For the $\eta_b$ mass shift, also a broken SU(5) symmetry from the $\Upsilon$ case 
has been studied within this limit. 
Our predictions for these cases at nuclear matter saturation density are, 
-6 to -9 MeV for $\Upsilon$, -31 to -38 MeV for $\eta_b$, 
and -8 to -11 MeV for $\eta_b$ with a broken SU(5) symmetry, where 
the corresponding mass shifts in the charm sector ones are, 
-5 to -21 for $J/\Psi$, -49 to -87 for $\eta_c$, and 
-17 to -51 for $\eta_c$ with a broken SU(4) symmetry. 
Thus, the bottomonium mass shifts are generally smaller than 
those of the corresponding charm sector in this limit, and the dependence on 
the cutoff mass value in the form factor is also smaller.
To see whether these or the other cases are closely realized in nature, 
further experiments are needed to get more information on the bottomonium-nucleon 
(bottomonium-(nuclear matter)) interactions as well as those for the charmonium.

For all cases of the predicted mass shifts for the $\Upsilon$ and $\eta_b$ mesons, 
the variations in the predicted values for a wide range of the cutoff mass values 
(from 2000 to 6000 MeV) in the corresponding form factors, are small --- less than 10 MeV, 
and this fact reduces some ambiguity in the predictions 
originating from the cutoff mass values.

In addition, we have also performed an initial study for the effects of a form factor 
on the lowest order $\Upsilon$ and $\eta_b$ mass shifts --- our predictions. 
The different form factor applied gives a clearer physics picture 
for the interaction ranges between the $\Upsilon$-$B$, $\eta_b$-$B$ and $\eta_b$-$B^*$. 
Using the cutoff mass values based on the physics picture of the form factor, 
the calculated $\Upsilon$ and $\eta_b$ mass shifts have turned out to give  
similar values with those for the predicted values of   
the $\Upsilon$ and $\eta_b$ mass shifts obtained using the original form factors.

Although the general results suggests that both $\Upsilon$ and $\eta_b$ shall form bound states with
all of the nuclei studied, the narrow gap between the values obtained for the bound state energies
may impose difficulties to the observability of the predicted bound states.
This has to do with the fact that we have ignored the possible widths of the mesons.
Then still remains to be tested the impact of the widths in our results.

In the future we plan to perform an elaborated study on the form factors appearing 
in the $\Upsilon$ and $\eta_b$ self-energy vertices, as well as 
those corresponding in the $J/\Psi$ and $\eta_c$. 
Furthermore, we plan to study the effect of the meson widths.


\begin{appendices}
\appendixpage
\noappendicestocpagenum
\addappheadtotoc

\chapter{Conventions}
\label{apx-conv}

\begin{enumerate}

\item A general four-vector is defined as the set of quantities

\begin{equation}
A^{\mu} = \left( A^{0}, A^{1}, A^{2}, A^{3} \right)
\equiv \left( A^{0}, \mathbf{A} \right).
\end{equation}

Then, the defition of the four-momentum is

\begin{equation}
P^{\mu} = \left( P^{0}, \mathbf{P} \right)
= \left( E, \mathbf{P} \right),
\end{equation}
where for a particle of rest mass $m$, $E = (\mathbf{P}^{2} + m^{2})^{1/2}$.

\item The metric tensors $g_{\mu \nu}$ and $g^{\mu \nu}$, responsible for
lowering and raising indices, are defined so that $g^{00} = g_{00} = 1$,
$g^{11} = g^{22} = g^{33} = g_{11} = g_{22} = g_{33} = -1$, with
all other components being zero.

So, if we apply $g_{\mu \nu}$ to $(A^{0}, A^{1}, A^{2}, A^{3})$, we have
$A_{0} = A^{0}$, $A_{1} = -A^{1}$ and so on. It is the same as to write
$A_{\mu} = g_{\mu \nu} A^{\nu}$.

\item The contravariant gamma matrices in the Dirac basis are

\begin{equation}
\gamma^{0} = 
\begin{pmatrix}
I  &  0\\
0  &  -I\\
\end{pmatrix},
\qquad
\gamma^{i} = 
\begin{pmatrix}
0  &  \sigma^{i}\\
-\sigma^{i}  &  0\\
\end{pmatrix},
\end{equation}
where $I$ is the identity matrix, $i = 1,2,3$, and $\sigma^{i}$ are 
the Pauli matrices

\begin{equation}
\sigma^{1} = 
\begin{pmatrix}
0  &  1\\
1  &  0\\
\end{pmatrix},
\quad
\sigma^{2} = 
\begin{pmatrix}
0  &  -i\\
i  &  0\\
\end{pmatrix},
\sigma^{3} = 
\begin{pmatrix}
1  &  0\\
0  &  -1\\
\end{pmatrix}.
\end{equation}

The defining property for the gamma matrices is the
anticommutation relation

\begin{equation}
{\gamma^{\mu}, \gamma^{\nu}} = \gamma^{\mu} \gamma^{\nu}
+ \gamma^{\nu} \gamma^{\mu} = 2g^{\mu \nu}.
\end{equation}

In addition, the fifth gamma matrix, $\gamma^{5}$, is given by

\begin{equation}
\gamma^{5} = i\gamma^{0} \gamma^{1} \gamma^{2} \gamma^{3} =
\begin{pmatrix}
0  &  I_{2}\\
I_{2}  &  0\\
\end{pmatrix}.
\end{equation}

\end{enumerate}

\chapter{The SU(3) group}
\label{apx-gluon}

The group SU(3) is formed by the set of unitary $3 \times 3$ matrices
with unitary determinant $\text{det} U = 1$.
The generators are the eight $3 \times 3$ matrices

\begin{equation}
\ T^{a}=\frac{ \lambda _{a}}{2},
\end{equation}
with $a=1, 2, \ldots , 8$, and $\lambda _{a}$ being the Gell-Mann matrices that form the matrix 
representation of the SU(3) group, in the same way as the Pauli matrices with respect to the 
group SU(2)

\begin{eqnarray}
\lambda _{1} =& \begin{pmatrix}
0  &  1  &  0\\
1  &  0  &  0\\
0  &  0  &  0\\
\end{pmatrix},
\qquad
\lambda _{2}= \begin{pmatrix}
0  &  -i  &  0\\
i  &  0  &  0\\
0  &  0  &  0\\
\end{pmatrix} \nonumber \\
\lambda _{3} =& \begin{pmatrix}
1  &  0  &  0\\
0  &  -1  &  0\\
0  &  0  &  0\\
\end{pmatrix}, 
\qquad
\lambda _{4}= \begin{pmatrix}
0  &  0  &  1\\
0  &  0  &  0\\
1  &  0  &  0\\
\end{pmatrix} \nonumber \\
\lambda _{5} =& \begin{pmatrix}
0  &  0  &  -i\\
0  &  0  &  0\\
i  &  0  &  0\\
\end{pmatrix},
\qquad
\lambda _{6}= \begin{pmatrix}
0  &  0  &  0\\
0  &  0  &  1\\
0  &  1  &  0\\
\end{pmatrix} \nonumber \\
\lambda _{7} =& \begin{pmatrix}
0  &  0  &  0\\
0  &  0  &  -i\\
0  &  i  &  0\\
\end{pmatrix},
\qquad
\lambda _{8}= \frac{1}{\sqrt{3}}\begin{pmatrix}
1  &  0  &  0\\
0  &  1  &  0\\
0  &  0  &  -2\\
\end{pmatrix}
\end{eqnarray}

The commutation relation 

\begin{equation}
\left[ T^{a},~T^{b} \right] =if^{abc}T^{c} 
\end{equation}
defines the structure constants $f^{abc}$, given by

\begin{eqnarray}
&f^{123} = 1, \nonumber \\
&f^{147} = -f^{156}=f^{246}=f^{257}=f^{345}=-f^{367}=\frac{1}{2} \nonumber \\
&f^{458} = f^{678}=\frac{\sqrt[]{3}}{2}
\end{eqnarray}  
with the rest being zero.

\chapter{Regge theory}
\label{apx-regge}

The Regge theory is a pre-QCD theory of the '60s that described nonperturbative hadronic processes.
The idea behind this theory is to treat the scattering process considering scattering amplitudes in 
the complex angular-momentum plane.

Considering a scattering between two particles, the partial-wave expansion of the
scattering amplitude in the s-channel is given by~\cite{Barone:2002cv}

\begin{equation}
T(s,t) = \sum_{l=0}^{\infty} (2l + 1) a_{l}(s) P_{l}(1 + 2s/t),
\end{equation}
where $s$ and $t$ are the Mandelstam variables, $l$ is the s-channel angular momentum, $P_{l}$ the Legendre polynomials, and $a_{l}$ are the expansion coefficients. Rewriting this as a contour integral in the complex $l$ plane, one obtains:

\begin{equation}
T(s,t) = \oint dl \frac{1}{sin \pi l} (2l + 1) a(l, s) P(l, 1 + 2s/t).
\end{equation}

The integration contour extends to infinity and circles the positive real axis clockwise. 
The argument l of a and P is no longer written as an index to indicate that it is a now complex variable and that
these functions have been analytically continued over the whole complex l plane.

In the high energy (Regge) limit, the integration contour is transformed so that the contour integral 
vanishes in this limit. The remaining residues of the poles are called Regge poles, and for each
pole there's a contribution 

\begin{equation}
T(s,t) \propto s^{\alpha (t)},
\end{equation}
where $\alpha (t)$ is the position of the pole in the complex $l$ plane, depending on $t$.
This is interpreted as an exchange of an object with angular momentum equal to $\alpha (t)$,
called ``Reggeon'', where $\alpha (t)$ is called its Regge trajectory.
In Regge theory, Regge trajectories are assumed to be

\begin{equation}
\label{reggetr}
\alpha (t) = \alpha (0) + \alpha ' t,
\end{equation}
with $\alpha (0)$ being called the Reggeon's Regge intercept, and $\alpha '$ its Regge slope.

\chapter{The Pomeron}
\label{apx-pomeron}

A Pomeron is a Reggeon (described by Regge theory), in the sense that it is an ``object''
which can be exchanged in an scattering processes.
It has its name after the Pomeranchuk theorem, which states that total cross sections of scattering
processes in which charged particles (``charged'' here reffers to any
type of charge, being either electric, color, etc.) are exchanged have to vanish asymptotically for high energy.
But in reality, these cross sections continue to rise slowly.
For this reason, Pomeranchuk postulated an object (not a particle), the Pomeron (PM),
which carries no charge, having the quantum numbers of the vacuum, to account for the 
continuing rise of total cross sections.

This account of the increasing total cross sections was made by introducing the Pomeron 
as an additional Regge trajectory~\cite{Jauch:1976ava}

\begin{equation}
\alpha (t) = 1 + \varepsilon + \alpha 't,
\end{equation}
and therefore, the Pomeron can be defined as a Reggeon with the intercept close to 1 
(cf. Eq.~\ref{reggetr}).

\chapter{Fock states}
\label{apx-fock}

Given an eigenvalue equation

\begin{equation}
\hat{n} \ket{\psi_{n}} = n \ket{\psi_{n}},
\end{equation}
where $n$ is the eigenvalue of $\hat{n}$ and $\ket{\psi_{n}}$ the corresponding eigenvector,
the complete set of orthogonal states may be written as

\begin{equation}
\braket{\psi_{n} | \hat{n} | \psi_{n}} = \braket{\psi_{n} | \hat{a}^{\dagger} \hat{a} | \psi_{n}}
= n \braket{\psi_{n} | \psi_{n}},
\end{equation}
with the number operator~\cite{Mandel:1995seg} $\hat{n} = \hat{a}^{\dagger} \hat{a}$ being Hermitian, and $\hat{a}^{\dagger}$ and $\hat{a}$ being the creation and annihilation operators.

Starting from the ground state $\ket{\psi_{0}}$,  all other states can be generated by successive
application of the creation operator $\ket{\psi_{n}} = (\hat{a}^{\dagger})^{n}\ket{\psi_{0}}$.
If we normalize these states according to $\ket{n} = \ket{\psi_{n}} / \braket{\psi_{n} | \psi_{n}}$,
we get the so called Fock states.
The $n$th Fock state is given by

\begin{equation}
\ket{n} = c_{n} (\hat{a}^{\dagger})^{n}\ket{0},
\end{equation}
where $c_{n}$ are the normalization constants.

\chapter{S-Matrix}
\label{apx-smat}

In a scattering process with given initial and final states, the asymptotic states
are defined by~\cite{Bjorken:1965sts}

\begin{eqnarray}
&\ket{i, \text{in}} \qquad \text{for}~ t \rightarrow -\infty \nonumber \\ 
&\ket{f, \text{out}} \qquad \text{for}~ t \rightarrow +\infty ,
\end{eqnarray}
where the ``in'' states are the ones created by creation operators evaluated at times
$-\infty$, and the ``out'' states those created by creation operators evaluated at times 
$+\infty$. The amplitude for the scattering involving those states is written

\begin{equation}
\braket{f,\text{out} | i, \text{in}}.
\end{equation}

The in and out states are assumed to be the same set of states but labeled differently (isomorphism),
which implies the existence of a unitary operator S called the S matrix, that can be defined such that

\begin{equation}
\ket{i, \text{in}} = S \ket{f, \text{out}},
\end{equation}
in a way that the amplitude can now be written in terms of the in and out states such as

\begin{equation}
\braket{f, \text{out} | i, \text{in}} = \braket{f, \text{in} |S| i, \text{in}} 
= \braket{f, \text{out} |S| i, \text{out}} \equiv \braket{f |S| i}.
\end{equation}

We say that S applies the out states on in states according to the above equation, and its matrix
elements are taken in the in space, the reason why the ``in'' index is dropped. The S operator
can be written as

\begin{equation}
S \equiv 1 + iT,
\end{equation}
where the identity in the first term is there to consider the possibility of no interaction 
(disconnected diagrams). But we are concerned with the part of the amplitude that do allow 
interactions, which is the second term.

The transition amplitude for a scattering of two particles into two other particles is then

\begin{equation}
\braket{p_{C} p_{D} | iT | p_{A} p_{B}} = \left( 2 \pi  \right) ^{4} \delta ^{4} \left( P_{A}^{'}+P_{B}^{'}-P_{A}-P_{B} \right) i{\cal M},
\end{equation}
where ${\cal M}$ is the Lorentz invariant transition amplitude with the overall momentum conservation 
delta function already factored out.

\chapter{Details on the nuclear potentials calculation}
\label{apx-nucp}

We can obtain the $\Upsilon$- and $\eta_b$-nuclear potentials for a given point $r$
inside a nucleus $A$ by fitting the cutoff dependent $\Upsilon$ or $\eta_b$ mass shifts to the
nucleus density profile data calculated within the QMC model, which gives us, for a given point of
$r$ (fm) from the center of the nucleus, the $\rho_{B}/\rho_{0}$ and $\rho_{B}$ for each nucleus $A$.

The fitting was done using an interpolation Fortran subroutine (SPLINE) available at~\cite{Forsythe:1997qy}.
For this routine we use the inputs $N$, which is the number of $r$ points for each nucleus corresponding
to the ``knots'' in the spline curve, the density $X = \rho_{B}/\rho_{0}$ associated with each value of 
the mass shift, which will be the abscissas in the curve, and $Y = \Delta m_{\Upsilon ,\eta_{b}}$ as
the ordinates of the knots.

The output is an array of the form ($r$,$V_A(r)$), giving us the value for the nuclear potential for each
$r$. We do this using the data for the mass shifts for each value of the cutoff $\Lambda_B$ to obtain
the potentials shown in the previous sections.
That said, this can also be done by using the Interpolation function in Wolfram Mathematica.

But since $V_A(r)$ needs to be in momentum space by going through a Fourier transform 
and then decomposed into partial waves, it's more practical to use an analytical expression for the potential,
instead of the numerical results we've obtained.

To find the analytical potential we use a Wolfram Mathematica program made by
Prof.~J.J.~Cobos-Mart\'{\i}nez, which find a fit of the Woods-Saxon type potential for $V_A(r)$.
The Woods-Saxon potential is of the form

\begin{equation}
V(r) = - \frac{V_{0}}{1 + exp \left( \frac{r - R}{a} \right)},
\end{equation}
where $V_{0}$, $R$ and $a$ are parameters to be determined by the program.
It uses the FindFit function to return these fitted parameters for the Woods-Saxon potential,
which will be the one used in the momentum space calculations for the bound states.

We are aware that this fitting may differ in some points from the original potential, specially for
lighter nuclei, since it is smoother. It remains to be checked how valid this approximation may be
for the nuclei considered in this study, which will be done in the near future.

\end{appendices}


\chapter*{International presentations at Conferences/Workshops}
\addcontentsline{toc}{chapter}{International presentations at Conferences/Workshops}

\thispagestyle{myheadings}

\noindent

\begin{enumerate}
\item (8-10 March, 2021) YITP international workshop on Hadron in Nucleus (HIN2020)
Panasonic hall in Yukawa Institute for Theoretical Physics, Kyoto University, Kyoto, Japan (Speaker).

\item (14-16 July, 2021) APCTP Focus Program in Nuclear Physics 2021, Seoul, Korea (Invited Speaker).

\item (26-31 July, 2021) 19th International Conference on Hadron Spectroscopy and Structure, Mexico City, Mexico (Speaker).

A contribution for the Proceedings can be found in [arXiv:2109.08636 [hep-ph]]. 

\item (5-10 September, 2021) 22nd Particles and Nuclei International Conference (PANIC2021)
Laboratory for Instrumentation and Experimental Particle Physics, University of Lisbon, 
Lisbon, Portugal (Speaker).
\end{enumerate}

\end{document}